\documentclass[twocolumn, pre]{revtex4}
\usepackage{bm,textcomp}
\usepackage{amsmath}
\usepackage{amsfonts}
\usepackage{amssymb}
\usepackage{graphicx}
\usepackage{subfig}
\usepackage{algorithm}
\usepackage{algorithmicx}
\usepackage{algpseudocode}
\usepackage{hyperref}
\usepackage{comment}

\newcommand{\mbeq}{\overset{!}{=}}
\newcommand{\umin}[1]{\underset{#1}{\operatorname{\bf{minimize}}}}
\newcommand{\umax}[1]{\underset{#1}{\operatorname{\bf{maximize}}}}
\newcommand{\uargmin}[1]{\underset{#1}{\operatorname{\bf{argmin}}}}

\newcommand{\prox}[1]{{\operatorname{\bf{prox}}}_{#1}}
\newcommand{\dotp}[2]{\langle#1,#2\rangle}
\newcommand{\norm}[1]{\left\lVert#1\right\rVert}
\newcommand{\qandq}{\quad\mbox{and}\quad}

\newcommand{\qwhereq}{\quad\mbox{where}\quad}
\newcommand{\enscond}[2]{\left\lbrace {#1}:{#2} \right\rbrace}

\newcommand{\bb}[1]{\mathbb{#1}}

\begin{document}

\setcounter{page}{1} 

\title{An Energy Based Scheme for Reconstruction of Piecewise Constant Signals observed in the Movement of Molecular Machines}

\author{Joachim Rosskopf$^{1}$ \textsuperscript{$\ddagger$}, 
        Korbinian Paul-Yuan$^{2}$  \textsuperscript{$\ddagger$}, 
        Martin B. Plenio$^{1}$, 
        Jens Michaelis$^{2}$}

\affiliation{$^{1}$ Institute of Theoretical Physics Ulm University, $^{2}$ Institute of Biophysics Ulm University}

\begin{abstract}%
{
Analyzing the physical and chemical properties of single DNA based molecular
machines such as polymerases and helicases requires to track stepping motion on
the length scale of base pairs. Although high resolution instruments have been
developed that are capable of reaching that limit, individual steps are oftentimes hidden by experimental noise which complicates data processing. 
Here, we present an effective two-step algorithm which detects steps in a high bandwidth
signal by minimizing an energy based model (Energy based step-finder, EBS).
First, an efficient convex denoising scheme is applied which allows compression
to tuples of amplitudes and
plateau lengths. Second, a combinatorial clustering algorithm formulated on a
graph is used to assign steps to the tuple data while accounting for prior information.

Performance of the algorithm was tested on Poissonian stepping data simulated
based on published kinetics data of RNA Polymerase II (Pol II). Comparison to
existing step-finding methods shows that EBS is superior in speed while 
providing competitive step detection results especially in challenging
situations.

Moreover, the capability to detect backtracked intervals in experimental data of
Pol II as well as to detect stepping behavior of the Phi29 DNA packaging motor
is demonstrated.  
}

{
  \textsuperscript{$\ddagger$} The authors contributed equally to this article.
}
\end{abstract}

\maketitle

\section*{Introduction}

Single molecule measurements of molecular motors allow to study the motion of
individual enzymes. The studies range from enzymes making comparably large steps
e.g. motor proteins like Myosin V \cite{myosinv} and Kinesin \cite{kinesinsteps}
to DNA based molecular machines which make steps on the scale of single
nucleotides \cite{bpsteprnapol, galburtbacktrack, smpolblock, jensphyslife}.
Experimental techniques to study these systems range from single molecule
fluorescence localization \cite{fiona} to optical and magnetic tweezers
\cite{magtwee}. Most of these measurements represent the underlying dynamics as
one-dimensional time series of positional changes. The enzymatic reactions which
fuel this motion appear as stochastic events resulting in step-like movements
\cite{motortheorist} obliterated by noise. Nowadays state of the art optical
tweezers experiments allow to study the movement of enzymes with a resolution
down to single base pairs \cite{bpsteprnapol, opttweereviewnew}.  For example,
studies on the $\varphi$29 bacteriophage ring ATPase
\cite{chemlaphi29, phagettest, phagedwells} used the information from step
detection data to propose a complete model of the mechanochemical cycle.
However, oftentimes analysis schemes rely on low pass smoothed data.

Indeed, the problem of finding steps is not only limited to studies of movement
of enzymes but appears in a wide range of biomolecular experiments from
fluorescence resonance energy transfer trajectories \cite{tjhahmm}, to steps in
membrane tether formation \cite{opferetal}, or the opening of ion channels
\cite{ionchannel}, just to name a few.

Consequently, there is a rich amount of signal processing techniques available
to recover piecewise constant signals from noisy data. Due to the stochastic
nature of enzymatic stepping the number of steps is often not known a priori.
Therefore, different step finding algorithms have been developed
\cite{stepdetcomp, milescuhmm, maxalittlegenmeth, maxlittlestepsbumps}.

One class of algorithms determine steps from single molecule data based on
statistical hypothesis testing in a moving window. A prominent example is the so
called t-test, which is based on the Student's t-test \cite{stepdetcomp}. In
this algorithm a step is recorded when the hypothesis that two normally
distributed random variables have the same mean is violated. The mean is
calculated with respect to a certain time window which is an input parameter
that can be eliminated by sweeping through various window sizes. Thus, the
t-test is conceptionally simple. However, for situations with small step-sizes
and as a result comparatively large noise, increasing window sizes are required,
limiting efficient step-detection.

Hidden Markov models (HMM) have been developed for situations with poor
signal-to-noise ratio \cite{mlesthmm, vshmm}. In HMM the signal is modelled as a
Markov process with transitions between discrete states obliterated by Gaussian
white noise. Thus, in the HMM analysis of stepping data, transition
probabilities of a Markov process are obtained from a maximum likelihood
estimation and the steps are reconstructed using the Viterbi algorithm
\cite{theviterbi}. A HMM for processive molecular motor data requires many
states to model the possible positions on the template, making it
computationally expensive. Performance can be improved by cutting the signal at
a predefined amplitude and transforming positions to periodic coordinates to
limit the necessary number of states \cite{vshmm}. HMMs proved to be excellent
tools for pattern recognition in many fields. However, in addition to being
computationally demanding they rely on assumptions about the hidden stepping
process and about the noise model.

Another popular class of step-finding algorithms reconstruct the underlying step
signal by successively introducing new steps until a stop-criterion is met
\cite{kerschi2, kvalgo}. One commonly used approach is developed by
Kalafut and Visscher (K \& V). It positions every new step such that the Bayesian
Information Criterion with respect to the noisy data is minimized \cite{kvalgo}.
It is a topic of current research, if this is a valid assumption for
change-point problems \cite{zhang07}. The algorithm does not require user input
and stops when the addition of new steps is unfavorable according to the
Bayesian Information Criterion.

The K \& V algorithm is a member of the larger class of step-finding methods
which minimize a certain energy function \cite{maxalittlegenmeth}. However,
since the K \& V algorithm only adds new steps and does not remove previously
found steps, it is not guaranteed that the global energy minimum is found
\cite{maxalittlegenmeth}. Finding the global minimum is possible, if these energy
functions are convex. In this case, efficient algorithms can be used  that yield
good approximations to the underlying step signal \cite{mlittlefastbayes}. However, for
poor signal to noise ratio, these convex energy functions are too simplistic to
optimally detect steps, resulting in an overfitting of the data, i.e. more
steps are detected than are actually present. Thus, if steps are hidden in noise
such algorithms behave as efficient filter-functions, and accurate
step-detection requires an additional second stage on the filtered, i.e.
denoised data \cite{maxlittlestepsbumps}.

Here, we present a novel two stage approach, termed Energy Based Stepfinding
(EBS), where both stages are based on the minimisation of energy functions.  In
a first stage, we denoise the signal  with a highly efficient and fast
optimization algorithm. The algorithm minimizes a convex energy function in a
process called total variation denoising (TVDN). We show that an optimal
denoising can be found making the process effectively parameter-free. There is
no further assumption about the noise necessary.  For actual step detection, we
proceed in the second stage of EBS with Combinatorial Clustering (CC) of the
denoised data into steps. Such an approach is already in common use in the
computer vision community \cite{Boykov01} and is both computationally efficient
and fast. The energy functions used in CC belong to a more general class which
allows the incorporation of prior knowledge such as the step size of the stepper
to make the algorithm more accurate. We tested EBS with simulated data that were
created based on experimental data of RNA polymerase II (Pol II) movement. We
compare the performance of EBS on the same simulated data to (i) a t-test, (ii)
to the variable stepsize HMM and (iii) to the K \& V algorithm.

The analysis reveals that EBS performs faster and more accurately. We
therefore applied the algorithm to detect steps in experimental data of the
bacteriophage $\varphi$29 packaging motor and to determine pauses of Pol II
transcription elongation in high resolution optical tweezers experiments.



\section*{Methods}

\subsection*{Energy Based Model}
\begin{figure*}
  \centering{
    \includegraphics[scale=.4]{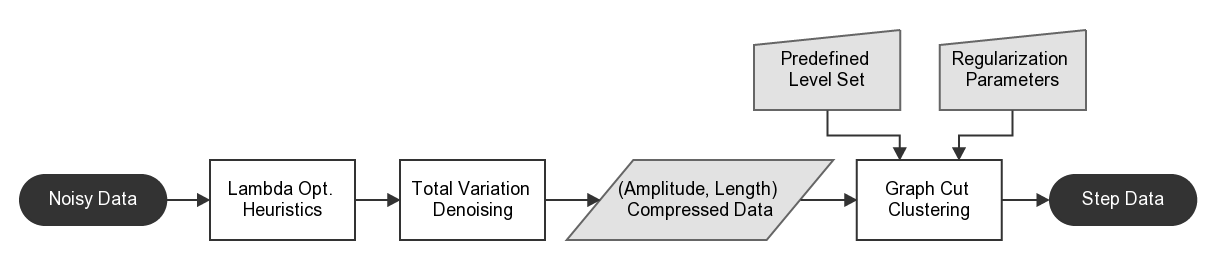}
  }
  \caption{Flowchart of the two stage process for finding steps in noisy stepping data. The input data is first denoised via solving the convex TVDN problem. This process requires no intervention, as the regularization parameter $\lambda$ is determined automatically. This results in  a lower dimensional, compressed representation of tuples (amplitude, length). This discrete data is then handed to a Graph Cut algorithm which solves a combinatorial clustering problem on a graph. The Graph Cut allows further customization by the use of regularization parameters $\rho_i$ and a pre-defined level set.}
   \label{fig:algorithmoverview}  
\end{figure*}

Starting from a large and noisy trajectory of motor protein movement we use an
energy based step detection (EBS). To reveal the
steps produced by the underlying biological system hidden in noise, one has to
identify piecewise constant parts in the data set. This is done by taking the
$N$-element noisy input data and creating an $N$-element output set of steps.
Therefore one needs to penalize variations within neighboring variables in the
signal. In contrast one needs to increase the energy if the free variables
deviate too much from the measured signal. This is reflected in the energy
function 
\begin{equation}
  E(\bm{x}, \bm{y}) = \sum_{i=1}^N D(|x_i - y_i|) + \sum_{i=1}^N S(|x_i - x_{i-1}|) 
  \label{eq:generalfunctional}
\end{equation}
where $\bm{y}$ and $\bm{x}$ are the $N$-element input data and output variable
vectors respectively. Minimizing the energy function is the conceptual baseline
of our approach. It consists of terms where variables interact with the input
data $D(\cdot)$, as well as nearest neighbor interaction between two adjacent
variables $S(\cdot)$.  Unfortunately, depending on the actual shape of these
terms the optimization problem can get prohibitively computationally expensive
\cite{Boykov01}.  One of the design goals of EBS was to work efficiently for
large data sets on commodity hardware. Therefore, we chose an approach 
which, in the first stage denoises and smoothens the signal by minimizing a
simple convex energy function, solving the TVDN problem. The result of this
stage is the set of denoised steps. Each step is characterized by its amplitude
and length. We call the combination of amplitude and length from now on a tuple.
The amount of tuples remains comparably low even for a significantly increased
sampling rate. This makes EBS well suited for high bandwidth data consisting of
a huge number of datapoints.  Afterwards we use this smaller set of tuples and
minimize a more sophisticated energy function in the CC stage, which is defined
on a discrete level set and incorporates a step height prior. A flowchart of
this two stage process is shown in figure
\ref{fig:algorithmoverview}. 

\subsection*{Total Variation Denoising}
\label{sec:convexdenoising}
In the first stage of EBS, we separate noise from the actual stepping signal.
This stage works on the full and noisy 1D input data set $\bm{y} = {y_1,...,y_N} \in
\bb{R}^N$, which can be quite large ($N > 10^7$ datapoints). To denoise we
minimize an energy function known as Total Variation Denoising (TVDN) problem
\cite{Rudin92}, 
\begin{equation} \begin{aligned} &\bm{x}^\star &= &\,\uargmin{x
      \in \bb{R}^N}\,q(\bm{x}, \bm{y}) + p(\bm{x}) \\ & &= &\,\uargmin{x \in
        \bb{R}^N}\, \frac{1}{2} \sum_{i=1}^N |x_i - y_i|^2 + \lambda
        \sum_{i=1}^{N-1}| x_{i}-x_{i+1}| \end{aligned} \label{eq:tvdnproblem}
\end{equation}
where the optimal solution $\bm{x}^\star$ represents the denoised signal. The
$\{y_i\}$ and $\{x_i\}$ are the $i$-th entry of the time-discrete input and 
solution vector, respectively. This optimal solution is a tradeoff between prior
knowledge that the enzymatic steps yield piecewise constant signals, which is
introduced by $p(\bm{x}) = \lambda \sum |x_{i}-x_{i+1}|$, a function which
penalizes introducing steps.  On the other hand the term  $q(\bm{x},
\bm{y})=\frac{1}{2} \sum |x_i - y_i|^2$ penalizes deviations of the resulting
solution from the input signal. The regularization parameter $\lambda$ is
important for the solution $\bm{x}^\star$ and controls the relative weight of
the two terms. The unique solution of this problem requires no assumption
about the characteristics of the noise. Therefore the denoising step works well
in case of Gaussian white noise as well as more complicated colored noise.

The energy function in Eq. \eqref{eq:tvdnproblem} is strictly convex, which means 
regardless of the input data $\bm{y}$ there exists one unique solution $\bm{x}^\star$ 
(see e.g. \cite{Boyd04}). We have applied a fast algorithm for solving the TVDN
problem (appendix) which can easily handle millions of data points
in a few milliseconds \cite{Condat13}. The algorithm scans forward through the
signal.  During this it tries to extend segments of the signal with the same
amplitude, until optimality conditions derived from the TVDN 
problem are violated. If this happens the method backtracks to a position 
where a new step can be introduced, re-validates the current segment 
until this position and starts a new segment (appendix).

An open problem in the context of TVDN for step detection is how to choose 
the regularization parameter $\lambda$ such that as few as possible true 
steps are lost (false negatives) but still the data is not overfitted 
(false positives). We propose a heuristic method to choose an optimal value for
$\lambda$, termed $\lambda_h$, automatically. To motivate these heuristics we
have a closer look to the two limits naturally imposed to $\lambda$.  For
$\lambda_{min} = 0$ the TVDN algorithm perfectly reproduces the 
input signal such that $\bm{x}^\star = \bm{y}$. On the other hand, the upper bound of 
sensible values is marked by $\lambda = \lambda_{max}$. Above this 
threshold the solution of Eq. \eqref{eq:tvdnproblem} is constant 
$x_i^\star = \mathrm{const}$ for all $i$. The value of 
$\lambda_{max}$ can be derived analytically from the underlying 
Fenchel-Rockafellar \cite{Rockafellar97} 
problem (appendix).

\begin{figure}
  \centering 
  \includegraphics[width=3.5in]{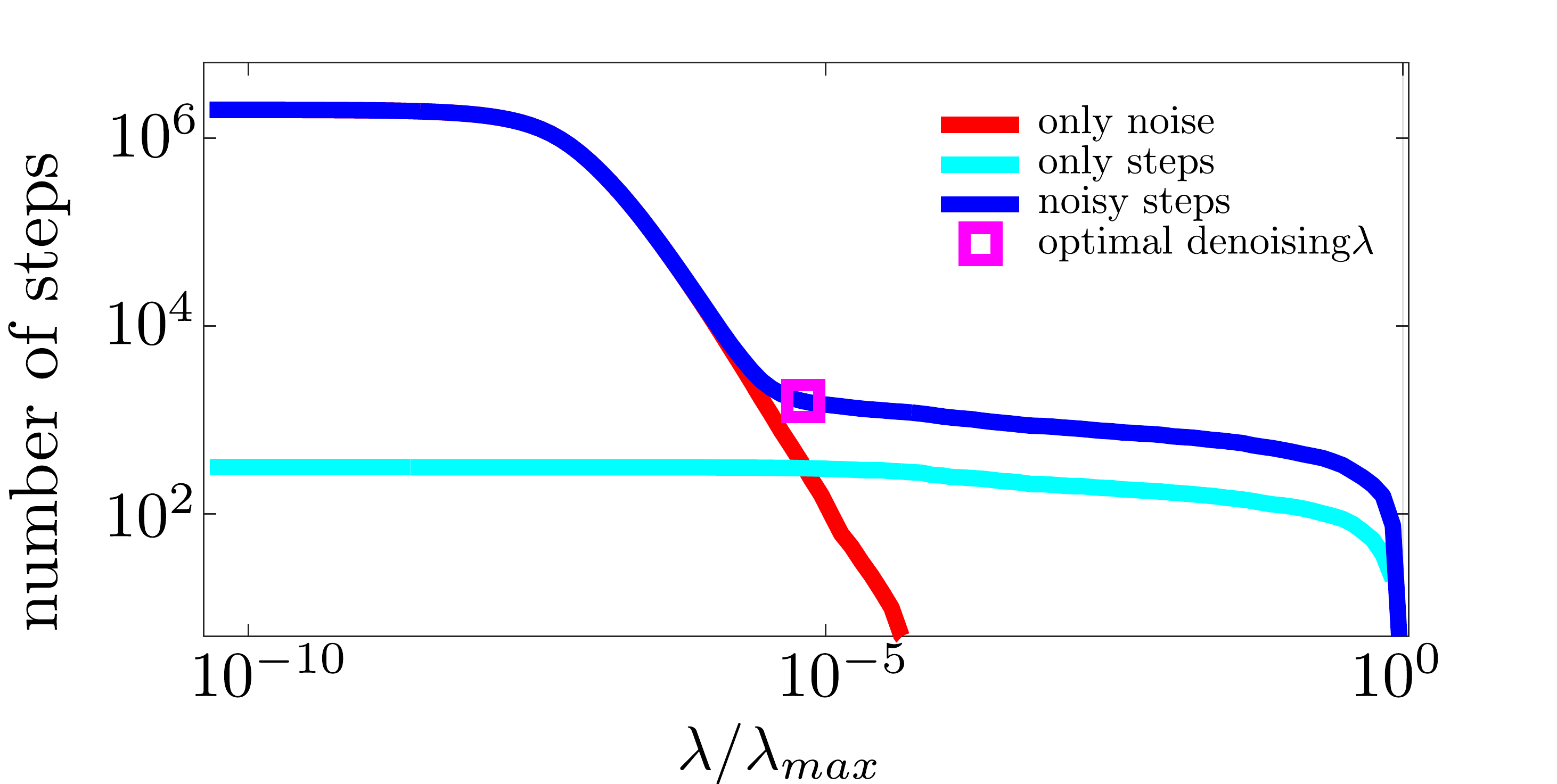}
  \caption{Breakdown of the TVDN model dependent on $\lambda / \lambda_{max}$.
  The number of produced steps and plateaus vs. different $\lambda$. For small
  $\lambda / \lambda_{max}$ solving the TVDN problem reproduces the input signal
  and the number of steps equals the number of data points. For big $\lambda /
  \lambda_{max}$ the number of steps is significantly lower. The point
  $\lambda_h / \lambda_{max}$ (magenta) before the number of steps increases suddenly  marks the value of the TVDN regularization parameter that we choose in our heuristic. Plotted is a constant signal with added Gaussian white noise (red), a signal with exponentially distributed dwell times and Gaussian white noise (blue), and the same signal without white noise (cyan).}
   \label{fig:tvdnheuristic}
\end{figure}

There exists a transition in TVDN while varying 
the regularization parameter from a stable minimization into the over fitting
regime. Thus, by lowering $\lambda$ from $\lambda_{max}$ to $\lambda_{min}$ one
observes a sudden increase of steps produced by TVDN (figure \ref{fig:tvdnheuristic}). 
This marks the point when the TVDN minimization breaks down and the solution
starts to fit noise. The breakdown also persists while varying sampling
frequency or rate of steps as well as signal to noise ratio (appendix).
To choose the optimal value of $\lambda$, $\lambda_h$, we use a
line-search algorithm which detects the sudden increase in the slope of the
number of steps in the resulting signal and uses a slightly larger value.
The sole input to this algorithm is the analytically determined value of
$\lambda_{max}$.  Therefore the $\lambda_h$-heuristic provides us with a stable
parameter-free means to choose an optimized TVDN regularization parameter.

\subsection*{Performance Characteristics of the Total Variation Denoising Algorithm}
Our implementation of 1D total variation denoising is based on the C code
published together with \cite{Condat13}. This publication also provides a detailed
outline of the TVDN algorithm, describtion of its working principles as well 
as the optimality condition it adheres to.

Typically TVDN is addressed by fixed-point methods
\cite{Vogel96}. These methods reach the minimal theoretically possible
algorithmic complexity \cite{Beck09}. A different kind of approach
\cite{Condat13} uses the local nature of the total variation denoising filter
and provides a very fast, memory efficient, non-iterative way to solve 
Eq. \eqref{eq:tvdnproblem}. Although the theoretical complexity of this 
algorithm is worse compared to fixed-point methods it actually achieves
competitive or even faster results on signals which exhibit piecewise constant
characteristics. For practical situation the complexity class of the algorithm
can be assumed to be $O(N)$. Thus for such signals denoising of $10^6$
datapoints takes around $30ms$ on a recent $2.5GHz$ processor.

After successful TVDN of the signal $\bm{x}^\star \in \bb{R}^N$ consists of $M$ steps.
A step is characterized by a discontinuity between two neighboring plateaus with
different  amplitude $a$. It is beneficial to represent the signal not in the
basis of indexed amplitudes $x^\star_i$, but instead to use tuples $(a, w)_j, j
\in {1, ..., M}$. Where $a_j$ is the amplitude and $w_j$ is the length of the
$j$-th plateau. By this change of representation the number of elements of the
data set is typically reduced from several millions to a few thousand.  This
increases computational efficiency due to the fact that the complexity of
following algorithms depends on the number of elements in the data set.
Therefore, a compressed signal consisting of tuples opens up the possibility to
apply sophisticated step-detection algorithms on the data. The problem can now
be cast as a Markov Random Field \cite{Li09} and can be tackled by a CC method
as will be presented in the following section.

\subsection*{Graph cut and $\alpha$-expansion used for combinatorial clustering to minimize energy functions}
\label{sec:cominatorialclustering}
As stated above, the input to the second stage of EBS are tuples of 
amplitude and corresponding length $(a, w)$ of the compressed signal. 
To reveal the actual steps, these tuples have to be clustered on a discrete set
of levels by minimizing an energy function. This means that a combinatorial
version of an energy function similar to Eq.\eqref{eq:generalfunctional} has to
be optimized.  The length of a plateau plays the role of a weighting factor
changing the contribution of a single tuple or a pair of tuples to the total
energy. With these modifications a general energy loss function takes the form,
\begin{equation}
\label{eq:combinatorialfunctional}
\begin{aligned}
&E(\bm{\xi}|\bm{(a,w)}) =  
	& &\sum_{i \in \mathcal{V}} \mathcal{Q}_i(\xi_i | a_i, w_i)  + \\ 
&	& &\sum_{(i,j) \in \mathcal{E}} \mathcal{P}_{i,i+1}(\xi_i,\xi_{i+1} | a_i, a_{i+1}, w_i, w_{i+1})
  \end{aligned}
\end{equation}
where the possible $\xi_i$ are taken from a set of levels
$\mathcal{L}$. The value of the
data term $\mathcal{Q}_i(\cdot)$ depends on deviations  of $\xi_{i}$ from the
input. The pairwise term $\mathcal{P}_{i,j}(\cdot, \cdot)$ encodes interaction
potentials between neighboring plateaus. Essentially the problem means to
cluster the tuples $(\bm{a}, \bm{w})$ to discrete levels, such that the joint
configuration $\bm{\xi}$ minimizes $E(\bm{\xi})$.

\begin{figure*}
  \centering 
  \subfloat[Build Graph Structure] {
    \includegraphics[width=2.3in]{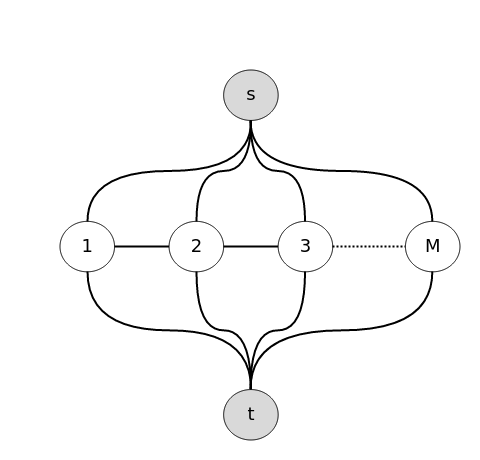} 
    \label{fig:gcproblem_structure} 
  } 
  \subfloat[Assign Capacities to Edges] {
    \includegraphics[width=2.3in]{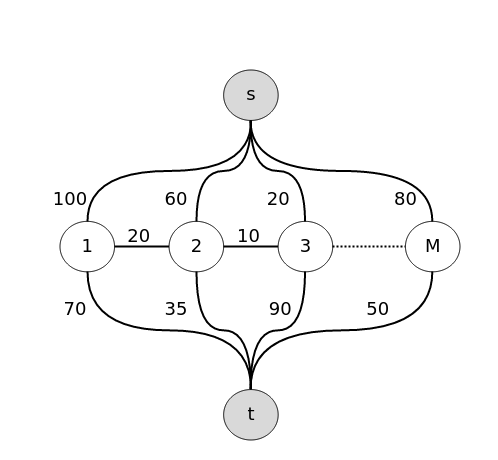} 
    \label{fig:gcproblem_capacities_edges} 
  }
  \subfloat[Find Optimal Cut] {
    \includegraphics[width=2.3in]{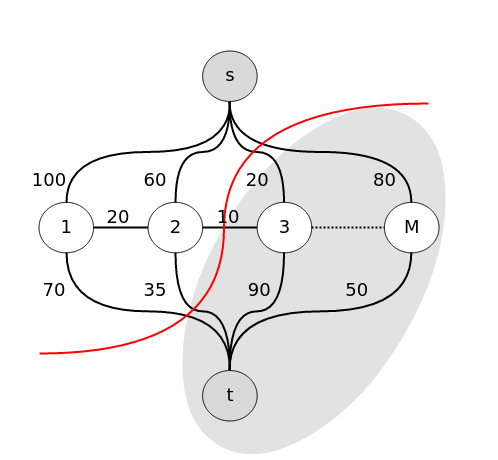} 
    \label{fig:gcproblem_optimal_cut} 
  }
    \caption{Cartoon representation of the graph cut algorithm.
      \ref{fig:gcproblem_structure}) the initial graph structure we use to model
      the step signal. The nodes $i \in \{1 \ldots M\}$ represent the variables
      $\xi_{i}$. \ref{fig:gcproblem_capacities_edges}) in a second step,
      energies are mapped to capacities of edges.
      \ref{fig:gcproblem_optimal_cut}) the
      Boykov-Kolmogorov Graph Cut algorithm \citep{Kolmogorov03} finds the max
      flow and cuts the graph into two subgraphs $\mathcal{G}=\mathcal{S} \cup
      \mathcal{T}$ where $\mathcal{S}$ is the part connected to $s$ and
      $\mathcal{T}$ is the remaining part connected to $t$.} 
    \label{fig:gcproblem} 
\end{figure*}

An elegant solution can be found by mapping the problem onto a graph
$\mathcal{G} = (\mathcal{V}, \mathcal{E})$, consisting of vertices $\mathcal{V}$
and edges $\mathcal{E}$. For the simple binary case, where the tuples have to be
assigned to only two levels, termed source $s$ and terminal $t$, both of these
levels as well as all tuples represent vertices  $\mathcal{V}$. $\mathcal{E}$
denotes the set of edges connecting the vertices (figure
\ref{fig:gcproblem_structure}) and each edge carries a capacity $c_{i} \geq 0$
(figure \ref{fig:gcproblem_capacities_edges}). Therefore there are two types of
edges, those connecting neighboring tuples and those edges connecting tuples to
levels. The capacities of the former are encoded in the pairwise term
$\mathcal{P}_{i,j}(\cdot, \cdot)$ and the latter are represented by the data
term $\mathcal{Q}_i(\cdot)$.
In the process of assigning a level $\xi_i$ to tuples the Graph Cut algorithm
solves the following binary decision problem: Is the assignment to level $t$
more favorable than assignment to level $s$ in terms of the energy function? In
the graphical representation this assignment is represented by a cut through
edges of neighboring tuples and edges between tuples and the $s$ and
$t$ level (figure \ref{fig:gcproblem_optimal_cut}).  

Due to the well known min cut/max flow theorem of graph theory the optimal
energy coincides with the smallest sum of capacities of the edges one has to cut
from the graph to disconnect $s$ from $t$ \cite{Papadimitriou98}. The cut splits
the graph $\mathcal{G}$ in two subgraphs: The part $\mathcal{S}$ which is
connected to the vertex $s$ and the part $\mathcal{T}$ which is connected to
$t$. The algorithm we apply solves this problem in polynomial time (appendix).

To make min cut/max flow useable for the above described assignment of multiple
different levels $\xi_i$ it has to be embedded into an outer procedure. For this
we use the $\alpha$-expansion algorithm \cite{Boykov01, Delong10}. It finds
provably good approximate solutions by iteratively solving Graph Cut problems on
graphs representing the binary decision whether to alter the previous assigned level
configuration or not \cite{Boykov01}. For a multi level problem new levels are added
successively in a random order. That means, once the graph has been optimized
for $i$ levels and the new $i+1$th level is introduced, $t$ corresponds to the
assignment to the predefined level set and $s$ to the new level. Again
capacities for all edges are computed. With the new graph cut, vertices in the
subgraph $\mathcal{S}$ get assigned their new level, the other vertices
connected to $\mathcal{T}$ keep their previously assigned level. After having
introduced all levels, in order to minimize the energy even more, the assignment
can be optimized by iteratively reintroducing the complete level set. This
iteration stops when the overall energy is not decreasing anymore (appendix).

In general finding the level configuration which coincides with minimal energy
requires at least nondeterministic polynomial time. Graph cut algorithms provide
the advantage to solve the problem in polynomial time, with the constraint to be
just applicable to energies which exhibit a strong local minimum
\cite{Kolmogorov07}. This is the case if the pairwise terms
$\mathcal{P}_{ij}$ of the energy function satisfy  
\begin{equation}
  \mathcal{P}_{ij}(\beta,
  \gamma) + \mathcal{P}_{ij}(\alpha, \alpha) \leq \mathcal{P}_{ij}(\beta,
  \alpha) + \mathcal{P}_{ij}(\alpha, \gamma) \label{eq:submodularity_condition}
\end{equation}
for arbitrary levels $\alpha, \beta, \gamma \in \mathcal{L}$. This is also known
as submodularity or regularity condition.\\ 

\subsection*{Energy function for step-finding}
\label{sec:energyfunc}

To perform CC we have to specify the energy function, Eq.(\ref{eq:combinatorialfunctional}) as well as the level grid. The levels can be chosen arbitrarily, and depending on the problem, provide an elegant way to introduce prior information. Often molecular motors move in discrete steps with known step-size. In this case, the spacing of the level grid can be chosen to match the known step-size. If such information is not known a priori or steps are expected to be nonuniform the levels have to be chosen with a refinement that corresponds to the required
numerical accuracy, i.e. with a sufficiently small spacing. 

To determine the relative importance of the terms $\mathcal{Q}$ and $\mathcal{P}$
in equation \ref{eq:combinatorialfunctional}, we introduce the parameters
$\rho_D$, $\rho_S$ and $\rho_P$ which regularize the detected steps.\\ 
The data terms $\mathcal{Q}_i$ penalizes deviations of the proposed
level amplitude $\xi_i$ to the original tuple amplitude $a_i$ at the vertex $v_i$
\begin{equation} \mathcal{Q}_i = \rho_D \cdot w_i \lvert \xi_i - a_i
  \rvert \end{equation}
where $\rho_D$ is a regularization parameter determining the importance of the
data term, and $w_i$ the weight of the current tuple. The most prominent plateaus are likely to be discovered by TVDN and contribute a tuple with a large weight. Thus, in order to preserve these plateaus the data term also depends on the weight $w_i$.

For the case of an equidistant level set $\mathcal{P}_{ij}$ consists of two
different terms, a smoothing term and a term that favors steps of a certain
size. The first and simpler pairwise energy uses a Potts Model \cite{Potts52} to
increase the energy whenever two assigned levels $\xi_i$ and
$\xi_{i+1}$ differ 
\begin{equation} 
  \mathcal{P}^{\mathrm{potts}}_{i,i+1} =
 \rho_S \cdot (w_i + w_{i+1}) (1 - \delta(\xi_i, \xi_{i+1}))
  \label{eq:smoothterm} 
\end{equation}
where $\delta(x, y) = 1$ if $x = y$ and $\delta(x, y) = 0$ else. Here $\rho_S$
is the smoothing parameter determining the energetic penalty for differing
adjacent levels. The Potts model satisfies the submodularity condition, Eq.
\eqref{eq:submodularity_condition},\cite{Kolmogorov07}. A larger regularization
parameter $\rho_S$ boosts clustering of the signal and therefore combines
steps. There is no other a-priori bias towards combining steps due to the CC 
algorithm itself.
\\ 

The second more sophisticated contribution to the pairwise term in
Eq.(\ref{eq:combinatorialfunctional}) favors level changes of specific size
between adjacent sites.

This second pairwise term is optional if step sizes are uniform and it serves
the purpose to introduce that prior information. Lowering the regularization
parameter $\rho_P$ gives rise to the introduction of new steps with a special
step height. The complete pairwise term $\mathcal{P}_{i,j}$ thus includes prior
information about step heights and is given by 
\begin{equation}
\begin{aligned}
&\mathcal{P}_{i,i+1} = 
	& &\mathcal{P}^{\mathrm{potts}}_{i,i+1} + \\
&	& &\rho_P (1 - \delta(\xi_i, \xi_{i+1})) (1 - \delta(|\xi_i - \xi_{i+1}|, \epsilon)),
	\label{eq:labelterm}
\end{aligned}
\end{equation}
with an expected average step height $\epsilon$ determined by the underlying
process. The depth of the jump height prior potential is given by the jump height parameter $\rho_P$.
In contrary to Eq. \eqref{eq:smoothterm} we chose this term to not depend on
the weights of the adjacent sites to regularize step sizes independently of the corresponding dwell times.

Note that not all pairwise terms constructed by Eq. \eqref{eq:labelterm} strictly 
fulfill the submodularity condition, Eq. \eqref{eq:submodularity_condition}.  
Therefore we applied an extension to the graph construction procedure proposed
in \cite{Rother05} to circumvent a submodularity violation (appendix).
The procedure truncates the energy until it satisfies
\eqref{eq:submodularity_condition}. The procedure is applicable to any energy
function and provides a provably good approximation for a single expansion
move. For the complete $\alpha$-expansion the procedure is applicable if most of
the terms are submodular \cite{Rother05}.  This is fulfilled by all signals
presented below: mean fraction of non-submodular terms $(0.27 \pm 0.08) \%$.




\subsection*{Simulation Method and Definition of Parameters for Algorithm Comparison}

In order to quantify positional and temporal accuracy of the steps detected by
EBS we use simulated data of noisy steps which are generated in a two stage
process. In a first stage we generate a piecewise constant signal according to
a simplified Pol II stepping model where a step is the product of an enzymatic
process with a certain net rate. This model contains an elongation state with
forward steps of $1 bp$ in size generated using an effective stepping rate
$k_{elong}$. We also account for backtracked states which can be entered by a
backward step of $1bp$  \cite{kashlrnap, blockbacktr} with a rate $k_{b,1}$. In
a backtracked state Pol II can step forward or backward by $1bp$ with the rates
$k_{f}$ or $k_{b}$ respectively.

Secondly, we simulate experimental noise including effects of confined Brownian
motion of trapped microspheres. To accurately reflect the experiment, we take
into account changes in tether length and tether stiffness due to the motion of
the enzyme. We apply a harmonic description of the trapping potentials and
assume that the DNA linker can be described by a spring constant $k_{DNA}$
determined by the worm like chain model (appendix).

In real experiments, the equilibrium position of the trapped microspheres is
influenced by drift which leads to colored noise characteristics on long
timescales. Sources of drift are, for example, pointing or power fluctuations
of the trapping laser or temperature drifts. To analyze the influence of drift
on the detected step signal we simulate drift as a confined Brownian motion
with a very slow time constant ($\sim10 min$) and a diffusion constant of $10
nt^2/s$. This represents a stochastically fluctuating base line which is added
to the simulated steps. Furthermore, the drift signal is assumed to be  small
enough not to affect kinetic parameters of the stepping simulation.  Using
these parameters, the simulation produces drifts of around $1nm$ on a timescale
of $\sim 1min$ (figure \ref{fig:exampledriftdata}) which can be even
outperformed by current high resolution instruments \cite{bpsteprnapol,
blockbacktr}.

We simulated a slow, an intermediate and a fast scenario with elongation rates
of $k_{elong} = 4.1 Hz$, $k_{elong} = 9.1 Hz$ and $k_{elong} = 25.1 Hz$,
respectively. For the slow scenario we generated $N=2.5 \cdot 10^{5}$ data
points with time increments corresponding to a $5kHz$ sampling frequency.
Simulated signals of the intermediate scenario consist of $N= 10^{5}$ data
points with $2kHz$ sampling frequency. The computed standard deviation in both
scenarios is $5.5 bp$ at the given sampling frequency. For the fast scenario we
chose  $N=5 \cdot 10^{4}$ data points and $1kHz$ sampling rate. Moreover in the
fast scenario we use higher noise amplitudes with a computed standard deviation
of  $10.0 bp$ at the $1kHz$ sampling frequency.

For EBS analysis of the noisy steps we have to choose the parameters $\rho_{D}$,
$\rho_{S}$ and $\rho_{P}$ as well as the level spacing for CC. Since our task is
to optimize Eq.(\ref{eq:combinatorialfunctional}) we are only interested in
relative values of the data and interaction function. Thus, we can arbitrarily set $\rho_{D} = 100$.
$\rho_{S}$ and $\rho_{P}$ are parameters that have to be defined by the user.
The smoothing parameter $\rho_{S}$ has to be large enough to cluster small
steps but small enough not to miss simulated steps. To this end, simulated data
can be used to optimize parameters such that as many steps as possible are
recovered but only few false positives are created (appendix). We
choose $\rho_{S}=2$, $\rho_{P}=50$ and use a level grid spacing of $1bp$ i.e.
the simulated step-size.

In order to compare different step-finding algorithms, we need to define a
criterion when a detected step occurs at the correct time-point. A detected step
is classified correct whenever its temporal position lies with $\pm
\Delta_{window}$ of the simulated step. We choose the window size
such that $\Delta_{window} = 1/(5\cdot k_{elong})$. This allows for a small
temporal shift of the detected steps with respect to a simulated step. The
window is small enough to minimize classification of a step as correct by chance
but large enough to make the step detection robust against numerical error.

The definition of correct steps is further used to introduce two quantities that
characterize step detection performance. The recall is defined by the
number of correct steps divided by the number of simulated steps and
provides information about the completeness of the  recovered steps. The
recall's value is only meaningful in combination with a second quantity called
precision. Precision is defined by the number of correct steps divided by the number of
detected steps, which is essentially the probability that a detected step is in
the above defined time interval around a simulated step.

\subsection*{Detecting backtracked regions}
In transcription elongation periods of forward motion are oftentimes interrupted
by backward steps. This so-called backtracking is important in-vivo for
regulating transcription and therefore it is desirable to accurately detect
backtracks in order to better understand regulation.  Dwell times between
detected steps are assigned to the set of backtracked states
when they lead to a backward step. A backtracked pausing interval ends at a
forward step that transfers Pol II back to the elongation state. At high noise
and for fast steps we do not expect that our method will perfectly find all
backtracking events present. For example, short backtracks can be omitted
resulting in a long dwell time between two forward transitions in the detected
steps. However, since the rates of backtracking are slow compared to elongation
rates, we can correct for the missed detection of a backward step by a
statistical hypothesis testing of dwell times, assuming that forward stepping
follows an exponential waiting time distribution (appendix). The
corresponding mean dwell time can be estimated from the dwell time
histogram of forward steps. Thus, dwell times which violate this hypothesis are
also considered as backtracked intervals, even if the actual backtracking step is not detected.

A typical method for this separation is a Savitzky-Golay filtered
velocity threshold pause detection (SGVT). SGVT finds backtracked regions in
Savitzky-Golay smoothed data from histograms of instantaneous velocities
\cite{ubiqupausing}. These histograms show a pause-peak around zero velocity and
an elongation-peak. One typically defines a velocity threshold by computing the
mean plus one (or two) standard deviation(s) of the pause-peak which is used to
characterize paused regions in transcription data. A sensible choice for typical
Pol II experiments of the Savitzky-Golay filter parameter is to use third order polynomials
and a frame size of 2.5s \cite{galburtbacktrack}. We will compare the performance of the SGVT algorithm to EBS in determining backtracks.



\newpage
\section*{Results \& Discussion}

\subsection*{Reliable implementation of EBS}
\begin{figure}
	\centering
  \subfloat[] {			
		    \includegraphics[width=1.75in]{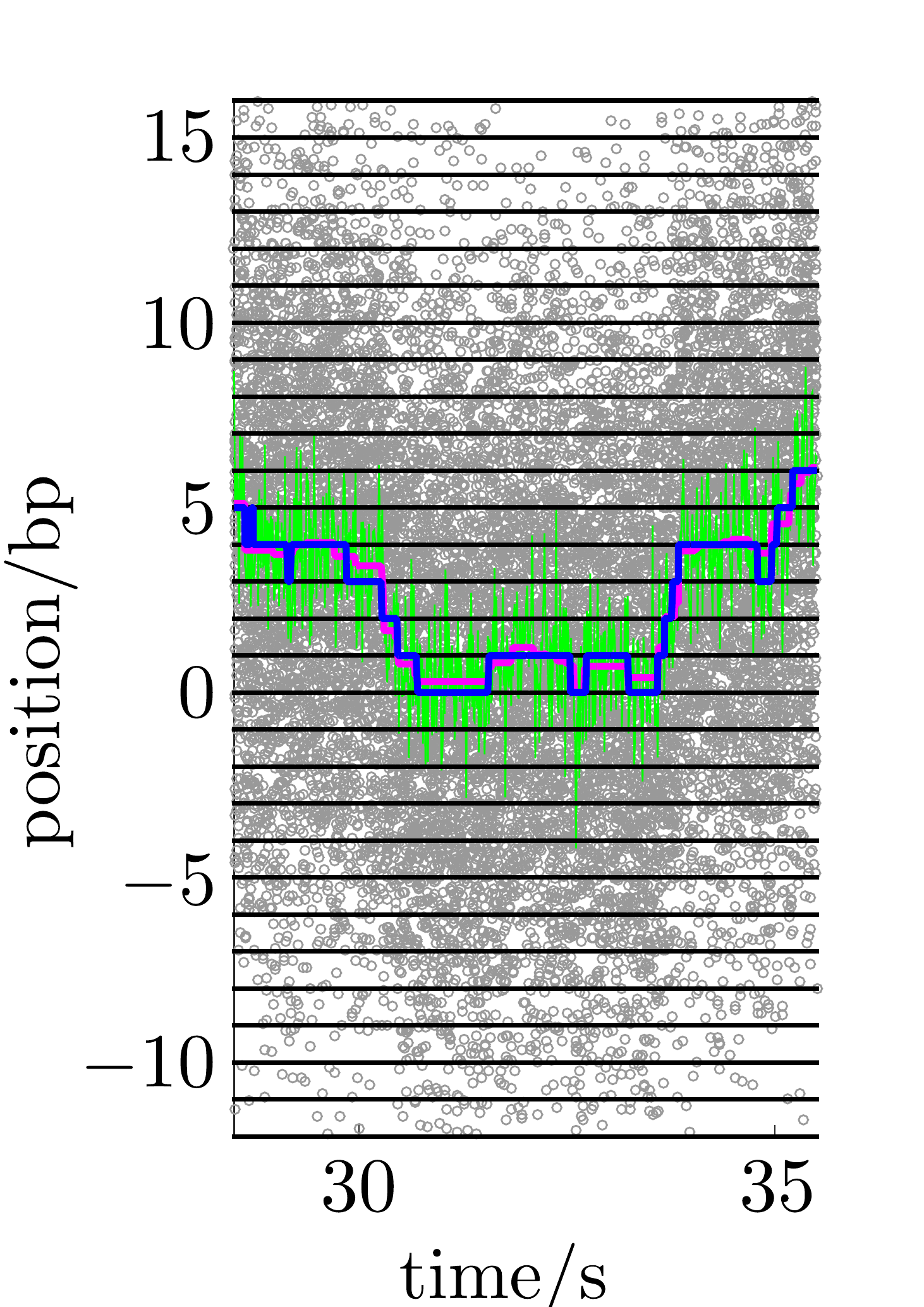}
		    \label{fig:tvdnstepsgcutresult}
  }
  \subfloat[] {
		    \includegraphics[width=1.75in]{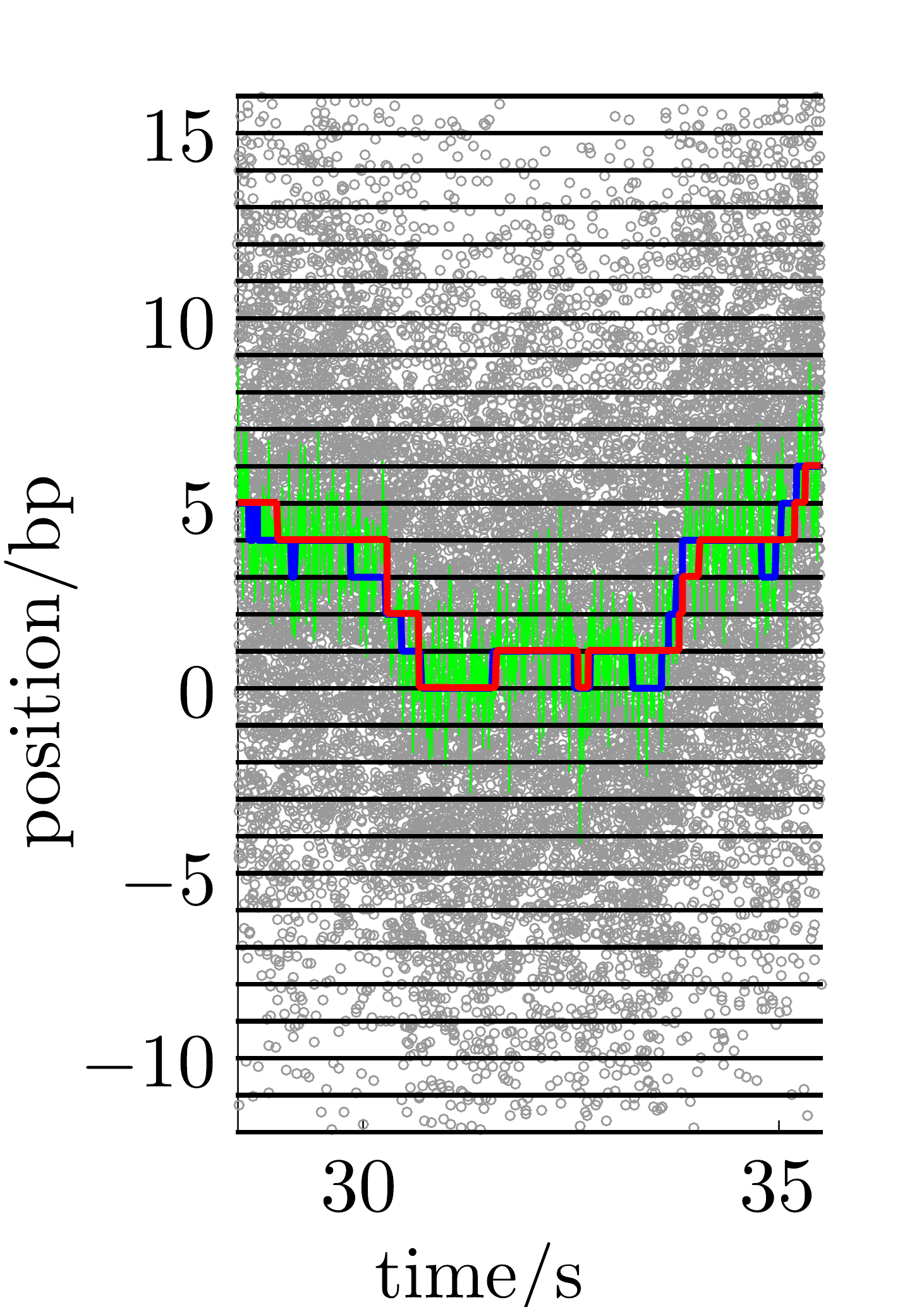}
			  \label{fig:ebsstepsgcutresult}
	}  
  \caption{EBS algorithm correctly detects steps in presence of high noise.
    \ref{fig:tvdnstepsgcutresult}, noise reduction after application of TVDN.
    \ref{fig:ebsstepsgcutresult}  step detection using combinatorial clustering.
    Shown is a zoomed in interval of the simulated noisy data (grey points),
    boxcar averaged noisy data (20 times reduced, green), simulated steps
    (blue), denoised signal from TVDN (magenta, \ref{fig:tvdnstepsgcutresult})
    and detected steps after application of combinatorial clustering by Graph
    Cut (red, \ref{fig:ebsstepsgcutresult}).}
  \label{fig:tvdngcutresult}
\end{figure}

We developed the EBS algorithm to determine steps in the trajectories of
molecular motors (the software package can be downloaded at
\url{https://github.com/qubit-ulm/pwcs}) and tested this algorithm on simulated
data of Pol II stepping using published rates (methods and appendix). We first
simulated data using the intermediate scenario (methods). We simulated a
trajectory of $50s$ (i.e. $10^5$ data points) resulting in $291$
steps. In our simulation noise amplitudes are much larger than the
$1bp$ steps of the simulated step signal (figure \ref{fig:tvdngcutresult}). TVDN
efficiently removes noise and produces a set of $587$ plateaus (figure
\ref{fig:tvdnstepsgcutresult}). The TVDN data approximates the simulated
step signal, but often decomposes a simulated step into several smaller steps.

How well a particular algorithm can detect steps is best tested by computing the
recall, i.e. the number of correct steps divided by the number of simulated
steps, and the precision, the number of correct steps divided by the number of
detected steps (methods). For ideal step detection both recall and precision
have to be close to one. A step finder which
exhibts low recall but high precision tends to underfit the simulated step
signal. On the other hand, high recall but low precision is a sign of
overfitting. Both, underfitting as well as overfitting are undesired since they
may significantly distort statistical properties calculated from the detected
step signal.

For the simulated data the computed recall of TVDN is $69 \%$, which is fairly
high. However, this comes at the cost of a low precision of $34 \%$.  In the
second step of EBS we use CC to cluster the denoised data to predefined
levels of integer multiples of the known step size of $1bp$. This results in a
total of $176$ found steps and thus many steps in the TVDN data are removed (figure
\ref{fig:ebsstepsgcutresult}). TVDN visually traces the simulated data very well
(figure \ref{fig:tvdnstepsgcutresult}), however overfits the signal, i.e. there
are many more detected plateaus than simulated steps.  In this example, the CC
algorithm performs much better, due to the high noise not all steps are
recovered (figure \ref{fig:ebsstepsgcutresult}). Some simulated steps were
missed or fused to steps of double size. Compared to the TVDN the computed
recall is slightly reduced, but the precision of $51 \%$ is much higher, showing
that the data is fitted more accurately. The quality of the performance of CC
depends on the value of the prior potential parameters $\rho_{S}, \rho_{P}$
tuned to optimize precision and recall (appendix).

\subsection*{Stability and Scalability of $\lambda_h$-Heuristics}
\label{sec:stability_lambda_heuristics}

\begin{figure}[t!]
  \centering{
    \includegraphics[width=3.5in]{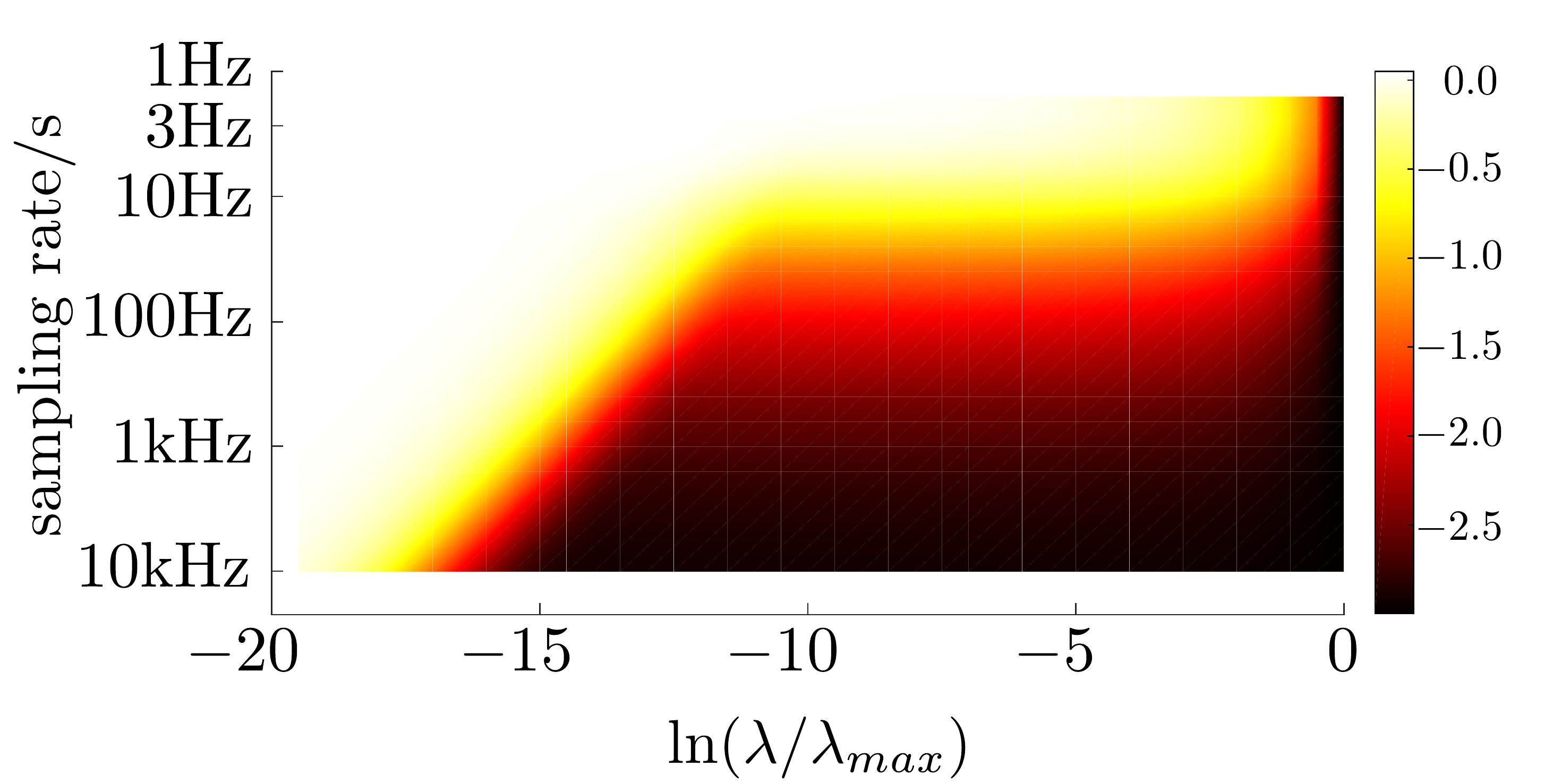}
  }
  \caption{Over fitting transition depends on sampling frequency. Each step
  signal has $1016$ poisonian distributed steps sampled with different
  frequencies and covered by noise. The signals are $100 s$ long. The colorbar
  shows the value of $\ln(n/N)$, where $N$ is the total number of data points
  and $n$ the number of denoised steps.}
\label{fig:phasesamplfreq}
\end{figure}

\begin{figure}[t!]
  \centering{
    \includegraphics[width=3.5in]{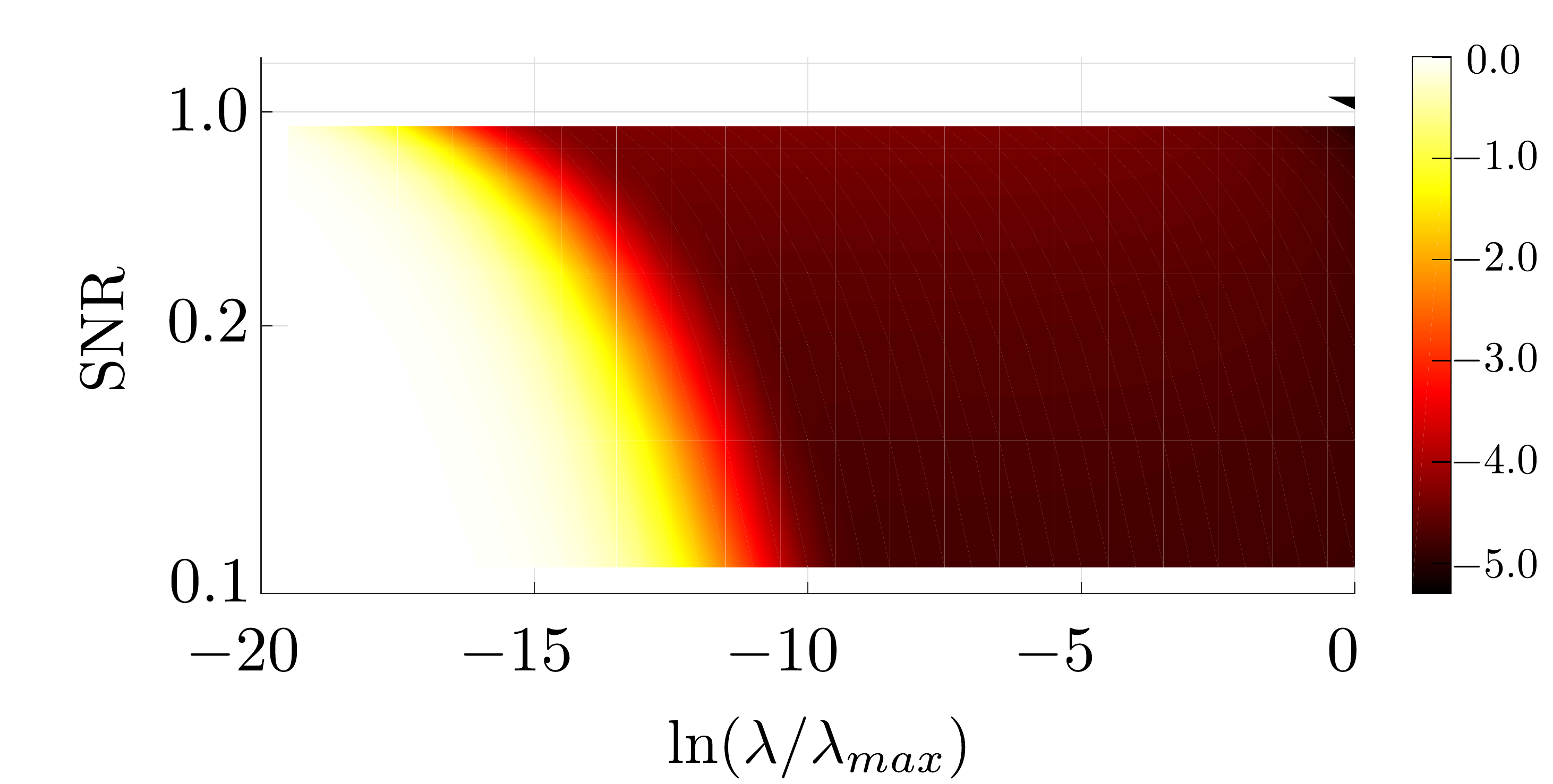}
  }
  \caption{Influence of SNR on over fitting transition. Each step signal has
  $980$ poissonian distributed steps with different noise amplitudes at $6kHz$.
Every signal is $100 s$ long and consists of $6\cdot 10^5$ data points. The
  scale of the colorbar is chosen analog to figure \ref{fig:phasesamplfreq}.}
\label{fig:phasesnr}
\end{figure}

The $\lambda_h$-heuristic is the starting point of finding steps which are
corrupted by noise and here we analyze the applicability of this scheme on
simulated data. In general we do not expect that this scheme returns good
results for arbitrarily large noise amplitudes or sampling frequencies on the
order of stepping rates. The dependencies on noise amplitudes and sampling
frequencies for Poisson distributed steps (forward stepping with rate constant
$10Hz$) covered by noise can be best summarized in the following phase diagrams
(figure \ref{fig:phasesamplfreq} and \ref{fig:phasesnr}). As for the data shown
in figure \ref{fig:tvdnheuristic} we compute the number of produced steps after
TVDN for different denoising parameter $\lambda$. For a signal of $100s$ length
with $1016$ Poisson distributed steps we vary the sampling frequency and keep
the standard deviation of noise constant at $4.4 bp$ (figure
\ref{fig:phasesamplfreq}). For each sampling frequency the number of produced
steps is normalized to the number of simulated data points. Furthermore, we vary
the standard deviation of noise and keep the sampling frequency constant at $6 kHz$
(figure \ref{fig:phasesnr}).

In the overfitting regime (white), the number of
steps of the denoised signal equals the number of data points. At
$\lambda/\lambda_{max} = 1$ the denoised signal is constant without any steps.
At a sampling frequency $f=10 kHz$ the number of steps as a function of
$\lambda$ has a clear transition between overfitting and underfitting and
resembles the data shown in figure \ref{fig:tvdnheuristic} (blue). 
As the sampling frequency is lowered the transition is shifted more and more torwards
$\lambda_{max}$.  Below a sampling frequency of $100 Hz$ the number of produced
steps are gradually increasing until there are as many steps as data points, as
was already observed for the data in figure \ref{fig:tvdnheuristic} (red curve).
If the sampling frequency is this low, the $\lambda_h$-heuristic is not
applicable anymore since TVDN breaks down and just imitates noise. At $100Hz$
there are on average $10$ data points for each plateau. Since steps are
poissonian distributed many steps have plateaus that consist of less than $10$
data points and are thus hardly distinguishable from noise.

For decreasing signal to noise ratio (SNR) we get a similar shift of the phase
boundary torwards $\lambda_{max}$ for worse SNR, figure \ref{fig:phasesnr}. 

\subsection*{The Influence of drift on EBS performance}
\label{sec:drift_subsection}

\begin{figure}
	\centering
   \subfloat[] {
		   \includegraphics[width=3.5in]{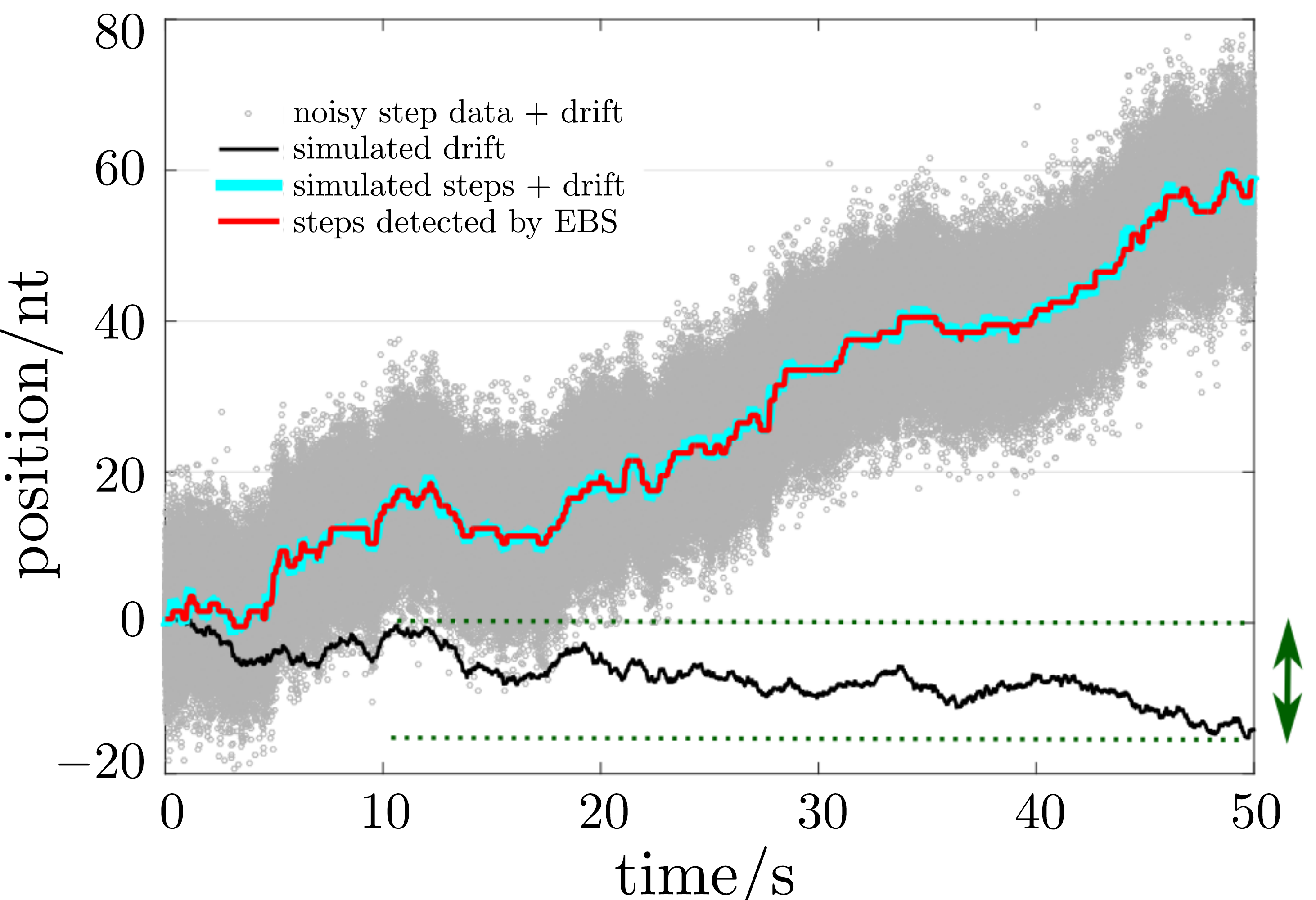}
			 \label{fig:exampledriftdata}
	  }
    \hfill
    \subfloat[] {	
		    \includegraphics[width=3.5in]{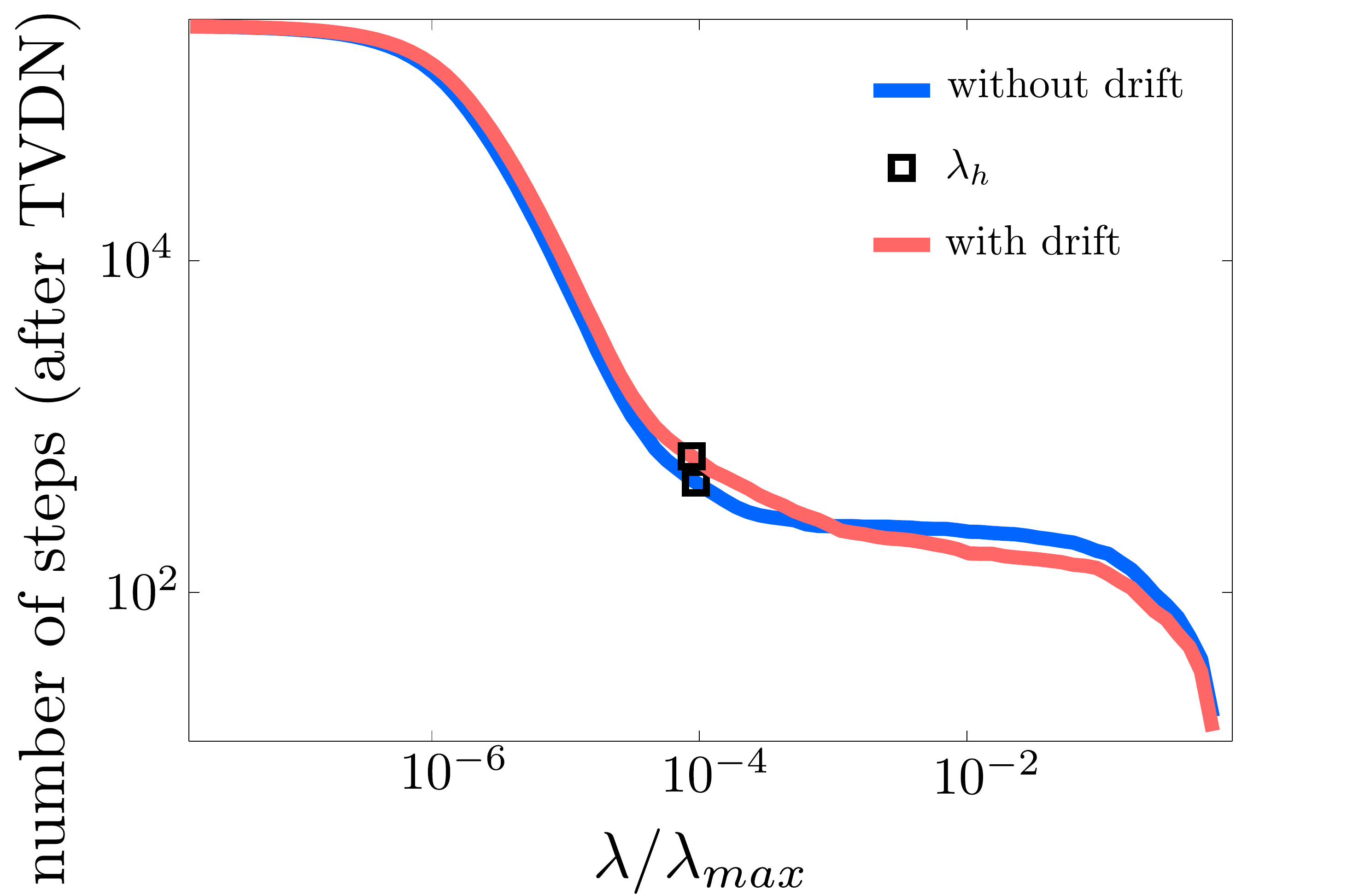}
		    \label{fig:tvdncurvedrift}
	  }
   \hfill
    \subfloat[] {	
      \includegraphics[width=3.5in]{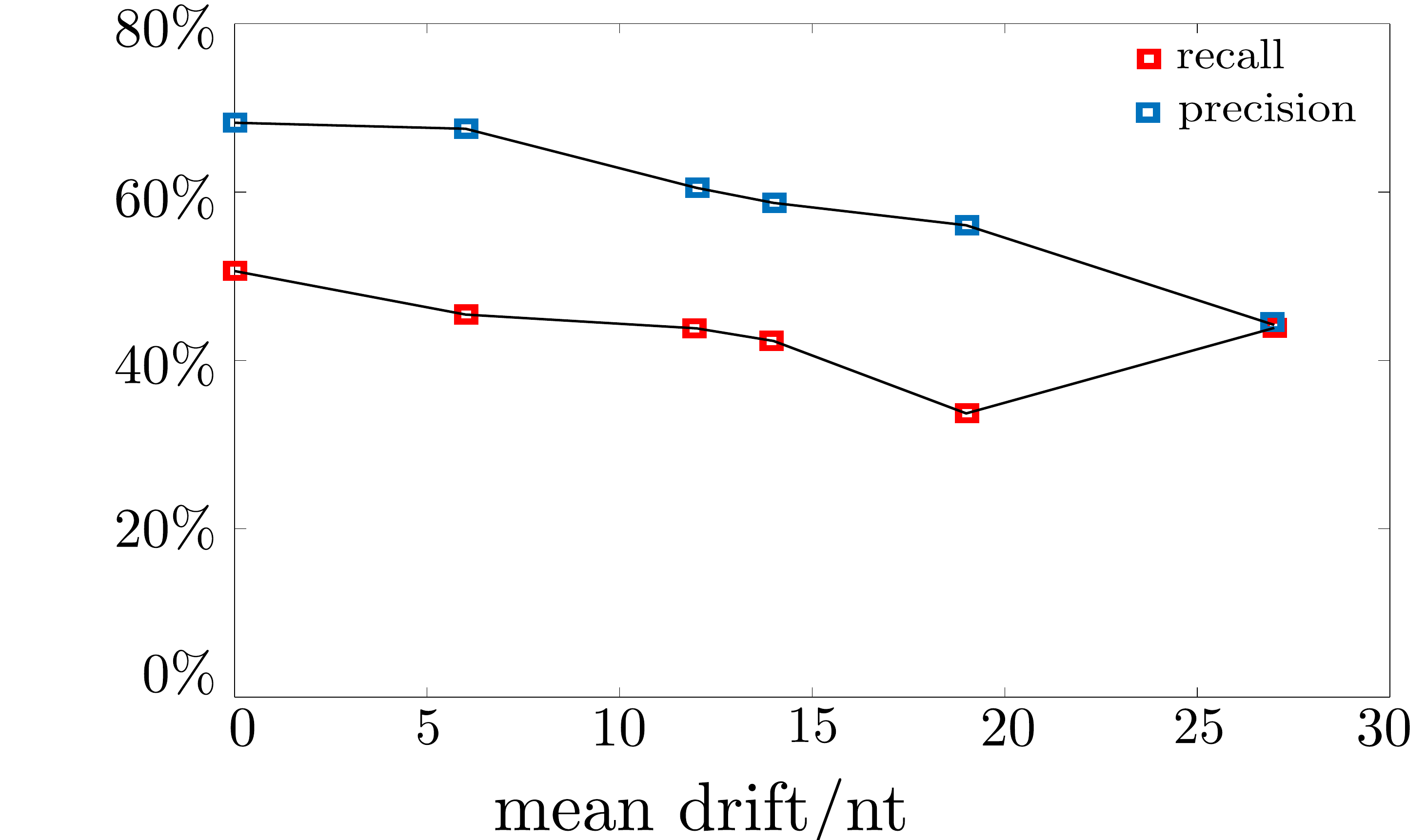}
		    \label{fig:corectlydetdrift}
    }
  \caption{Detection of steps in noisy signals corrupted by drift.
    \ref{fig:exampledriftdata}) Shown is the noisy signal with drift (grey
    dots), the drift (black), simulated steps + drift (cyan) and the detected
    steps (red). The drift is modeled by an Ornstein Uhlenbeck process ($1/k =
    9 min$, $D=10nt^2/s$) and mimics low frequency instrumental fluctuations
    around an equilibrium position. Here, over the simulated time interval of
    $50 s$ the signal has a drift of $16.6 nt$ (peak-to-peak, indicated by green
    arrow dotted lines). \ref{fig:tvdncurvedrift}) TVDN heuristic
    of noisy signals with simulated drift is preserved. In the intermediate
    regime where $\lambda_{h}$ is located, the $\lambda$-heuristic curve of the
    noisy signal with drift (red) is slightly above the curve of the same signal.
    \ref{fig:corectlydetdrift}) With increasing drift steps are detected less precisely. 
    Shown is the precision (blue) and recall (red) of 
    peak to peak difference of the drift 
    (mean of 25 signals, simulated according to the slow scenario, SEM-bars
    (not shown) are smaller than the size of the squares.)}
  \label{fig:driftcorrsteps}
\end{figure}

Actual measured data exhibits drift stemming from the instrument. To analyze the influence of drift on step detection performance of EBS, simulated data (slow scenario) is used with different amplitudes of a stochastic drift. An example of such a drift can be seen in figure \ref{fig:exampledriftdata}. It produces a deviation of $16.6 nt$ in a time interval of $40s$ compared to the simulation without drift (figure \ref{fig:exampledriftdata}, green arrow). This slightly influences the $\lambda$-versus-number-of-steps curve of the TVDN heuristic (figure \ref{fig:tvdncurvedrift}). In an intermediate regime of $\lambda/\lambda_{max}$ the additional low frequency
fluctuations produce slightly more steps when drift is present (figure
\ref{fig:tvdncurvedrift}, red) compared to the same noisy step signal without
drift (figure \ref{fig:tvdncurvedrift}, blue).

Although EBS can eliminate most of these drift induced TVDN steps
some false positives remain which decreases the algorithms step detection
precision. We analyzed the influence of drift on the precision by successively increasing the
diffusion constant $D$ of the drift simulation ($D \in
\{0.0,1.7,5.9,10.0,34.0,117.0 \} nt^2/s$). For every value of $D$, 25 signals
were simulated according to the slow scenario and the mean precision of step
detection was plotted against the mean peak-to-peak difference of the drift of
each signal (figure \ref{fig:corectlydetdrift}). For relatively small drift
($6nt$) precision decreased by only $1 \%$ and thus the influence of such a
drift is rather negligible. Only the comparably large drift of $27 nt$ decreases
the precision to $44 \%$ and thus introduces much more false positives than for
signals without drift.

\subsection*{EBS outperforms existing algorithms}

In the following we compare the performance of the EBS algorithm to commonly
used algorithms for detecting steps in the trajectory of motor proteins namely,
a t-test \cite{stepdetcomp},(using an implementation that sweeps over different window sizes, \cite{phagettest}), the Kalafut and Visscher algorithm (K \& V),
\cite{kvalgo}, (using the implementation from \cite{maxlittlestepsbumps}) and the variable
stepsize hidden Markov model (HMM) \cite{vshmm}. 

\begin{figure}
	\centering
  \subfloat[] {	
	    \includegraphics[width=3.4in]{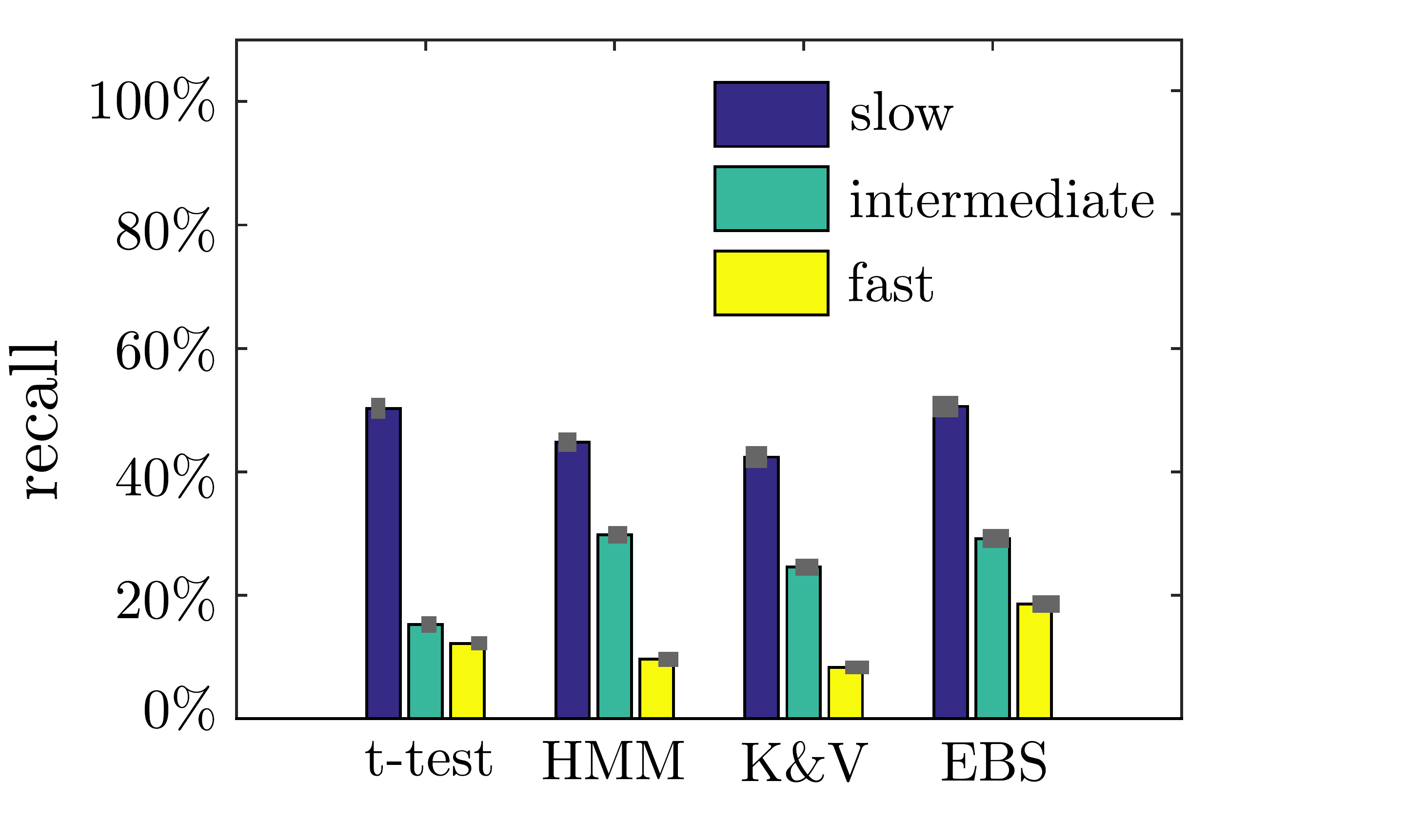}
		  \label{fig:stepsrecall}
  }
  \hfil
  \subfloat[] {	
		  \includegraphics[width=3.4in]{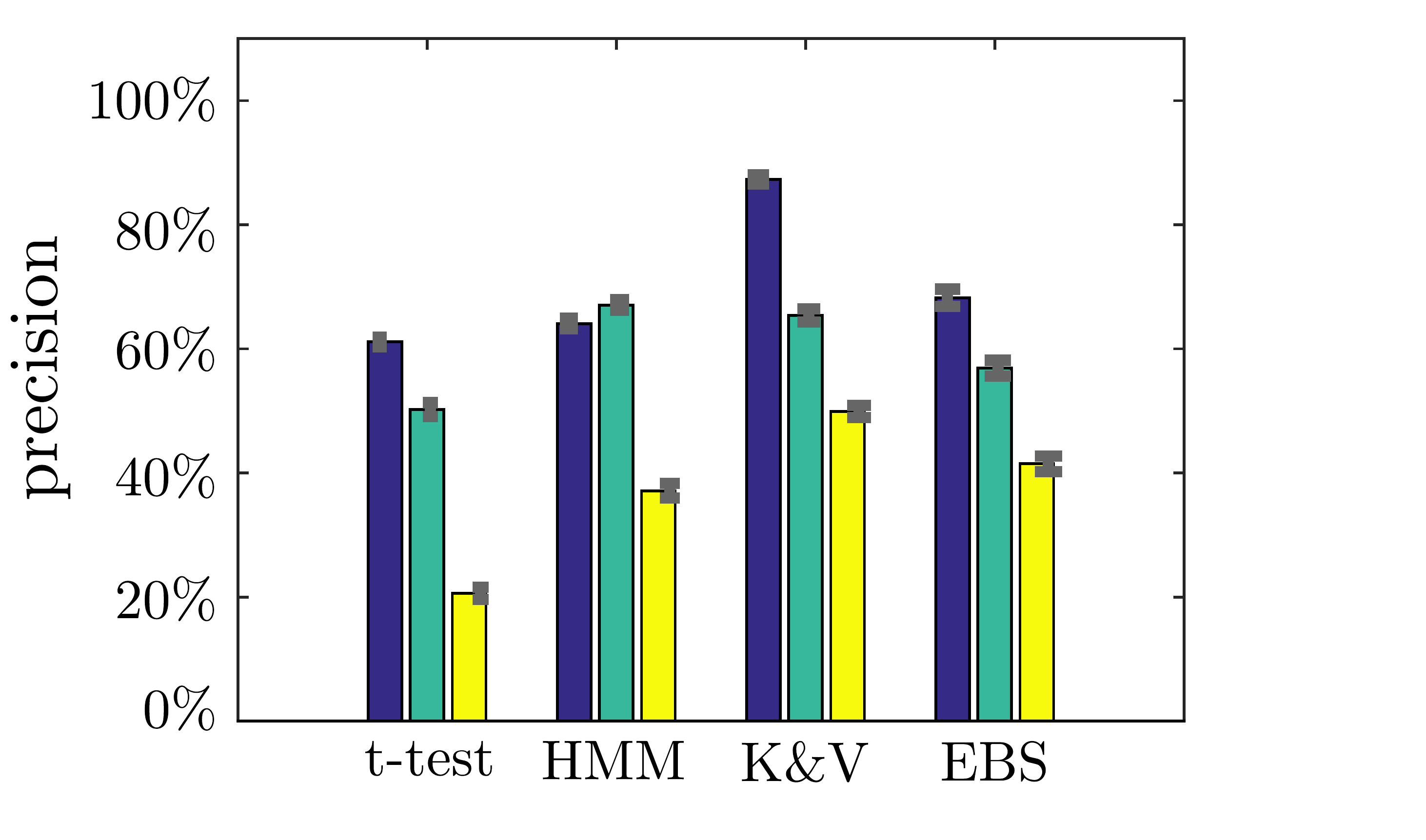}
 			\label{fig:stepsprecision}
  }
  \hfil
  \subfloat[] {	
		  \includegraphics[width=3.4in]{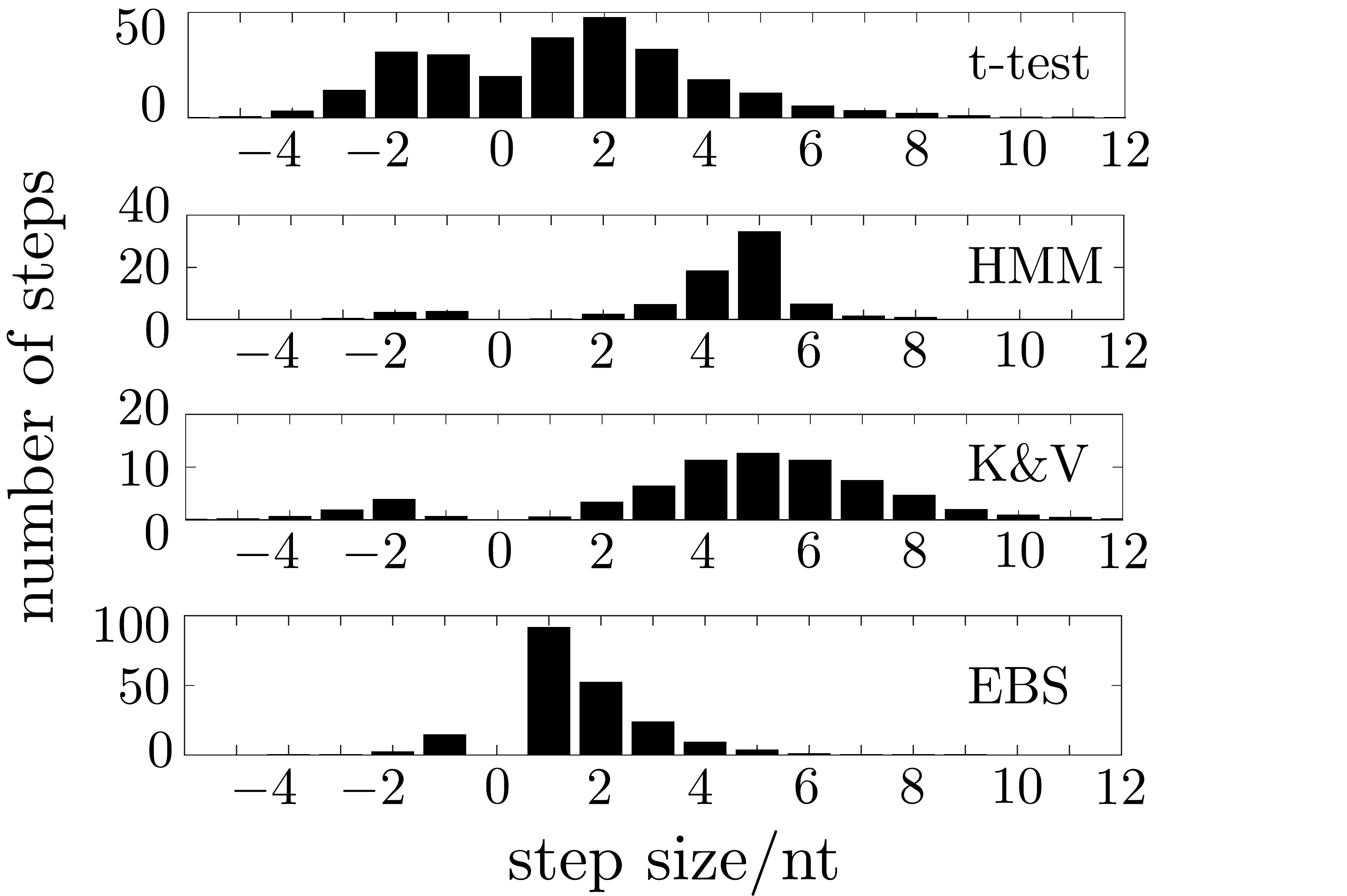}
 			\label{fig:corrsizestephist}
	}  
  	\caption{Performance of step detection algorithms with respect to slow
      scenario (blue bars), intermediate scenario (green bars) and fast scenario
      (yellow bars). (\ref{fig:stepsrecall}) shows in percent of the total
      number of simulated steps the number of detected steps.
      (\ref{fig:stepsprecision}) percentage of correct steps among the simulated
      steps. Error bars are SEM. (\ref{fig:corrsizestephist}) shows average step
    size histograms with $1bp$ binning of the detected steps of the fast
  scenario for t-test, HMM, K \& V and EBS (from upper to lower histogram).}
  \label{fig:compperformance}
\end{figure}
 
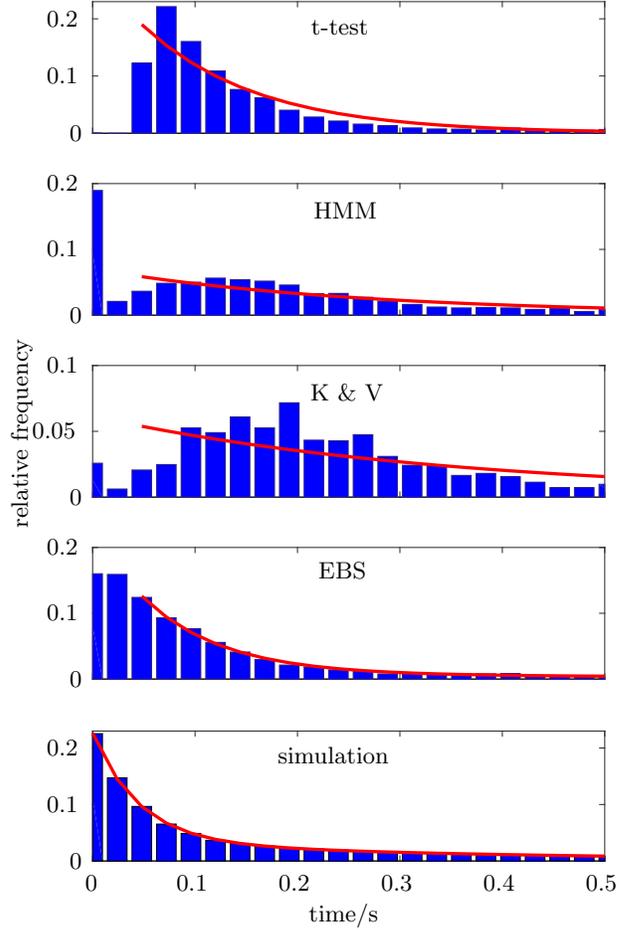
\begin{figure}[h]
\begingroup%
  \makeatletter%
  \providecommand\color[2][]{%
    \errmessage{(Inkscape) Color is used for the text in Inkscape, but the package 'color.sty' is not loaded}%
    \renewcommand\color[2][]{}%
  }%
  \providecommand\transparent[1]{%
    \errmessage{(Inkscape) Transparency is used (non-zero) for the text in Inkscape, but the package 'transparent.sty' is not loaded}%
    \renewcommand\transparent[1]{}%
  }%
  \providecommand\rotatebox[2]{#2}%
  \ifx\svgwidth\undefined%
    \setlength{\unitlength}{252bp}%
    \ifx\svgscale\undefined%
      \relax%
    \else%
      \setlength{\unitlength}{\unitlength * \real{\svgscale}}%
    \fi%
  \else%
    \setlength{\unitlength}{\svgwidth}%
  \fi%
  \global\let\svgwidth\undefined%
  \global\let\svgscale\undefined%
  \makeatother%
  \begin{picture}(1,1.58023033)%
    \put(0,0){\includegraphics[width=\unitlength]{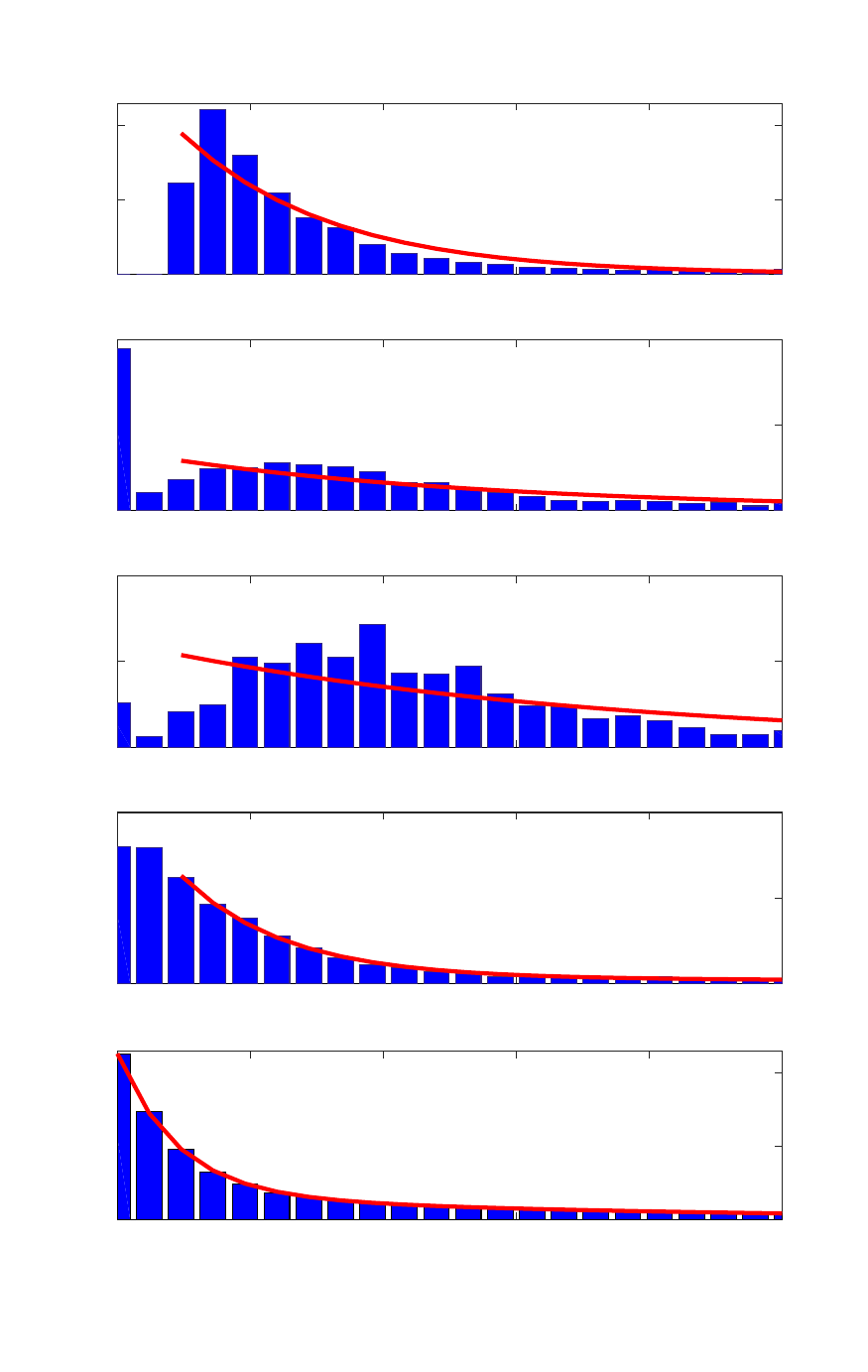}}%
    \put(0.1006768,1.25324419){\makebox(0,0)[lb]{\smash{0}}}%
    \put(0.06664107,1.33886205){\makebox(0,0)[lb]{\smash{0.1}}}%
    \put(0.06664107,1.4244799){\makebox(0,0)[lb]{\smash{0.2}}}%
    \put(0.46218538,1.41224308){\makebox(0,0)[lb]{\smash{t-test }}}%
    \put(0.1006768,0.98095835){\makebox(0,0)[lb]{\smash{0}}}%
    \put(0.06664107,1.07941885){\makebox(0,0)[lb]{\smash{0.1}}}%
    \put(0.06664107,1.17787936){\makebox(0,0)[lb]{\smash{0.2}}}%
    \put(0.46583206,1.13718381){\makebox(0,0)[lb]{\smash{HMM}}}%
    \put(0.1006768,0.7086725){\makebox(0,0)[lb]{\smash{0}}}%
    \put(0.04476096,0.807133){\makebox(0,0)[lb]{\smash{0.05}}}%
    \put(0.06664107,0.90559351){\makebox(0,0)[lb]{\smash{0.1}}}%
    \put(0.46096982,0.86489796){\makebox(0,0)[lb]{\smash{K \& V}}}%
    \put(0.1006768,0.43638665){\makebox(0,0)[lb]{\smash{0}}}%
    \put(0.06664107,0.53484716){\makebox(0,0)[lb]{\smash{0.1}}}%
    \put(0.06664107,0.63330766){\makebox(0,0)[lb]{\smash{0.2}}}%
    \put(0.47312543,0.59805637){\makebox(0,0)[lb]{\smash{EBS}}}%
    \put(0.1242587,0.1307944){\makebox(0,0)[lb]{\smash{0}}}%
    \put(0.26040163,0.1307944){\makebox(0,0)[lb]{\smash{0.1}}}%
    \put(0.41356242,0.1307944){\makebox(0,0)[lb]{\smash{0.2}}}%
    \put(0.56672321,0.1307944){\makebox(0,0)[lb]{\smash{0.3}}}%
    \put(0.719884,0.1307944){\makebox(0,0)[lb]{\smash{0.4}}}%
    \put(0.87304479,0.1307944){\makebox(0,0)[lb]{\smash{0.5}}}%
    \put(0.45853869,0.08460306){\makebox(0,0)[lb]{\smash{time/s}}}%
    \put(0.1006768,0.1641008){\makebox(0,0)[lb]{\smash{0}}}%
    \put(0.06664107,0.2486616){\makebox(0,0)[lb]{\smash{0.1}}}%
    \put(0.06664107,0.3332224){\makebox(0,0)[lb]{\smash{0.2}}}%
    \put(0.03841175,0.67363568){\rotatebox{90}{\makebox(0,0)[lb]{\smash{relative frequency }}}}%
    \put(0.41234734,0.32090827){\makebox(0,0)[lb]{\smash{simulation }}}%
  \end{picture}%
\endgroup%


 \caption{Dwell time distribution of detected steps in the fast scenario.
    Displayed are dwell time histograms (blue bars) of the t-test, HMM, K \& V and EBS
    algorithm (from upper to lower panel). For better comparison, the histogram
    of simulated dwell times is depicted in the lowest panel. Each dwell time
    histogram is fitted by a double exponential decay (red line) which yields
    the following rates ($k_{pause}/k_{elong}$ in $Hz$): t-test ($0.017/8.9$),
    HMM ($0.088/3.9$), K \& V ($0.01/2.8$), EBS ($1.3/13$), simulation
    ($2.8/22$). Note that in case of the detected dwell times the first two
    bars are not taken into account since steps with short dwell times are
    likely to be skipped by step detection algorithms.} 
  \label{fig:dwelltimedist}
\end{figure}

\begin{figure}[h]
\begingroup%
  \makeatletter%
  \providecommand\color[2][]{%
    \errmessage{(Inkscape) Color is used for the text in Inkscape, but the package 'color.sty' is not loaded}%
    \renewcommand\color[2][]{}%
  }%
  \providecommand\transparent[1]{%
    \errmessage{(Inkscape) Transparency is used (non-zero) for the text in Inkscape, but the package 'transparent.sty' is not loaded}%
    \renewcommand\transparent[1]{}%
  }%
  \providecommand\rotatebox[2]{#2}%
  \ifx\svgwidth\undefined%
    \setlength{\unitlength}{252bp}%
    \ifx\svgscale\undefined%
      \relax%
    \else%
      \setlength{\unitlength}{\unitlength * \real{\svgscale}}%
    \fi%
  \else%
    \setlength{\unitlength}{\svgwidth}%
  \fi%
  \global\let\svgwidth\undefined%
  \global\let\svgscale\undefined%
  \makeatother%
  \begin{picture}(1,0.57142857)%
    \put(0,0){\includegraphics[width=\unitlength]{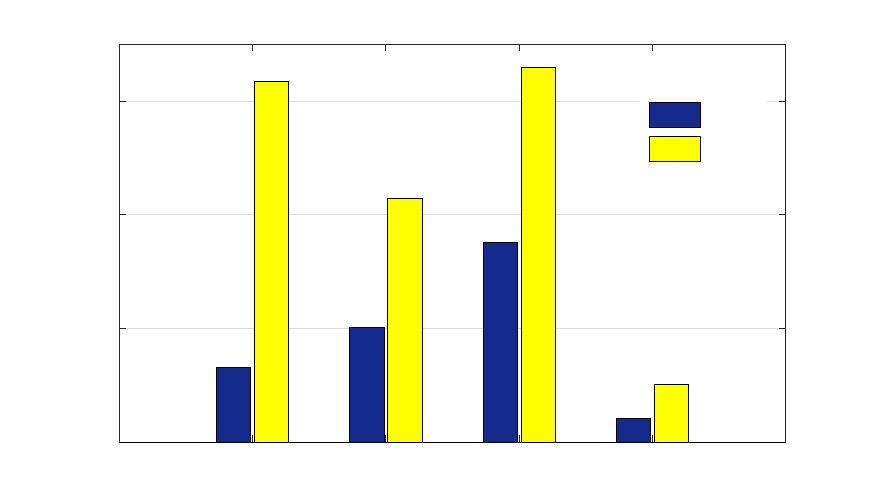}}%
    \put(0.25522732,0.03194034){\makebox(0,0)[lb]{\smash{t-test}}}%
    \put(0.40480564,0.03194034){\makebox(0,0)[lb]{\smash{HMM}}}%
    \put(0.55531301,0.03194034){\makebox(0,0)[lb]{\smash{K \& V}}}%
    \put(0.71511097,0.03194034){\makebox(0,0)[lb]{\smash{EBS}}}%
    \put(0.10992267,0.05739653){\makebox(0,0)[lb]{\smash{0}}}%
    \put(0.08390905,0.18746463){\makebox(0,0)[lb]{\smash{0.1}}}%
    \put(0.08390905,0.31753273){\makebox(0,0)[lb]{\smash{0.2}}}%
    \put(0.08390905,0.44760083){\makebox(0,0)[lb]{\smash{0.3}}}%
    \put(0.06269492,0.06529175){\rotatebox{90}{\makebox(0,0)[lb]{\smash{Kullback-Leibler Divergence}}}}%
    \put(0.80627013,0.4317223){\makebox(0,0)[lb]{\smash{slow}}}%
    \put(0.80627013,0.39287096){\makebox(0,0)[lb]{\smash{fast}}}%
  \end{picture}%
\endgroup%


  \caption{Shows the Kullback-Leibler divergence of dwell time histograms of the
    detected dwell time distribution with respect to the simulated one.} 
 		\label{fig:dwelltimeKL}
\end{figure}

 
In order to quantitatively compare the results of the algorithms, we chose the slow, intermediate and fast
scenarios (methods). To get statistically meaningful results we simulated 25
time traces for each scenario. Input parameters of the step-detection algorithms
were adjusted once for each simulation scenario (appendix). After
the analysis the detected steps were compared to the simulated input steps by
computing recall and precision according to our criterion of correctly recovered
steps (methods and figure \ref{fig:compperformance}).

For the slow scenario, around half of the simulated steps could be recovered by
each of the four algorithms (figure \ref{fig:stepsrecall}). While the K \& V
algorithm recovers the fewest of the simulated steps (recall: $42 \%$),
the much larger precision ($87 \%$) shows that there are comparably few false
positives among the detected steps. The other algorithms exhibit a somewhat smaller
precision (t-test $61 \%$, HMM $64 \%$ and EBS $68 \%$), but a
higher recall (t-test: $50 \%$, HMM: $45 \%$, and EBS: $51 \%$). This means that
more detected steps are misplaced or shifted with respect to the simulated
steps. Hence, for these conditions all four algorithms work well and recover a
similar amount of steps in a close vicinity of the simulated steps. However,
the K \& V algorithm is a little more conservative towards placement of new
steps thus increasing the precision but lowering the recall. 

In the intermediate scenario stepping rates are faster which clearly
reduces the recall for the t-test ($15 \%$). This effect is less dramatic for
the HMM ($30 \%$), K \& V ($26 \%$) and EBS ($30 \%$). The precision of HMM ($67
\%$) and K \& V ($65 \%$) are at a similar level followed by EBS ($57 \%$) and
t-test ($50 \%$).

The fast scenario exhibits even faster steps and higher noise amplitudes and
thus is the most difficult simulation setting considered here. The t-test
recovers $12 \%$ of the simulated steps at around half the precision of the
other algorithms showing the worst performance. The performance also decreased
for the other three algorithms. However, compared to HMM ($10 \%$) and K \& V
($8 \%$), EBS recovers approximately twice as many correct steps ($19 \%$) at a
comparable precision (HMM: $37 \%$, K \& V: $50 \%$ and EBS: $42 \%$).

The correct timing of a detected step, as described by the computed values of
precision and recall is only one important aspect of step detection. It is also
important to test whether the recovered step-size distribution resembles the
simulated stepping behaviour (figure \ref{fig:corrsizestephist}). For the fast
scenario, due to the lower bandwidth and faster stepping
rates the algorithms do not reproduce the simulated step size well. Here, all
algorithms tend to fuse $1bp$ steps to steps of larger size which explains the
smaller number of found steps compared to number of simulated steps (figure
\ref{fig:stepsrecall}, \ref{fig:stepsprecision}). While the
t-test is showing a broad distribution of step sizes and both HMM and K \& V
detect mostly steps of size larger than $2bp$, the step size distribution
obtained by EBS resembles the expected distribution most closely. In contrast
for the slow scenario step-size histograms show a majority of the expected
$1bp$ steps for all algorithms considered here (appendix, figure
\ref{fig:addstepsizehist}). Therefore, in comparison with the fast scenario it
becomes evident how much the noise influences the step size distributions.
Compared to the other algorithms, the denoising stage of EBS is the most
robust.

Important statistical properties of the underlying chemical cycle of a motor
protein are often obtained from the distribution of dwell times, i.e. the
duration between adjacent steps. To analyze the quality of the detected dwell
times in the fast scenario, we compute dwell time histograms ($25 ms$ binning)
from the detected steps of each algorithm and compare them to the distribution
of simulated dwell times (figure \ref{fig:dwelltimedist}).  While the EBS
derived dwell time histogram has a similar shape than the actual simulation
input, the other algorithms fail to recover the general shape of the histogram.
This observation is also reflected in the rate constants of a double
exponential fit to the dwell time distribution (figure \ref{fig:dwelltimedist},
red curve). Here, rate constants extracted from the steps detected by EBS
deviate about a factor of two from those extracted directly from the simulated
distribution. In contrast, the rate constants determined by the other
algorithms deviate by several orders of magnitude when determining the pausing
rate and are also considerably worse compared to EBS in determining the
elongation rate.

How well the detected dwell time distributions reproduce the simulation can
also be quantified by the Kullback-Leibler divergence (figure
\ref{fig:dwelltimeKL}). As expected the Kullback-Leibler divergence of EBS is
smaller compared to the other algorithms.  Moreover, due to the slower stepping
rates and smaller noise amplitudes in the slow scenario, dwell time histograms
of detected steps are more similar to the distribution of the simulation than
in the fast scenario (figure \ref{fig:dwelltimeKL}, blue bars).

In summary, with properly adjusted parameters, none of the algorithms overfits
the highly noisy data since precision exceeds recall in all four cases and thus
there are fewer detected steps than simulated steps (figure
\ref{fig:stepsrecall} and \ref{fig:stepsprecision}). Nevertheless, the low
recall performance means that step detection accuracy is strongly compromised for the
lower bandwidth signal of the fast scenario and directly extracting information
of the underlying enzymatic cycles of elongation from dwell time fluctuations
would result in errors.

\subsection*{EBS is orders of magnitude faster than competing algorithms}

Moreover, we also compared the run-times for all four algorithms on signals
which contain $2.5 \cdot 10^5$ data points and $\sim 600$ simulated steps. We chose
rate constants according to the intermediate scenario and recorded the
respective run times ($t_{run}$). EBS is the fastest algorithm with run times of
$\sim 5 s$ . The t-test is $150$ times , the K \& V: $500$ times and the HMM:
$1000$ times slower (appendix, table \ref{tab:dataptsruntime} ).
EBS is fast enough, that even very high bandwidth signals with $10^{7}$ data
points ($\sim 900$ simulated steps) can be compressed very quickly, yielding a
run time of only $\sim 3 min$ (appendix). Therefore, EBS can
process much more data points at comparably short run time and is essentially
limited only by the available memory size (appendix). The ability
to quickly process a large number of data points can be used to increase the
accuracy of step-finding when the signal is sampled with higher rates. For
example when using kinetics of the intermediate scenario, the recall can be
increased at similar precision from $30 \%$ with $2kHz$ sampling rate to $40\%$ with $200 kHz$ (appendix).

In summary, in the slow and intermediate scenario the algorithms under
consideration perform similarly in the total number of steps found as well as in
the number of correct steps. In the fast scenario where elongation rates are
faster, bandwidth is lower and noise amplitudes are higher EBS shows better
results. Moreover when using EBS, the results of the fast scenario could be improved by higher sampling rates which gives more data points for
each plateau while still preserving comparably short run times. Thus, the EBS
method especially excels for signals obtained from long measurement time, high
bandwidth and poor signal to noise ratio.



\subsection*{EBS detects sub-steps in experimental data of $\varphi 29$ DNA packaging}

Experimental data of Pol II transcription at saturating nucleotide
concentrations yield rates comparable to the fast scenario. For these conditions
the EBS as the best performing algorithm would be able to correctly detect only
$\sim 19 \%$ of all simulated steps. Therefore, in order to better test the
step-finding properties of EBS on actual experimental data one would need to
reduce the stepping rates, or apply the algorithm to a motor protein with larger
step size. A prominent example for such a process is the packaging of DNA by
the bacteriophage $\varphi 29$ motor, which makes steps of $10 bp$ which
consist of a burst of four steps with a size of $2.5 bp$ each
\cite{phagettest}. 

We have applied EBS to experimental stepping data of $\varphi 29$ recorded with
a bandwidth of $2.5 kHz$ using opposing forces of around $5pN$
\cite{phagettest,phagedwells}. We used $2.5 bp$ for the level grid spacing as
well as for the jump height prior ($\epsilon$ in Eq.(\ref{eq:labelterm})). The
standard deviation of the experimental noise at this sampling frequency was
found to be $\approx 3.8 bp$. For this motor at low forces of a few pN a fast
burst of four $2.5bp$ steps is followed by a long dwell time (figure
\ref{fig:pauseregions} a). The presence of $2.5 bp$ steps had previously been
identified at large forces leading to a slow down of the $2.5 bp$ steps
\cite{phagettest}. At the forces of $5pN$ the previously applied t-test had
failed to resolve the $2.5bp$ steps. In contrast, some of the steps are detected
by EBS (figure \ref{fig:pauseregions} and appendix figure \ref{fig:phi29example}).

\subsection*{EBS detects pausing of Pol II}

While the EBS algorithm is not able to determine a large fraction of steps of
Pol II at saturating nucleotide concentrations given published noise levels, it
can be used to investigate pausing of the enzyme.

\begin{figure*}[t!]
  \includegraphics[width=7in]{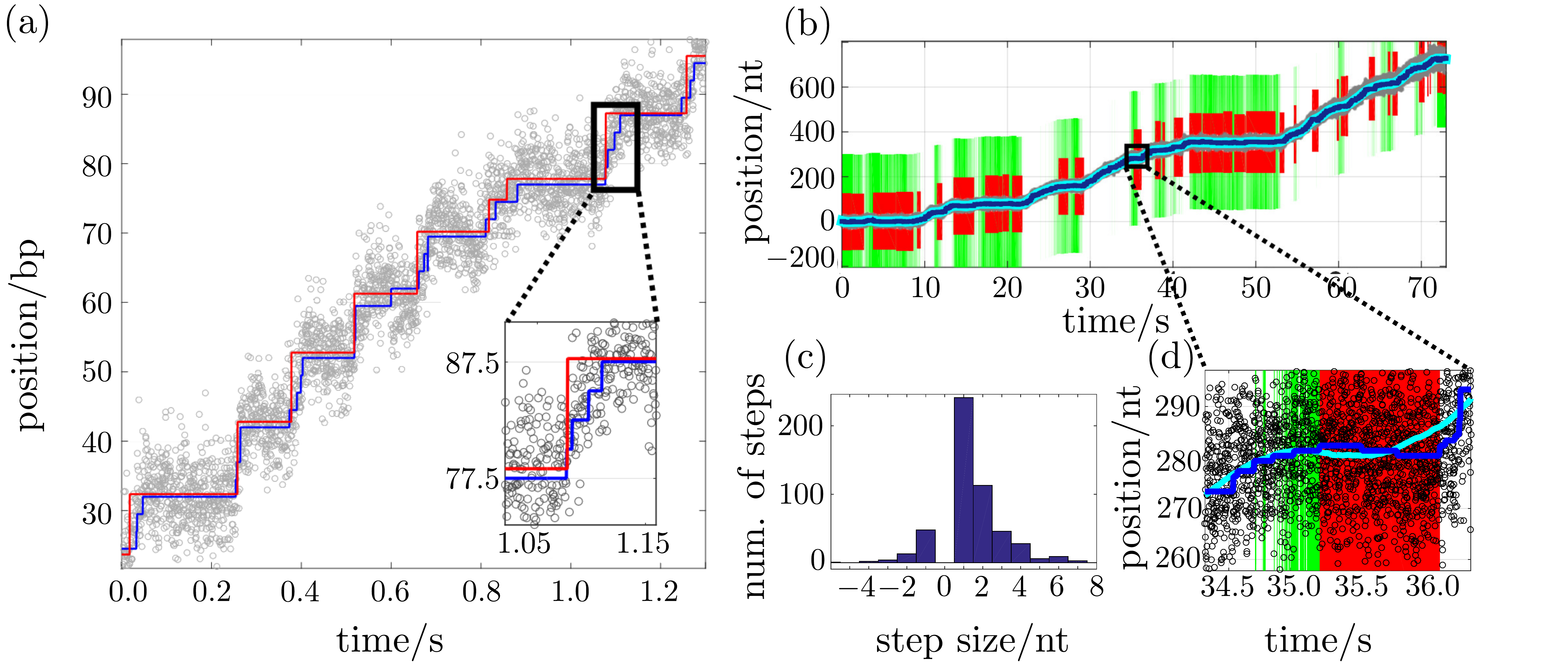}
  \caption{(a) $2.5bp$ substeps in $\varphi 29$ bacteriophage data (circles)
  measured in an optical tweezers experiment, EBS (blue) and t-test (red). (b)
Paused regions in experimental Pol II transcription elongation data. Shaded
regions indicate paused intervals found by the SG-filtering method with a
velocity threshold of two standard deviations of the pause peak (green) and EBS
(red). 1kHz sampled transcription data (grey), SG filtered data of polynomial
order 3 and frame size: 2.5s (cyan) and step detection result of our method
(blue). (c) Step size histogram of detected steps by EBS in the Pol II data shown in
(b). (d) Zoom into a detected pause. Shown is EBS signal (blue), SG filtered
signal (black), measured data at 1kHz (black circles) and paused regions (SG:
green, EBS: red).}
\label{fig:pauseregions}
\end{figure*}

We use EBS to detect pauses (methods) in experimental data from single molecule
transcription elongation data of Pol II (M. Jahnel, S. Grill Lab). Further we
compare the pauses extracted from EBS step data to the result of SGVT (methods)
which is a commonly applied method from the literature \cite{galburtbacktrack}.
The signal consists of $N \sim 7\cdot 10^4$ data points and was recorded with a
sampling frequency of $1kHz$.  The noise amplitude has an estimated average
standard deviation of $\sim 10$ $bp$. Thus the experimental data is comparable
to the fast scenario. We used both SGVT as well as EBS to detect pauses and
backtracks of the enzyme (methods, figure \ref{fig:pauseregions}). When
comparing the results from both algorithms one finds that most long pauses do
overlap, while differences are observed for the detected short pauses.

In order to get a better understanding of how well the two algorithms perform,
we again use simulated data with parameters for stepping rates and sampling
frequency according to the fast scenario (appendix). In accordance
with previously published discussions on backtracked pauses
\cite{shorttranscriptpause} we distinguish long ($t>t_{p}$) and short pauses
($t<t_{p}$) by a time scale  $t_{p}=1/\sqrt{k_{f}\cdot k_{b}}=0.8 s$. All
simulated long pauses were found by EBS ($100 \%$) and the total length of long
pauses compared to simulated long pauses was $113 \%$. Also the SGVT  found
almost all long pauses ($98 \%$) with $94 \%$ of the total duration of simulated
long pauses. Both methods did not falsely assign long pauses and thus the result
of finding long pauses in step detected data and in SG filtered data largely
agrees.

However concerning short pauses, EBS outperforms SGVT in recall ( EBS:
$61\%$, SGVT: $38\%$) and precision (EBS: $92 \%$, SGVT:
$57\%$).

Especially for experiments with near base pair resolution and slow
elongation rates (i.e. $k_{elong} \sim k_{f},k_{b}$), SG filtered data is not
suitable to distinguish between pauses and natural waiting times of elongation
and hence step-detection becomes the only option. For these
experiments pause-detection accuracy is very high and allows the analysis of
dwell time fluctuations. This provides further insights into enzymatic reaction
cycles such as DNA sequence dependent dynamics \cite{wangseqdep}.
\\



\section*{Conclusion \& Outlook}

We have presented a novel energy based step finding scheme comprised of a
denoising stage that uses TVDN followed by a CC analysis. The CC stage uses a
Graph Cut algorithm and provides the possibility to include prior information.
For biomotors with unknown step size, CC can be performed without step size
prior terms. If the detected steps exhibit a dominant step size, a second
application of EBS with this prior information can improve results. The EBS
algorithm outperforms current schemes for detecting steps or pausing events in
time trajectories of molecular motors.  In case of high-noise data it had the
highest recall with comparable precision. The higher step detection performance
of EBS is also reflected in the step size and dwell time distributions which
better reproduce the simulated distributions. In particular, for the fast
scenario where the recall is rather low, further analysis of the dwell time
distribution returns useful rate constants in contrast to the rates extraced
from dwell time distributions of the competing algorithms. In addition EBS is
much faster than competing algorithms.

In particular the high computational speed of EBS becomes an advantage when
multiple executions of the algorithm are necessary. One example for an extension
of EBS with multiple executions, is an iteratively adapting level grid which
could be used for signals with unknown step heights. Similar schemes are already
available for HMMs \cite{tjhahmm} and were successfully applied to FRET data
\cite{baerbl}.  For EBS, this could be implemented by methods from Multi-Model
Fitting \cite{Isack12}. Another example could be to expand EBS to allow for
drift correlation. As is, EBS is relatively insensitive to drift so that drifts
on the order of $10 bp/min$ have a negligible effect on step-finding
(results \& discussion) \cite{bpsteprnapol}. However, one could explicitly correct for drift by
using a decorrelation scheme as previously developed \cite{corrnoise}. Again
this would necessitate multiple execution of EBS.

A reason the proposed EBS method exhibits competitive performance is a favorable
representation of information in the signal, which led to the two stage process.
Here, we have used TVDN to build a fast and unbiased denoising scheme while
still preserving the step features of the underlying signal. This was possible
by using a drastically improved algorithm for solving the one dimensional TVDN
problem which allows us to choose the regularization parameter $\lambda_h$
automatically. In fact, TVDN with this $\lambda_h$ performs often very well in
tracing the actual signal even under noisy conditions. Consequently, if TVDN is
used as first stage, the choice of the regularization parameter is very
important and can significantly influence the performance of further steps.

Previously, TVDN has been applied in a step detection algorithm of the rotary
flagella motor movement \cite{mlittlefastbayes, maxlittlestepsbumps}. The
method to determine the parameter $\lambda$ developed here could be directly
applied to this problem thus increasing the accuracy of the denoising scheme.
Nonetheless, a more rigorous theoretical examination of the sudden change from
over- to underfitting of TVDN which led to our heuristics remains to be done.
Donoho et al. \cite{Donoho09} have reviewed the observation that sudden
break-downs of model selection or robust data fitting occur in high-dimensional
data analysis and signal processing.
They further refined this finding for Compressed Sensing in \citep{Donoho11},
which is a class of $l_1$ regularized convex optimization problems. It remains
an interesting question if similar theoretical statements can be established for
TVDN.

There exist different ways to solve the subsequent clustering problem
for step detection. For example when step sizes are uniform and the signal is
periodic, such as for the above mentioned rotary bacterial flagella motor, a
Fourier transform-based technique with nonlinear thresholding in frequency space
can be used \cite{maxlittlestepsbumps}.

In contrast the presented CC algorithm is broadly applicable to non-periodic signals.
We found that our implementation of CC is very well suited to cluster the output
of the compression since it provides a framework to include prior information
and it applies to a broad class of step signals including steps with non uniform
sizes. Further there are comparably fast algorithms available to solve relevant
energy functions. In fact, we found that our algorithm scaled approximately
quadratically in the number of tuple and linearly in the size of the predefined
level set in our applications (appendix). The penalizing energy
scheme can be extended in an intuitive way to other prior information.  For
example, a histogram prior could yield a global energy term
that favors certain step sizes and dwell time histograms.

The adjustment of the regularization parameters of the $\rho_i$ energy function
can be guided by comparing results with simulated stepping data.
This choice is not dependent on noise due to the preceeding application of TVDN.
Further by using weights in the energy terms the regularization parameters can
be applied to different datasets of the same underlying stepping process.

Both, the TVDN stage as well as the clustering stage, provide the possibility to
harness parallelization to gain speedups. A long high bandwidth trajectory could
be divided into smaller time-intervals, which could then be treated in parallel.
Of course one would need to find a way to take care of the boundaries between
the intervals, e.g. by shifting the time intervals and merging the data. This
extension would also make a quasi online processing of measurement data possible,
where new intervals are successively ingested.

EBS was successfully applied to detect pauses by Pol II as well as $2.5bp$ steps
in the packaging of DNA by the bacteriophage $\varphi 29$ motor. However, while
some steps could be found, the larger the noise and smaller the step-size the
fewer correct steps are found. To make fully use of the advantages of EBS 
higher bandwidth data is needed.  Moreover, shorter tether length, smaller beads
or stiffer handles provided by DNA origami \cite{dnaorigamibeams} increase
resolution and thus improve step detection.

In summary, the EBS method fills the gap of tools which are able to handle high
bandwidth data with many data points as well as very noisy data under quite
general assumptions.  Regardless of the difference in TVDN and Graph Cut the
energy based model provides an intuitive access for the user of the method.



\section*{Author Contributions}
J. R., K. P.-Y., M. B. P. and J. M. designed research and wrote the manuscript. K. P.-Y. and J. R. developed algorithms and analyzed data.

\section*{Acknowledgement}
We thank Marcus Jahnel for providing the experimental Pol II data, Gheorghe Chistol for providing the experimental $\varphi 29$ phage packaging data and Jeffrey R. Moffitt and Gheorghe Chistol for MATLAB code of the t-test algorithm.\\
This work was supported by the EU Integrating project SIQS, the ERC Synergy
grant BioQ as well as the ERC starting grant Remodeling and an Alexander von Humboldt Professorship.\\
Unless otherwise stated computations were performed on the computational resource bwUniCluster funded by the Ministry of Science, Research and the Arts Baden-W\"urttemberg and the Universities of the State of Baden-W\"urttemberg, Germany, within the framework program bwHPC.

\newpage

\section*{APPENDIX}


\subsection*{Determination of $\lambda_{max}$ in TVDN}
In this section we want to show, how to determine the value of $\lambda_{max}$
in TVDN analytically. The $\lambda_{max}$ value determines the value of the
regularization parameter $\lambda$ in equation \eqref{eq:tvdnproblem} above
which the solution $x^\star$ remains constant and therefore contains no steps
anymore. 
The information in the following subsections is twofold: First derive general
expressions for the Fenchel-Rockafellar-Dual problem and the forward-backward
splitting applied to TVDN, second we then derive a condition for $\lambda_{max}$
from the Fenchel-Rockafellar dual problem and provide an analytical solution. 
Furthermore we give hints on the special (tridiagonal) structure of the involved
operators. The definitions in the next sections follow the work of
\cite{Bauschke11}.

\subsubsection*{Fenchel-Rockafellar Dual Problem and Forward-Backward splitting}
Independent of the problem of our work, we start with a function $f(x)$ which is
convex, proper, and lower semi-continuous. Then 
\begin{equation}
\forall u \in \bb{R}^n, \quad f^*(u) = \umax{x \in \bb{R}^N} \dotp{x}{u} - f(x)
\end{equation}
is called it's Legendre-Fenchel dual function \cite{Bauschke11}. $f^*$ is also
convex, and it holds $(f^*)^* = f$. A further specialization is useful in the
context of our work, as the TVDN problem consists of a minimization of two
composed convex functions \begin{equation}
\umin{x \in \bb{R}^n} f(x) + g(A(x))
\end{equation}
where $A \in \bb{R}^{(p \times n)}$ and the convex functions $f:\bb{R}^n \rightarrow \bb{R}$ and $g:\bb{R}^p \rightarrow \bb{R}$. We assume, that $f^* \in C^1$ and therefore there exists a Lipschitz continuous gradient.

Due to the Fenchel-Rockafellar theorem, covered in Chapter 15 of \cite{Bauschke11}, the following problems are equivalent:
\begin{equation}
\umin{x \in \bb{R}^n} f(x) + g(A(x)) =
  - \umin{u \in \bb{R}^p} f^\dagger( -A^\dagger u ) + g^\dagger(u)
\end{equation}
where $\dagger$ denotes the adjoint function. The unique solution of the primal problem $x^{\star}$ can be recovered from a solution of the dual problem $u^{\star}$, which has not to be necessarily unique.
\begin{equation}
x^\star = \nabla f^\dagger( -A^\dagger u^\star )\,.
\label{eq:primal_dual_connection}
\end{equation}
We use an additional assumption which is not a constraint for the TVDN problem:
$g$ is simple. That means, one can compute a closed-form expression for the
so-called proximal mapping \begin{equation}
\prox{\gamma g}(x) = 
	\uargmin{z \in \bb{R}^n} \frac{1}{2} \norm{x-z}^2 + \gamma g(z) \,\, \forall \gamma > 0.
\end{equation}
Further due to Moreau's identity $g^\dagger$ is also simple \cite{Bauschke11}.

Now having the connection between primal and dual problem at hand, this means,
one has to solve again a composite problem of a convex and a simple
function 
\begin{equation}
\umin{u \in \bb{R}^P} F(u) + G(u)
\end{equation}
with $F(u) = f^\dagger( -A^\dagger u )$ and $G(u) = g^\dagger(u)$.

A typical method to do Proximal Minimization is Forward-Backward splitting (see
eg. Chapter 27 of \cite{Bauschke11}). The dual update is given by
\begin{equation}
u^{(\ell+1)} = \prox{\gamma G} \left( u^{(\ell)} - \gamma \nabla F( u^{(\ell)} ) \right)\,.
\label{eq:fenchel_moreau_rockafellar_dual_update}
\end{equation}
In this update step $\gamma < L/2$, where $L$ is the Lipschitz constant. The primal iterates are given by:
\begin{equation}
x^{(\ell)} = \nabla F( -A^\dagger u^{(\ell)} )\,.
\end{equation}
The above general statements and theorems are taken from the tool set of Convex
Analysis. For further background see e.g. \cite{Rockafellar97} or
\cite{Bauschke11}. In the following we discuss more problem specific
expressions.

\subsubsection*{Application to Total Variation Denoising}
In a continuous picture the total variation of a smooth function $\phi: \bb{R} \rightarrow \bb{R}$ is defined as
\begin{equation}
J(\phi) = \int \norm{\nabla \phi(s)} d s\,.
\end{equation}
In the discretized version one has to consider a discretized gradient operator
$A : \bb{R}^n \rightarrow \bb{R}^p$ with $p = n - 1$.  
\begin{equation}
J(x) = \norm{Ax} = \sum_i u_i
\end{equation}
where $u_i = x_{i+1} - x_i$ and therefore $A$ taking the following form:
\begin{equation}
A = \begin{pmatrix}
1 & -1 & 0 & \dots & 0 \\ 
0 & 1 & -1 &  & \vdots \\ 
\vdots &  & \ddots & \ddots &  \\ 
 &  &  &  & -1 \\ 
0 & \dots  &  &  & 1
\end{pmatrix}\,. 
\end{equation}
Using this and taking into account that the Divergence and Gradient operator are
minus adjoint of each other ($\dotp{\nabla f}{g} = -\dotp{f}{\nabla \cdot g}$)
the adjoint of the discrete gradient operator $A^\dagger$ is minus the discrete
divergence: \begin{equation}
A^\dagger = 
\begin{pmatrix}
-2 & 1 & 0 & \dots & 0 \\ 
1 & -2 & 1 &  & \vdots \\ 
0 & 1 & \ddots & \ddots &  \\ 
\vdots &  & \ddots &  & 1 \\ 
0 & \dots &  & 1 & -2
\end{pmatrix}\,. 
\end{equation}
Therefore the divergence highly resembles a typical laplace filter from signal
processing. This leads for a single entry to $u_i - u_{i - 1} = x_{i+1} - 2 x_i
+ x_{i-1}$. For the deviation of $\lambda_{max}$ we assume the boundary
conditions that $u_0 = 0$ and $u_n = 0$. 

For noise removal (and to get the connection to eq. \eqref{eq:tvdnproblem}) the
following problem has to be solved 
\begin{equation}
x^\star = \uargmin{x \in \bb{R}^n} \frac{1}{2}\norm{x-y}^2 + \lambda J(x),
\end{equation}
To make use of the material so far choose the following composition
\begin{equation}
f(x) = \frac{1}{2}\norm{x-y}^2 \qandq
	g(u) = \lambda \norm{u}.
\end{equation}
After that one has to translate $f(x)$ and $g(x)$ into their dual
representations $f^\dagger(u)$ and $g^\dagger(u)$ by using the following
relations 
\begin{itemize}
  \item For $f(x) = 1/2 \norm{Ax - y}$ and $A \in \bb{R}^{n \times n}$ can be inverted then
  \begin{equation}
    f^\dagger(u) = \frac{1}{2} \norm{(A^\dagger)^{-1} u + y}^2
  \end{equation}

  \item For $f(x) = \norm{ x }_p = \sum_i \left( |x_i|^p  \right)^{\frac{1}{p}}$ is a $p$-norm:
   Then the dual function corresponds with the indicator function $\iota_C$ of the convex set $C$:
   \begin{equation}
     f^\dagger(u) = \iota_{\norm{\cdot} \leq 1} \qwhereq \frac{1}{q} + \frac{1}{p} = 1
   \end{equation}
\end{itemize}
Using that we get the following dual representation of the dual functions $F(u)
+ G(u)$ for the TVDN problem in the Fenchel-Moreau-Rockafellar formulation.
\begin{equation}
\begin{split}
&F(u) = \frac{1}{2} \norm{y - A^\dagger u}^2 - \frac{1}{2}\norm{y}^2 \qandq \\  
	&G(u) = \iota_{C}(u) \qwhereq C = \enscond{u}{\norm{u}_\infty \leq \lambda}.
\end{split}
\end{equation}
The solution to the dual problem $u^\star$ can be obtained by solving
\begin{equation}
u^\star \in \uargmin{ \norm{u} \leq \lambda } \norm{y - A^\dagger u}\,, 
\end{equation}
and by applying eq \eqref{eq:primal_dual_connection} the solution to the primal problem $x^\star$
\begin{equation}
x^\star = y - A^\dagger u^\star\,.
\label{eq:primal_solution}
\end{equation}

What is missing for concrete expression for the forward backward iterations is
first a closed form for the gradient of $F$, which is given by
\begin{equation}
\nabla F(u) =  A (A^\dagger u - y)\,.
\end{equation}
Secondly it is possible for the proximal operator of $G$, which is the
orthogonal projection on the set $C$ 
\begin{equation}
\prox{\gamma G}u = \frac{u}{ \max(1,\norm{u}/\lambda) }.
\end{equation}
\begin{equation}
\gamma < \frac{2}{\norm{A^\dagger A}} = \frac{1}{4}.
\end{equation}
Inserting the above statements into the general dual update step from eq. \eqref{eq:fenchel_moreau_rockafellar_dual_update}, one gets the following expression:
\begin{equation}
\begin{split}
&u^{(l+1)} = \prox{\gamma G}\left(u^{(l)} - \gamma \nabla F(u^{(l)}) \right) \\
& = \frac{u^{(l)} - \gamma \nabla F(u^{(l)}} {\max \left( 1, \frac{\norm{u^{(l)} - \gamma \nabla F(u^{(l)} }}{\lambda} \right) } \\
& = \frac{u^{(l)} - \gamma A (A^\dagger u^{(l)} - y)} {\max \left( 1, \frac{\norm{u^{(l)} - \gamma A (A^\dagger u^{(l)} - y) }}{\lambda} \right) } \\
\end{split}
\label{eq:proximal_iteration}
\end{equation}.

\subsubsection*{Derivation of $\lambda_{max}$ from the Proximal Iteration}
Finding a maximal regularization parameter $\lambda$ is equal to finding a a
criterion, such that the dual iterations remain constant $\forall l$
\begin{equation}
u^{(l+1)} \mbeq u^{(l)}\,.
\label{eq:iter_const_condition}
\end{equation}
By using eq. \eqref{eq:primal_solution} one can see, that this will lead to a
steady state solution $x_i^\star = \mathrm{const} \,\, \forall \, i$. 
For simplicity assume $\tilde{\lambda} = \lambda / \gamma$. Starting from the
proximal iteration we find that in case of $\tilde{\lambda} \leq \norm{u^{(l)} -
\nabla F(u^{(l)})}$ the problem simplifies to 
\begin{equation}
\begin{split}
&u^{(l+1)} = u^{(l)} - A y + AA^\dagger u^{(l)} 
\end{split}\,.
\end{equation}
To satisfy the constant condition from eq. \eqref{eq:iter_const_condition} the
$u^{(l)}$ has to be in the solution of: \begin{equation}
AA^\dagger u = A y\,.
\label{eq:iter_steady_solution}
\end{equation}
The shape of $AA^\dagger$ is the following
\begin{equation}
A A^\dagger = 
\begin{pmatrix}
-3 & 3 & -1 & \dots & 0 \\ 
1 & -3 & 3 & \ddots & \vdots \\ 
0 & 1 & \ddots & \ddots & -1 \\ 
\vdots &  & \ddots & -3  & 3 \\ 
0 & \dots &  & 1 & -2
\end{pmatrix} 
\end{equation}
Linear equations with a tridiagonal affine transform $AA^\dagger$
can be efficiently solved for example an algorithm proposed by Rose
\cite{Rose69}.

Still missing is a treatment of the primal iteration step $x^{(l+1)} = y -
A^\dagger u^{(l+1)}$. The connection to the $\lambda$ in the original TVDN
problem is given such that, the Karush-Kuhn-Tucker conditions are still
valid for our steady state solution \eqref{eq:iter_steady_solution}. 
This means, that every $u^{(l)}$ in the dual solution has to satisfy 
\begin{equation}
  u^{\star}_k \in [-\lambda, \lambda]\,.
\end{equation}
To ensure this, we have to choose 
\begin{equation}
\lambda_{\mathrm{max}} = \norm{u}_\infty
\end{equation}
which gives as a clear statement how to choose a maximal lambda.

\subsection*{Algorithm Implementing the $\lambda_h$-Heuristics}
As outlined in the Methods section of the paper, we use a sudden increase of
resulting steps when decreasing the regularization parameter $\lambda$ in the TVDN
problem shown in eq. \eqref{eq:tvdnproblem} from $\lambda_{max}$ to determine $\lambda_h$.  
In the following, we want to describe the heuristic method, we used to choose
the value of $\lambda_h$.  Starting point for the algorithm is the value of
$\lambda_{max}$ on a curve like the one depicted in figure \ref{fig:tvdnheuristic}. The
iterative method shown in algorithm  \ref{algo:lambda_heuristic} approximates the
point of steepest ascent in an $\lambda$-$n$ diagram, where $n$ is the number of
steps, by searching an interval where the slope exceeds the slope of the secant
of $\{0, \lambda_{max}\}$. The function $N(\lambda)$ counts the number of steps
after the TVDN minimization for a given value of $\lambda$. \\
This simple method gave us stable results for a variety of our test signals,
either simulated or experimentally gathered. In the following section we have a
closer look into the stability of the effect of sudden increase of steps.
\begin{algorithm}[H]
\begin{algorithmic}[1]
\State $\lambda,\,n \gets \lambda_{max}, N(\lambda_{max})$
\State $\lambda^+,\, n^+  \gets \frac{\lambda_{max}}{2},\, N(\lambda_{max} / 2)$
\State $\delta_{\mathrm{start}} \gets \frac{|N(0) - N(\lambda_{max})|}{\lambda_{max}}$
\While {less than max. iterations}
	\State $\delta \gets \frac{|n^+ - n|}{\lambda^+ - \lambda}$ 
	\If {$\delta > \delta_{\mathrm{start}}$} 
		\State \textbf{break}		    
	\EndIf
	\State $\lambda,\, n \gets \lambda^+,\, n^+$
	\State $\lambda^+,\, n^+ \gets \frac{\lambda^+}{\rho},\, N(\lambda^+)$
\EndWhile
\State \Return $\lambda_h \gets \lambda^+$
\end{algorithmic}
\caption{Outline of our line-search algorithm to determine $\lambda_h$}
\label{algo:lambda_heuristic}
\end{algorithm}

\subsection*{Mapping of Energies on Edge-Capacities}

In the process of assigning a level $\xi_i$ to vertex $v_j$ the above mentioned
Graph Cut algorithm solves a binary decision problem, whether the assignment of
a new level is more favorable in terms of the energy loss function or not. The
binary outcome of the decision is reflected in the graph structure by
introducing two special vertices, where $t$ is associated with keeping the old
and $s$ with assigning the proposed level. The energy values of the data term
$\mathcal{Q}_i$ as well as the pairwise term $\mathcal{P}_{i,i+1}$ and their
different combinations of keeping the current level or assigning a new level 
are mapped to capacities of edges in the Markov Random Field. In this section this mapping is
explained stemming theoretical foundations outlined by Kolmogorov et. al. 
in \cite{Kolmogorov04}. The Graph Cut algorithm solves this problem in polynomial time for a certain set of useful energy functions \cite{Boykov04, Kolmogorov03, Kolmogorov04}.

\begin{figure}[t!]
  \centering{
    \includegraphics[scale=.35]{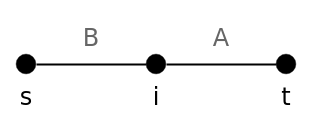}
  }
  \caption{Situation in an Markov Random Field concerning a single variable $v_i$ 
           and edges $A$, $B$ to special variables $s$ and $t$ relevant 
           for the data term.}
  \label{fig:mrf_energy_f1}
\end{figure}
In figure \ref{fig:mrf_energy_f1} the situation for the data term is depicted.
Here the mapping is easy, as the energy for a single variable $v_i$ for the
current level $E_0$ is mapped to the Edge $A$. The energy $E_1$ for a new level
is mapped to the edge $B$.

\begin{figure}[t!]
  \centering{
    \includegraphics[scale=.35]{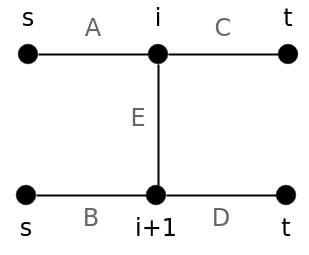}
  }
  \caption{Two neighbouring variables in a Markov Random Field and edges relevant for the
  pairwise term. Here the two $s$ vertices represent the same vertex in the
graph and are just drawn seperated to make the diagram look nicer. The same is
true for the $t$ vertices.}
  \label{fig:mrf_energy_f2}
\end{figure}
The situation for the pairwise term $\mathcal{P}_{i,j}$ is more complicated and
depicted in figure \ref{fig:mrf_energy_f2}.
Here two variables $v_i$ and $v_{i+1}$ are involved which leads to four
different energy combinations $E_{0,0}$, $E_{0,1}$, $E_{1,0}$, $E_{1,1}$ are
possible. Here $E_{0,0}$ is associated with the energy value if both variables get
assigned a new level. In contrast $E_{1,1}$ represents the energy of both
variables keeping their current levels. The two other combinations represent the
case when one variable keeps the current label and the other gets the new level
assigned. $E_{0,1}$ the variable $i+1$ keeps its level, for $E_{1,0}$ this is
the case for the variable $i$.

The four energies can be represented in the following way
\begin{equation}
  \begin{Vmatrix}
    E_{0,0} &E_{0,1} \\
    E_{1,0} &E_{1,1}
  \end{Vmatrix}
  =
  \begin{Vmatrix}
    a &b \\
    c &d
  \end{Vmatrix}
  = 
  \begin{Vmatrix}
    a &a \\
    d &d
  \end{Vmatrix}
  +
  \begin{Vmatrix}
    0   &b-a \\
    c-d & 0
  \end{Vmatrix}
  \, .
\end{equation}

The first summand on the right hand side is mapped to terminal capacities. This
means that the capacity $a$ is associated with the edge $C$, and the capacity
$d$ with the edge $B$. The second summand maps to the edge $E$ and gets the
capacity $b-a+c-d$. 

At this point the above mentioned strategy to circumvent a violation of
submodularity is applied if $E_{0,0} + E_{1,1} > E_{0,1} + E_{1,0}$. Then in
turn $E_{0,1}$ and $E_{1,0}$ is increased and $E_{0,0}$ is decreased by a small
amount until the submodularity condition Eq. \eqref{eq:submodularity_condition} is satisfied.
Details and limitation of this approach can be found in \cite{Rother05}.

\subsection*{$\alpha$-Expansion Algorithm Outline}
Finding a solution $\bm{\xi}^\star$ that minimizing
eq.\eqref{eq:combinatorialfunctional} is a problem that is in general NP-hard to
solve for $|\mathcal{L}| \geq 3$. The iterative $alpha$-expansion algorithm
outlined in algorithm \ref{algo:alpha-expansion} finds provably good approximate
solutions to this problem.  
\begin{algorithm}[H]
\begin{algorithmic}[1]
\State $\xi^\prime \gets $ arbitrary labeling of sites
\While {not converged}
	\ForAll {$\alpha \in \mathcal{L}$} 
		    \State $\xi^\alpha \gets \uargmin{\xi} E(\xi, \xi^\prime)$ \label{algo:alpha-expansion:argmin}
		    \If {$E(\xi^\alpha) < E(\xi^\prime)$} 
				\State $\xi^\prime \gets \xi^\alpha$		    
		    \EndIf
	\EndFor
\EndWhile
\end{algorithmic}
\caption{$\alpha$-Expansion outline}
\label{algo:alpha-expansion}
\end{algorithm}
In each iteration the algorithm updates or moves the current labeling
$\xi^\prime$ if it has found a better configuration. To achieve this, in each
iteration, a new, randomly chosen label $\alpha \in \mathcal{L}$ is introduced
and each site $v_i$ has the choice to stay with the previous label or adopt the
new proposed label $\alpha$. The binary optimization problem is solved via a
Graph Cut (line \ref{algo:alpha-expansion:argmin} of algorithm
\ref{algo:alpha-expansion}). This step is called $\alpha$-expansion due to the
fact, that the number of nodes with the label $\alpha$ assigned could grow
during this phase. The outer iteration stops if no new label assignments
happened within two cycles. The $\alpha$-expansion algorithm was initially
published by Boykov et al. in \cite{Boykov01}.

\subsection*{Relaxation of Submodularity Condition by Truncating Energy}
The submodularity condition \eqref{eq:submodularity_condition} imposes structure on the
energy minimization problem which allows stronger algorithmic results. In this
sense the concept of submodularity plays a similar role for discrete,
combinatorial clustering as convexity plays for continuous optimization.

The Max-Flow/Min-Cut algorithm we use for minimization relies on that the supplied
energy function satisfying the submodularity condition. Unfortunately, the
pairwise term \eqref{eq:labelterm} does not strictly satisfy the submodularity
condition \eqref{eq:submodularity_condition}. Therefore we adopted a truncation scheme
proposed by Rother et al. in \cite{Rother05}. The truncation procedure for a
single term can be summarized as follows: Either $\mathcal{P}_{ij}(\beta,
\gamma)$ decreased or $\mathcal{P}_{ij}(\beta, \alpha)$ or
$\mathcal{P}_{ij}(\alpha, \gamma)$ are increased until the submodularity
condition is satisified. This procedure is applicable to any energy function,
and provides a provably good approximation for a single expansion move. 
The authors of \cite{Rother05} limit suitability for the case only a limited amount
of terms are non-submodular.  
\\
In principle there exist more sophisticated Graph Cut algorithms that alter the
mapping of the combinatorial values of the energy to capacities of the edges of
the graph \cite{Kolmogorov07}. In the same work, the authors compare performance
of their  more complicated optimization scheme for non-submodular energies 
to truncating the energy as we did.
For a small percentage of terms violating the submodularity condition no severe
degradation of the performance was found so we stayed with the simpler method,
as it is more accessible and easier to reason about. The implications of
non-submodular terms highly depend on the underlying dataset and the chosen
pairwise energy function.  If, like in case of our label prior term
\eqref{eq:labelterm} the non-submodular case is a rare event, the simple
truncation procedure has a positive impact. Thus, the submodularity violation is a problem that has rather theoretical implications than practical importance for our applications.





\subsection*{Scaling of Graph Cut}

\begin{figure}
	\centering
  \subfloat[ ] {
		    \includegraphics[width=3.4in]{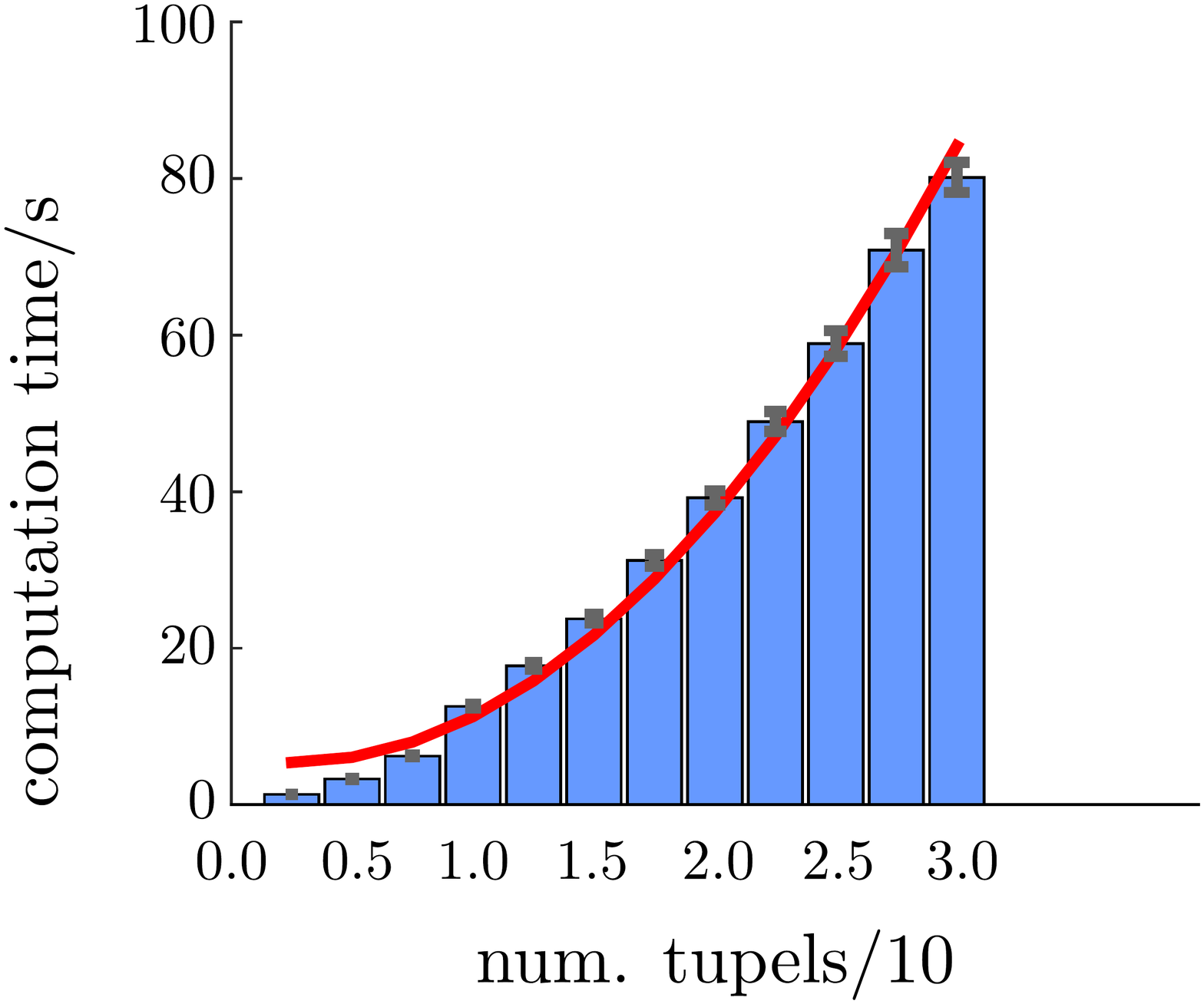}
		    \label{fig:scalingnodes}
	}
  \hfil
  \subfloat[ ] {	
		    \includegraphics[width=3.5in]{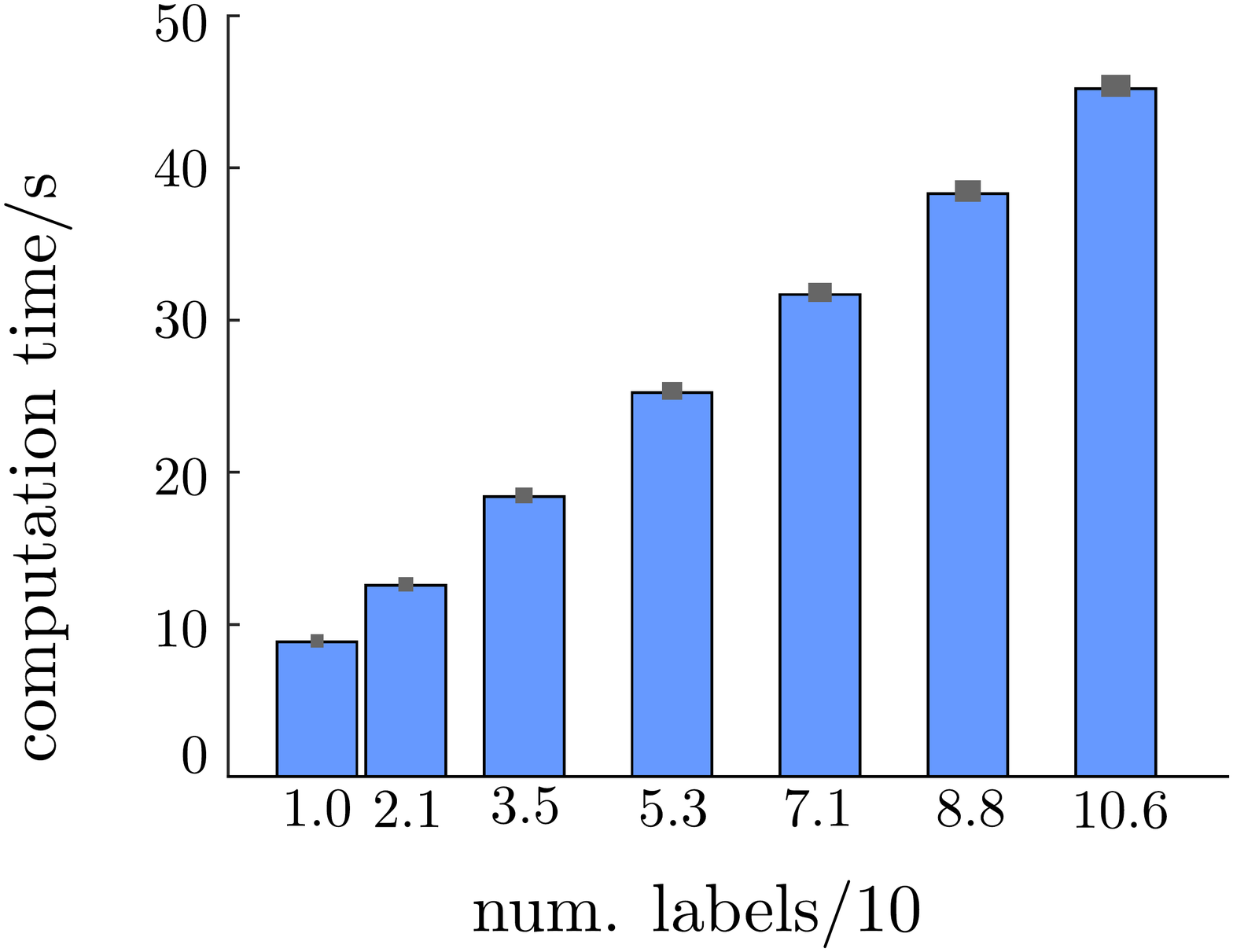}
 			  \label{fig:scalinglabels}
  }	
  	\caption{Mean run time performance of combinatorial clustering stage for 10
      simulated noisy step signals. \ref{fig:scalingnodes} Graph Cut computation
      time versus number of tuples for a label grid of 800  levels. Second order
      polynomial fit to computation times (red curve). \ref{fig:scalinglabels}
    Graph Cut computation time versus label grid size for a fixed system size of
  750 tuples. Linear scaling of performance with increasing number of levels in the label grid set. The error bars are SEM.}
  	\label{fig:gcutscaling}
\end{figure}

When analyzing high-bandwidth noisy time traces of the movement of molecular
motors the CC  step often limits run time performance. Most of all, perfomance
is influenced by system size, i.e. the number of tuples and number of levels in
the label grid set. To analyse the scaling behaviour for these two influences
numerically, we simulated 10 noisy Poisson step signals for each system size and
label grid set and record computation times. fig. (\ref{fig:gcutscaling}) shows
mean and standard deviations as error bars.  In fig.(\ref{fig:scalingnodes})
system size was increased from 250 to 3000  tuples and the number of levels
offered to the combinatorial  optimization problem was kept constant to around
800 levels. In this case computational time is expected to scale mostly with the
complexity of the Boykov-Kolmogorov max flow algorithm which has a worst case
complexity of $\mathcal{O}(|edges| \cdot |nodes|^{2} \cdot C)$
\cite{Kolmogorov03}. Where $C$ is the cost of the minimal cut, $|edges| $ and
$|nodes|$ are respectively the number of edges and nodes in the graph. For the
type of graphs considered here, for each additional tuple in the input data set we have to add two edges which would give a worst case complexity of roughly $\mathcal{O}(N^{3} \cdot C)$. However, computation times fit well to a quadratic function meaning that for our signals the scaling behaviour is better than the worst case complexity (figure \ref{fig:scalingnodes}, red curve).\\
The second case is shown in fig.(\ref{fig:scalinglabels}). When the system size
is fixed (here: $750$  tuples ) and the number of labels increases (here: from $10^{3}$ to ca. $10^{4}$) by refining the label grid subsequently, the corresponding run times increase linear. This is in agreement with the theory behind multi label graph cut problems \cite{Kolmogorov03}. The $\alpha$-expansion offers new labels one by one in a random order until all labels were used and the iteration stops. Thus the observed linear scaling in the number of labels is also expected from theory.

It is important to point out that due to the TVDN compression the expression above is an improvement for this type of step signals (high bandwidth, number of data points $\sim 10^5$ but comparably few steps $<1000$ ) compared to the Fourier transform accelerated HMM implementation \cite{vshmm}: $\mathcal{O}(m\cdot n^{2} N \cdot log_{2} m)$ where $m$ is the number of position states, $n$ the number of molecular states and $N$ the number of data points. Moreover, the direct comparison of run times and memory consumption given in the main text shows that our algorithm is advantageous regarding computational resources compared to existing algorithms.

\subsection*{Comparison of Graph Cut and Markov Chain Monte Carlo}

\begin{figure}[t!]
  \centering{
    \includegraphics[width=3.5in]{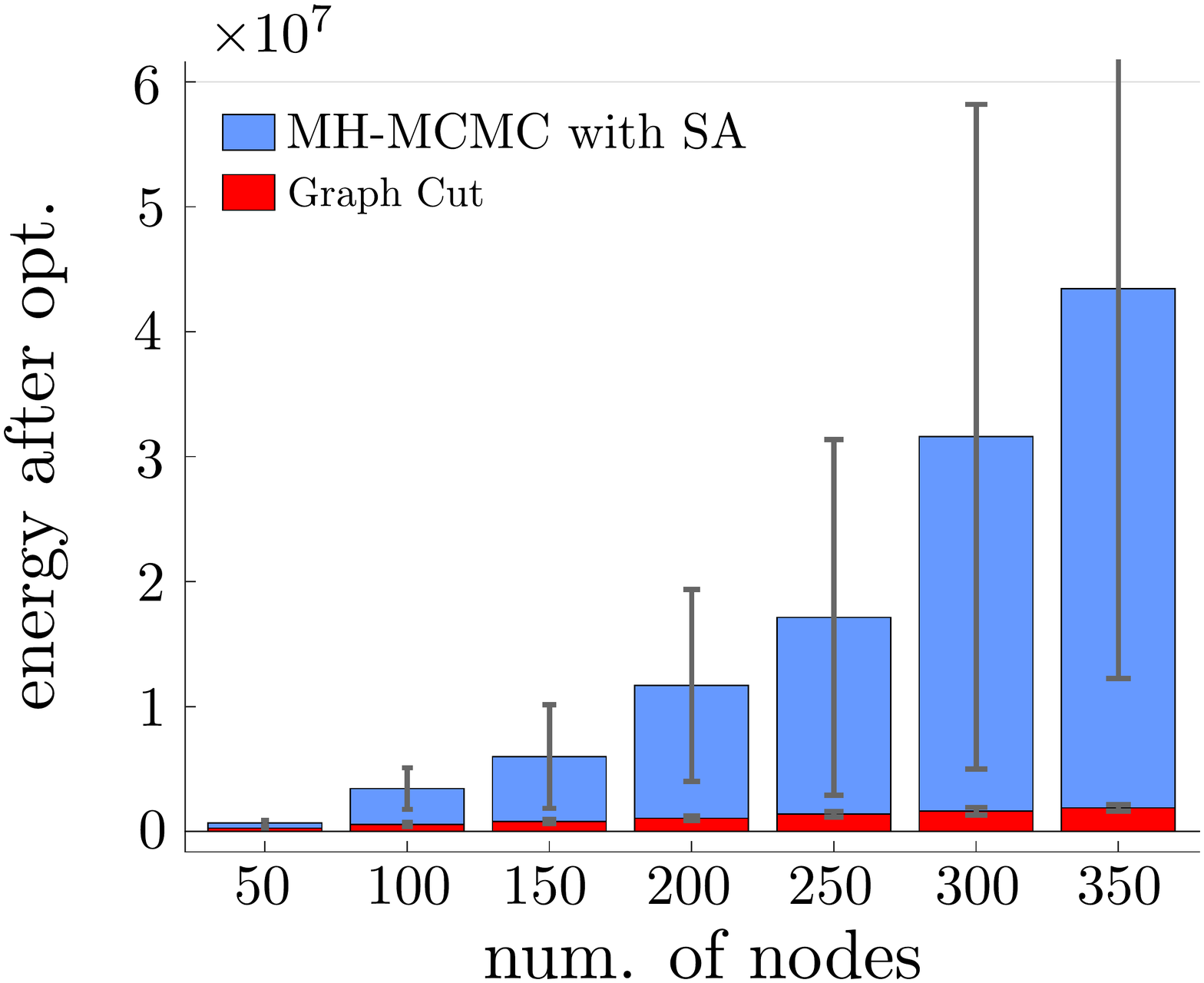}
  }
  \caption{Graph Cut MCMC comparison. Energies of optimal solutions with increasing system size for MCMC and Graph Cut algorihms.}
  \label{fig:rgraphcutmcmc}
\end{figure}

Since Markov Chain Monte Carlo (MCMC) methods are standard techniques to optimize an energy functional with Pott's model terms like Eq.(\ref{eq:labelterm}), we compare the Graph Cut method with a Metropolis Hastings (MH) sampling and simulated annealing (SA) optimization algorithm \cite{simanealopt}.
In each iteration we randomly generate a proposal assignment of labels. The new assignment of a site is accepted or rejected according to the standard  MH rules. Moreover a logarithmic temperature schedule is used for SA. The temperature parameter is introduced as commonly done: $p(\mathbf{x}) \propto exp(-E(\mathbf{x})/T)$. If an accepted proposal has smaller energy than all previous ones it becomes the new configuration that minimizes Eq.(\ref{eq:combinatorialfunctional}). 
To compare the quality of the step detection result we computed the energy,
Eq.(\ref{eq:combinatorialfunctional}) with prior terms Eq.(\ref{eq:labelterm})
for the energy minimizing solutions of Graph Cut and MCMC method,
Fig.(\ref{fig:rgraphcutmcmc}) . For a system size  below $350$ tuples, computation times of the Graph Cut algorithm were always below $10 s$. Since MCMC is computationally more complex longer computation times were used for MCMC, i.e. $45 min$ which allowed for $12$ iterations of a SA temperature cycle. Each cycle consists of $20$ subsequent cooling steps and in each step we iterate through $20000$ proposals.
Inspite of the significantly higher computational cost the MCMC solutions always have higher energies compared to Graph Cut and the excess energy increases for larger systems. This shows that MCMC returns increasingly worse solutions compared to the Graph Cut technique when the number of input data grows for fixed computation time. As expected, Graph Cut shows an approximately linear increase in energy with linearly increasing system size.

To conclude, the plain MCMC algorithm used here is conceptually simpler than Graph Cut but computationally more expansive and also less suited to cluster the denoised steps optimally according to an energy functional. This finding in one dimension is not surprising, since similar observations had been made in 2D image analysis \cite{Boykov01}.

\subsection*{Noisy step simulations}

\begin{figure}[t!]
  \centering{
    \includegraphics[scale=.68]{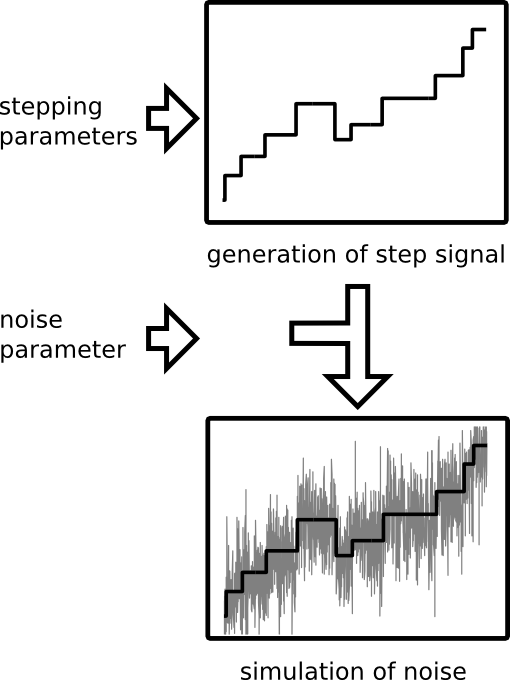}
    }
    \caption{Step and noise simulation procedure. State transition model and corresponding stepping rates determine the probability distribution from which a piecewise constant signals (first inset) is sampled. In a second step noise is simulated with the piecewise constant signals as the mean (second inset). }
    \label{fig:datasimulationmodel}
\end{figure}

\begin{figure}[t!]
  \centering{
    \includegraphics[scale=.35]{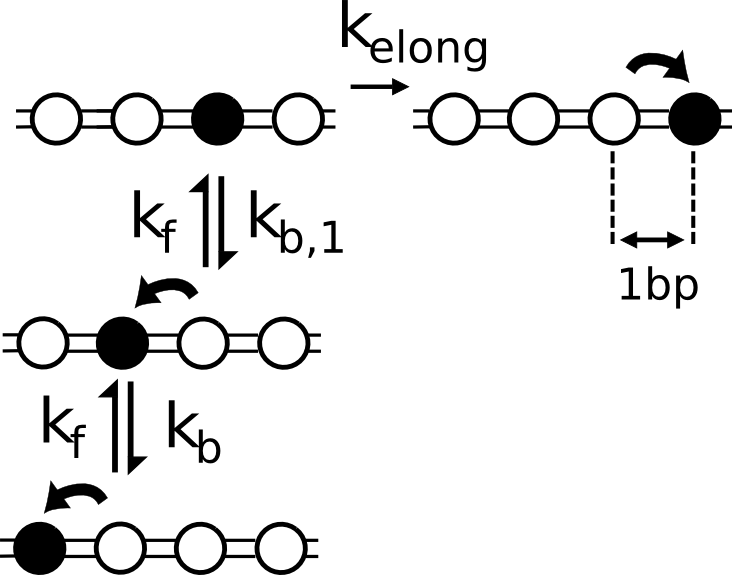}
    }
	\caption{Simplified stepping model of RNAP II with an elongation and backtracked states.}
    \label{fig:pol2stepgeneration}
\end{figure}

Single base pair steps are typically exceeded by noise fluctuations and most of the time it is not possible to judge by eye whether an algorithm correctly positioned steps. Therefore simulated data is necessary to show and compare the performance of step detection algorithms. We generate noisy steps in two stages as outlined in Fig.(\ref{fig:datasimulationmodel}).
First, we generate a piecewise constant signals according to a simplified version of the linear ratchet model of Pol II \cite{completedissect}. This model contains elongation and backtracked states and reproduces the ability to pause \cite{kashlrnap, blockbacktr}, but does not accurately reflect the temporal order of translocation and other
enzymatic processes. During elongation, $1bp$ forward steps are generated with an effective rate of $k_{elong}$. This effective rate includes the process of translocation, NTP insertion and pyrophosphate release. In our model catalysis, bond formation and $PP_i$ release are summarized by a rate $k_{3}$. Furthermore, the NTP-binding net rate is $k_{2} = c_{NTP}\cdot k_{3}/K_{D}$ and the translocation net rate $k_{1} = k_{+}\cdot k_{2}/(k_{-} + k_{2})$. $c_{NTP}$ is the NTP concentration, $K_{D}=9.2 \mu M$ the dissociation constant, $k_{+}=88 Hz$ is the forward translocation rate of Pol II and $k_{-}=680 Hz$ backward translocation. The values of these constants are known from experiments \cite{completedissect}.
The elongation rate is then determined by $k_{elong} = (1/k_{1} + 1/k_{2} + 1/k_{3})^{-1}$. \\
With a rate of $k_{b1}=5Hz$ the motor makes a backward step of identical size as the forward step and thus enters the backtracked state. The enzyme can further backtrack by a rate $k_{b}=1.3 Hz$ or return to the original state with a rate $k_{f}=1.3 Hz$ (figure \ref{fig:pol2stepgeneration}).\\
The rates corresponding to a forward step ($k_{+}$, $k_f$) or backward step ($k_{b1}$, $k_{b}$, $k_{-}$ ) are modified under external forces according to $k(F) = k(0) \cdot exp(\pm F \cdot 0.17nm / (k_{b}T))$, where $k_{b}T = 4.11 pN\cdot nm$ and the plus sign in the exponent applies to rates of forward steps. Simulations were computed for an assisting force of $6.5 pN$. At this force forward and backward diffusion rates are $k_{b} =3.8 Hz$, $k_{b} =1.0 Hz$ and $k_{f} = 1.7 Hz$, in accordance with the kinetic model. For numerical simulation purposes the rates above are divided by the simulation's time increment to yield dimensionless quantities.\\
The transitions between elongation and backtracked states are generated using the Gillespie stochastic simulation scheme \cite{gillespie} for a single enzyme. Dwell times are sampled from an exponential distribution according to the respective rates.\\
In a second step, we simulated experimental noise including effects of confined brownian motion of trapped micro spheres. To accurately reflect the experiment, we take into account changes in the  tether length and in the tether stiffness due to motion of the enzyme. We apply a harmonic description of the trapping potentials and assume that the DNA linker can be described by a spring constant $k_{DNA}$ determined by the worm like chain model \cite{wlcdna}. \\
To formulate the equation of motion of two trapped micro spheres tethered by DNA we choose the coordinate system such that the enzyme moves in $x$-direction. Furthermore we assure that drag coefficients $\gamma_{i}$ and the trapping stiffness $k_{trap,i}$ are identical in both traps. With this the effective DNA length $x$ can be described by the following equation.
\begin{equation}
\gamma\dot{x} = -k \cdot x +  F_{T}(t)
\label{eq:trapeqsimplemain}
\end{equation}
where $k = k_{trap} + 2k_{DNA} $, $k_{DNA}$ is the DNA stiffness and $\gamma $ is the drag coefficient. $F_{T}(t)$ is the thermal force which is treated as gaussian white noise: $\langle F_{T}(t)\rangle = 0$ and $\langle F_{T}(t) F_{T}(t')\rangle = 2k_{B}T\gamma \delta(t-t')$.
Eq.(\ref{eq:trapeqsimplemain}) describes a so called Ornstein-Uhlenbeck process and can be solved and simulated by standard techniques of stochastic differential equations \cite{kloedenplaten} which is shown in the next subsection.
Eq.(\ref{eq:trapeqsimplemain}) was derived for the static situation without positional changes. However, a molecular motor which is attached between micro spheres by a DNA-tether will change the tether length during its activity. Thus, $k_{DNA}$ is also changing and can be computed using the worm-like chain model \cite{wlcdna}.\\
In the simulations we use a trap stiffness of $k_{trap} = 0.25 pN/nm$, a drag coefficient of $\gamma = 0.8\cdot 10^{-5} pN\cdot s/nm$ corresponding to beads with $850 nm$ diameter and an initial length of  $L=3kbp$ for the DNA tether.

We simulated a slow, an intermediate and a fast scenario which differ by stepping speed, sampling frequency, number of data points and noise amplitudes. Sampling frequencies and number of data points of the slow scenario are $f=5kHz$ and $N = 2.5 \cdot 10^{5}$ points, for the intermediate scenario: $f=2kHz$ and $N=10^{5}$ points and for the fast scenario: $f = 1kHz$ and $N = 5 \cdot 10^{4}$. The elongation rate $k_{elong}$ of the slow scenario $k_{elong} = 4.1 Hz$ can be expected at a NTP concentration of $c_{NTP} = 7mM$. Since backtracking becomes more likely at these NTP concentrations we limited analysis to simulated data that shows a net forward translocation. This  excludes analysis of simulated data which exhibits only backtracked states. Elongation rates of the intermediate ($k_{elong} = 9.1 Hz$) and fast scenario ($k_{elong} = 25.8 Hz$) are expected at $c_{NTP} = 20mM$ and $c_{NTP} = 1000mM$ respectively. The standard deviation of noise amplitudes are directly computed from the noisy input data. This is done by subtracting the simulated step signal from the noisy steps and computing the standard deviation of the remaining signal. In both scenarios, slow and intermediate, the computed standard deviation is $5.5 bp$ at the given sampling frequency. For the fast scenario we choose  $N=5 \cdot 10^{4}$ data points and $1kHz$ sampling rate. Moreover in the fast scenario we use higher noise amplitudes with a computed standard deviation of  $10.0 bp$ at the $1kHz$ sampling frequency.\\
Finally, for all three scenarios 25 data sets were simulated and analyzed.
Table \ref{tab:overviewSim} gives an overview over the simulation parameters.
\begin{table*}[!bht]
\begin{center}
\caption{Overview of simulation parameters. Shown is elongation rate $k_{elong}$, corresponding NTP concentration $c_{NTP}$ and rate constants of the backtracking state. Moreover, the standard deviation of noise $\sigma_{n}$, sampling frequency $f$ and number of data points $N$ is given.}
\label{tab:overviewSim}
\begin{tabular}{ccccccccc}\hline
scenario: & $k_{elong}/Hz$ &  $c_{NTP}/mM$ &$k_{b1}/Hz$ & $k_{b}/Hz$ & $k_{f}/Hz$   & $\sigma_{n}/bp$  & $f/kHz$ & $N$ \\ \hline
slow  &  $4.1$ & $7$ & $3.8$ & $1.0$ & $1.7$ &$\sim 6$ & $5$ & $2.5 \cdot 10^{5}$ \\
intermediate  &  $9.1$ & $20$ & $2.3$ & $1.0$ & $1.7$ &$\sim 6$ & $2$ & $1 \cdot 10^{5}$ \\
fast  &  $25.8$ & $1000$ & $2.3$ & $1.0$ & $1.7$ & $\sim 10$ & $1$ & $5 \cdot 10^{4}$ \\  \hline
\end{tabular}
\end{center}
\end{table*}

\subsection*{Simulating beads in a harmonic optical trap}

As described above we account for confined brownian motion of trapped beads in a dual trap optical tweezers. A harmonic description of trapping potentials is applied and we assume the DNA linker can be described by a WLC model with a spring constant $k_{DNA}$. In the following we briefly show the derivation of eq.(\ref{eq:trapeqsimplemain}) and its solution. We focus on the x-coordinates of two beads trapped in different optical traps and tethered by DNA. The equation of motion of such a system of reads \cite{diffdualtrap}:
\begin{equation}
\boldsymbol \gamma \dot{\mathbf{x}} = -\boldsymbol \kappa \mathbf{x} + \boldsymbol F_{T} (t)
\label{eq:trapeqomot}
\end{equation} 
where $\mathbf{x} = (x_{1},x_{2})$ is the x-coordinate of first and second bead. Furthermore drag coefficient, stiffness and thermal force are:
\begin{equation}
\begin{aligned}
& \boldsymbol \gamma = \begin{pmatrix} \gamma_{1} & 0 \\ 0 & \gamma_{2} \end{pmatrix}, \\ 
& \boldsymbol \kappa = \begin{pmatrix} k_{trap,1}+k_{DNA} &  -k_{DNA} \\ -k_{DNA} & k_{trap,2}+k_{DNA} \end{pmatrix}, \\
& \boldsymbol F_{T}(t) = \begin{pmatrix} F_{T,1}(t) \\ F_{T,2}(t) \end{pmatrix} \nonumber
\end{aligned}
\end{equation}
The thermal force fulfills the gaussian white noise properties: $\langle \xi_{i}(t)\rangle = 0$ and $\langle F_{T,i}(t) F_{T,j}(t')\rangle = 2k_{B}T\gamma \delta_{ij}\delta(t-t')$
The relative coordinate $\tilde{x} = x_{2} - x_{1}$ which will be called x in the following can be simplified by assuming that $\gamma_{1} = \gamma_{2} = \gamma$ and $k_{trap,1} = k_{trap,2} = k_{trap}$ to:
\begin{equation}
\dot{x} = -2 \pi f_{c} \cdot x + \frac{1}{\gamma} F_{T}
\label{eq:trapeqsimple}
\end{equation}
Where $f_{c} = \left(k_{trap} + 2k_{DNA} \right) / 2 \pi \gamma $ is the corner frequency of the system. $k_{DNA} = k_{DNA}(F,L)$ depends on force and length of the DNA tether and is calculated from the wormlike chain model \cite{stretchdna}. During enzyme stepping $k_{DNA}$ has to be updated repeatedly with respect to the external parameters force $F$ and length $L$.
Eq.(\ref{eq:trapeqsimple}) describes a so called Ornstein-Uhlenbeck process (OU) and can be solved and simulated by standard techniques of stochastic differential equations \cite{kloedenplaten}. From Eq.(\ref{eq:trapeqsimple}) it can be seen that for timescales slower than the corner frequency $f_{c}$ noise behaves essentially as white noise. For faster timescales than $f_{c}$ noise rather has characteristics of brownian motion. In the following, the simulation of an Ornstein-Uhlenbeck process is described. 
We rewrite Eq.(\ref{eq:trapeqsimple}) in Ito form:
\begin{equation}
dx_{t} = -k \cdot x \cdot dt + \sqrt{2D} \cdot dW_{t}
\label{eq:itoharmonic}
\end{equation}
where $k=(\kappa+2k_{DNA})/\gamma$, $D=k_{B}T/\gamma$ is the diffusion constant and  $dW_{t}$ infinitesimally describes brownian motion. For a finite time interval $\Delta W_{t} = \int_{t-h}^{t} dW_{t'}$ describes a standard normal distributed random variable $\mathcal{N}(0,h)$, with standard deviation $\sigma = \sqrt{h}$.\\
Eq. (\ref{eq:itoharmonic}) just describes a gaussian random variable with mean $\mu$ and variance $\sigma^{2}$ \cite{simbrowndyn}:
\begin{equation}
x_{t} \in \mathcal{N}(\mu,\sigma^{2}) = \mathcal{N} \left(x_{t-1}e^{-k t}, \frac{D}{k} \left( 1 - e^{-2k t}
\right) \right),
\label{eq:oudrawgauss}
\end{equation}
and a random path can be straightforwardly simulated starting from an initial position $x_{0}$.

\subsection*{Details of algorithm comparison}

\begin{figure}
  \subfloat[] {	
		    \includegraphics[width=3.4in]{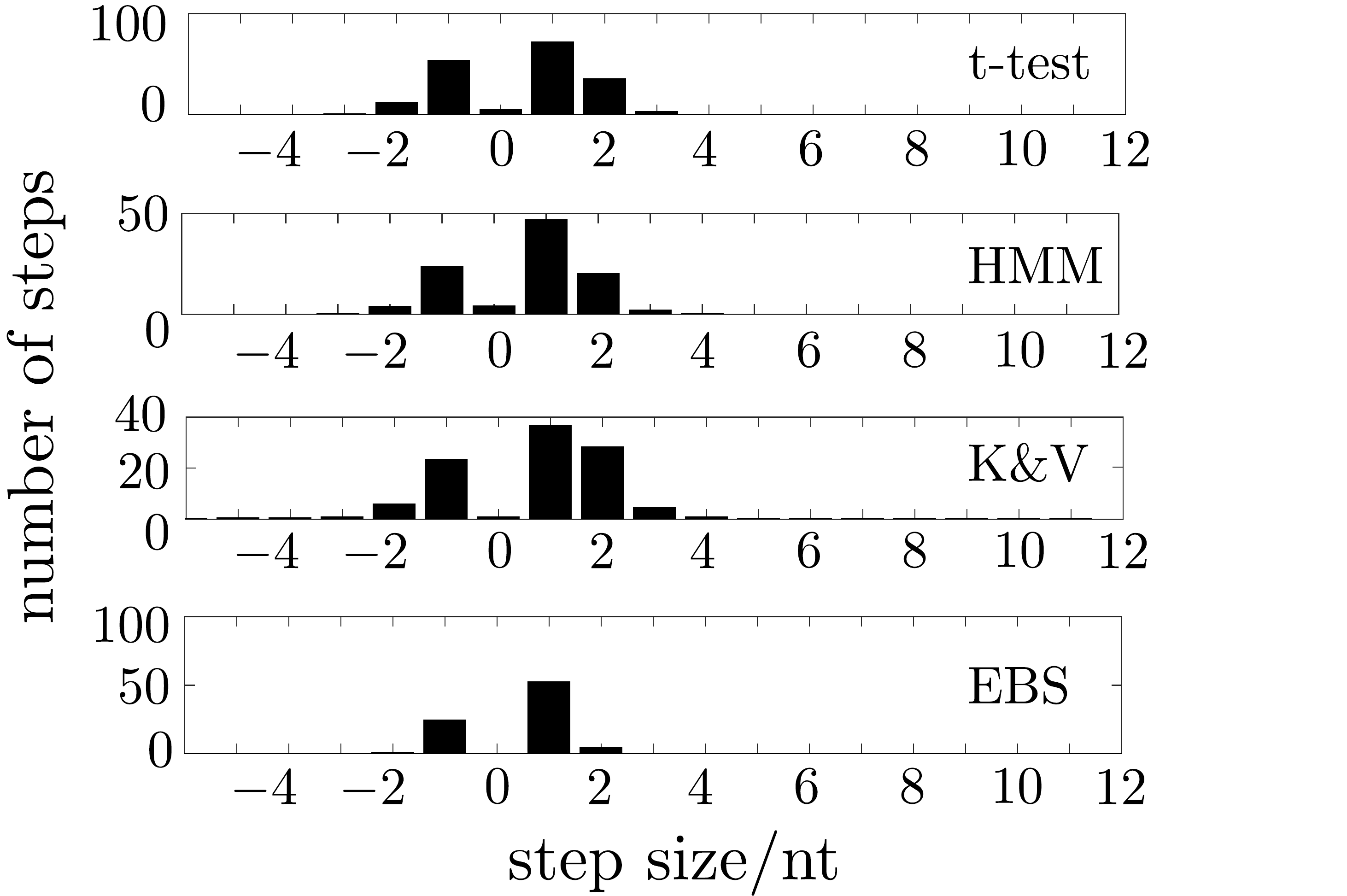}
		    \label{fig:addstepsizehisteasy}
  }
  \hfill 
  \subfloat[] {	
		     \includegraphics[width=3.4in]{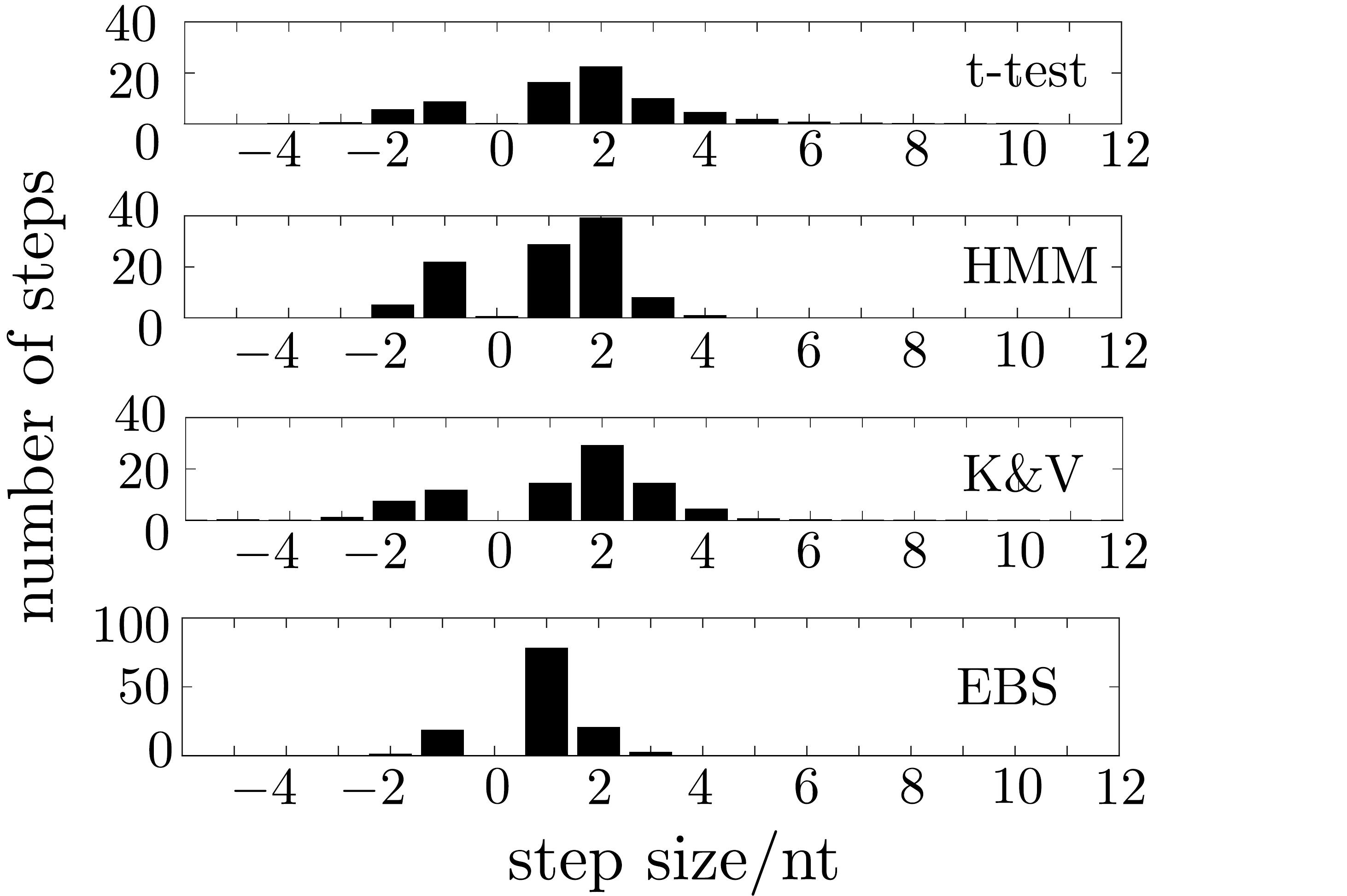}
 			 \label{fig:addstepsizehistmedium}
	}  
  	\caption{Histogram of step sizes of the slow scenario \ref{fig:addstepsizehisteasy} and the intermediate scenario \ref{fig:addstepsizehistmedium} for t-test, HMM, K \& V  and EBS (from upper to lower histogram).}
  	\label{fig:addstepsizehist}
\end{figure}

To achieve best results for the three simulation scenarios we need to tune the external parameters of the t-test, the HMM and the EBS algorithm. While the latter is described in the main text (only for the slow scenario we used $\rho_S = 1.5$ instead of $\rho_S = 2$) we will  briefly explain how to adapt the other two algorithms to yield as many correct steps as possible but also to have a large fraction of correct steps among the found steps.\\
For the t-test a minimum step size of $0.3 nm$ and a shortest dwell of $10 ms$ was used. Moreover, the t-test threshold was $0.01$, the binomial threshold $0.005$ and the maximal number of iterations was $100$.\\
The HMM analysis was conducted with maximally $100$ iterations for maximum likelihood estimation of transition probabilities. More iterations did not give better results and fewer iterations ($\leq 10$) could not optimize the log-likelihood properly (data not shown). For the slow scenario $85$ states were used and for the intermediate and fast scenario we used $140$. To prevent memory overflow in the intermediate scenario, we performed box car averaging to reduce the number of data points by a factor of two. Furthermore, a grid spacing of $1/2$ $bp$ was used which proved to be better than a $1bp$ spacing. Since the HMM level grid has to be aligned by using noisy data as an input, a two times smaller grid spacing showed better results. In contrast the situation is advantegous in case of combinatorial clustering which constructs the level grid on already denoised data.\\
The algorithm from Kalafut and Visscher has the great advantage that it works completely without parameters that have to be adjusted by the user.\\
To complete the performance results of the algorithm comparison (figure \ref{fig:compperformance}) the step size histograms of the step detection result for easy and intermediate scenario are given, fig.(\ref{fig:addstepsizehist}). For slow stepping rates the step size histograms resemble the simulated step size $\pm 1bp$ quite well, fig.(\ref{fig:addstepsizehisteasy}). However, a deviation from the $1bp$ steps can be already seen in the intermediate scenario for the t-test and HMM, fig.( \ref{fig:addstepsizehistmedium}). For EBS the majority of the detected steps are $1bp$ in size in both scenarios. The more difficult situation in the intermediate scenario is reflected by the larger fraction of $2bp$ steps (figure \ref{fig:addstepsizehistmedium}, red).\\

Table \ref{tab:dataptsruntime} summarizes computational speed and memory consumption of the algorithms for test runs with $100 s$ of temporal length and rate constants according to the intermediate scenario. We simulated data containing $\sim 900$ simulated steps and successively increased the number of data points by increasing the sampling frequency. For each sampling frequency the standard deviation of noise was constant. EBS processed $2\cdot 10^7$ data points, simulated with $200kHz$ sampling frequency in $4min$. The other algorithms had comparably long run times and we restricted computation times to $\sim 50 min$ and memory consumption to a limit of $2.38GB$. EBS can process much more data points at comparably short run time and is essentially limited only by the size of available memory. The high bandwidth signals can be compressed very well by TVDN to a few thousand plateaus, which yields shorter run times for the subsequent CC. 
In contrast the t-test becomes slower at $>10^{5}$ data points since it has more possibilities to adapt window sizes. A limiting factor in case of a lot of data points for the HMM beside processing speed is memory consumption of the Viterbi reconstruction. Taken together the more efficient computation of EBS compared to the other algorithms allows for the analysis of high bandwidth data. This in turn increases can increase the performance of step-finding. \\
To show how much the performance of step detection improves we use the intermediate scenario but increase the bandwidth (i.e. number of data points) from $2kHz$ to $200 kHz$. In order to have the same standard deviation of noise at $2kHz$ sampling rate, the noise amplitude of the high bandwidth signal is increased accordingly. This increases precision and recall from ($\sim 30\%$ and $\sim 60 \%$) at the lower bandwidth to  ($\sim 40\%$ and $\sim 60 \%$) at the higher bandwidth. For CC the same set of parameters is used for low and high bandwidth signal (methods). Due to the very fast denoising stage and the efficient compression to tuples run times are still below $3 min$ for $10^{7}$ data points and $200-300$ steps.

\begin{table}[!bht]
\begin{center}
\caption{Comparison of computation efficiency of the different step-finding algorithms. $\sim 900$ simulated steps on commodity hardware (i7-2600, $3.6GHz$ CPU Ubuntu System, $4GB$ memory). Corresponding run times were recorded in matlab for the signal of the given size. Peak memory usage, i.e. resident set size (RSS) was measured with Linux's proc information system.}
\label{tab:dataptsruntime}
\begin{tabular}{lccccc}\hline
 & t-test &  HMM  & K \& V  & EBS\\ \hline
 data points/$10^5$ & $2.5$ &  $2.5$ &  $2.5$ &  $2.5$ \\ 
run time/$s$ & 708 & 4858 & 2555 &  5  \\ 
peak RSS/$GB$ & 0.17  & 2.38 & 0.14   & 0.15 \\ \hline
\end{tabular}
\end{center}
\end{table}

\subsection*{Remarks on Example of TVDN and combinatorial clustering}

\begin{figure}[t!]
  \centering{
    \includegraphics[width=3.4in]{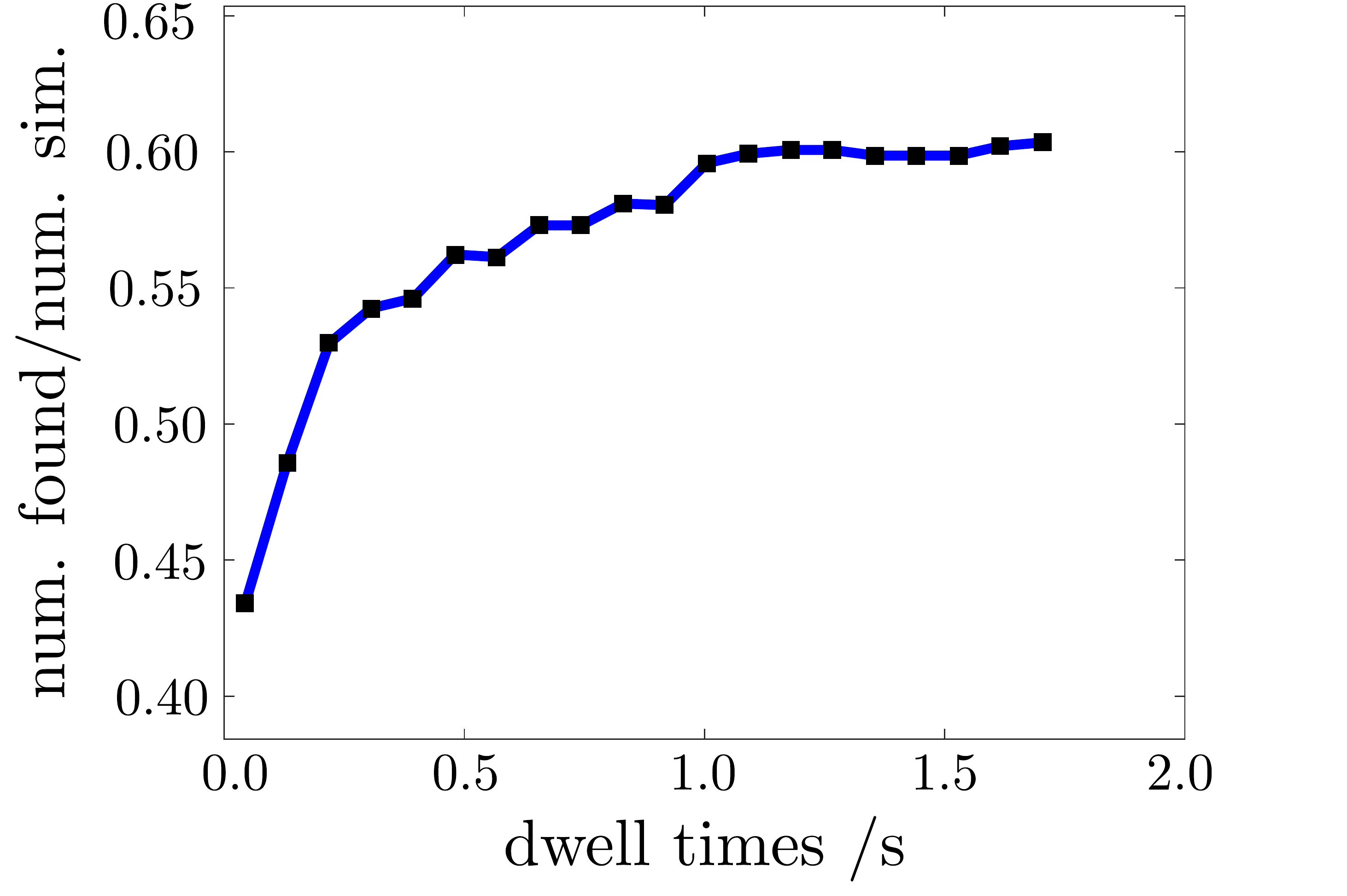}
  }
  \caption{Cumulative fraction of found steps for the EBS in the intermediate scenario. Plotted is the cumulative number of found steps/simulated steps of the signal plotted in Fig.(\ref{fig:tvdngcutresult}) for different dwell times. The number of detected steps compared to simulated ones is smaller for short dwell time steps. Dwell time histograms with a binning of $87.3 ms$ were determined for the detected and simulated steps respectively. The number of detected steps for each dwell time was divided by the corresponding number of simulated steps and cumulatively summed up. In total $60.5\%$ of the number of simulated steps were found.}
  \label{fig:tvdnebsexamplemissingsteps}
\end{figure}

In order to get temporal information of the missing steps in the example given in the Results \& Discussion section, fig.(\ref{fig:tvdngcutresult}), we compare the dwell time histograms of the simulated and detected steps. The cumulative fraction of found steps for a certain dwell time shows that steps with short dwell times are omitted with higher probability, fig.(\ref{fig:tvdnebsexamplemissingsteps}).

\subsection*{Effect of prior information in combinatorial clustering}

\begin{figure*}
	\centering
    \subfloat[] {	
		     \includegraphics[width=2.1in]{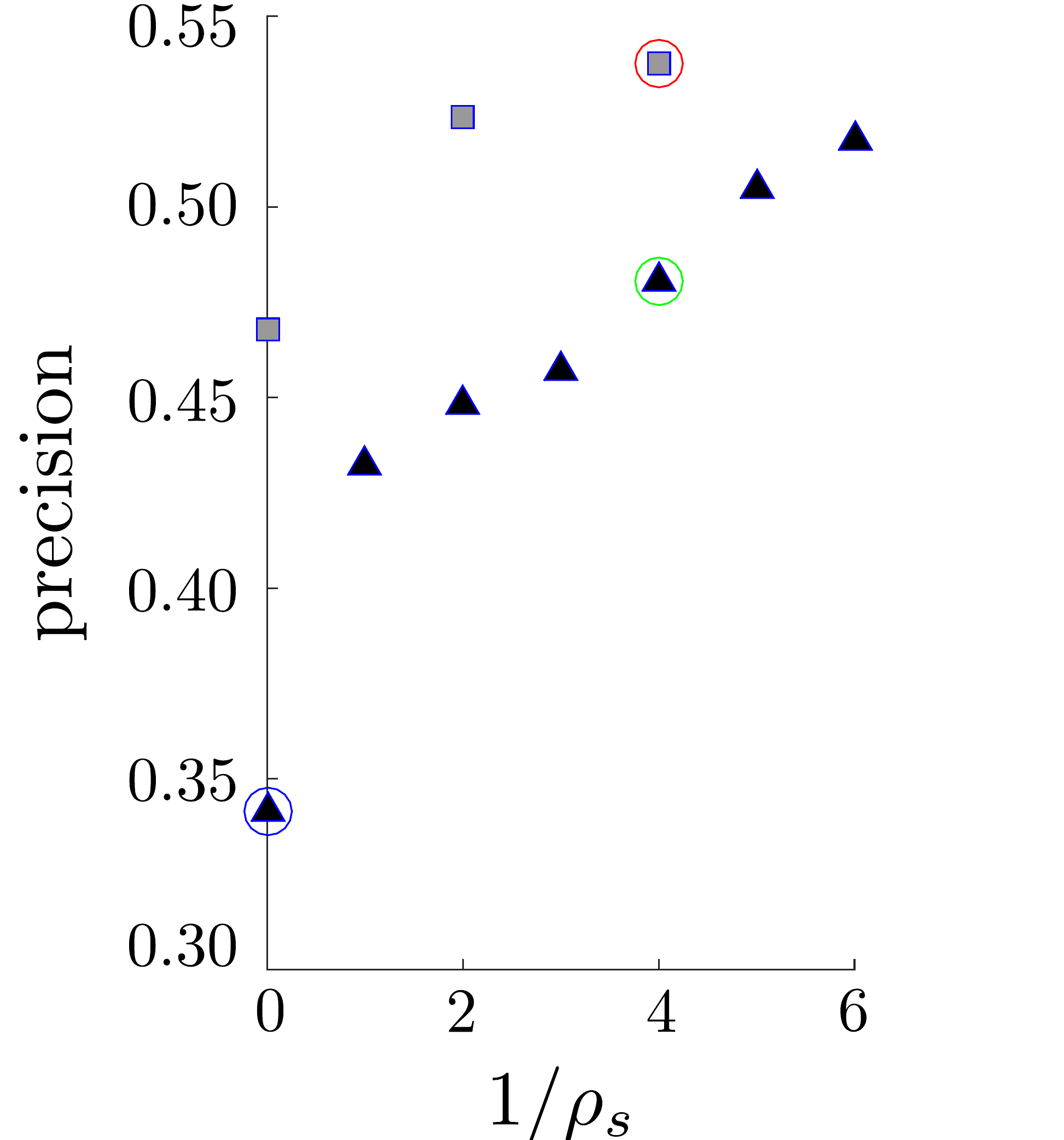}
			 \label{fig:effectoflabel}
	  }
    \subfloat[] { 
		     \includegraphics[width=2.1in]{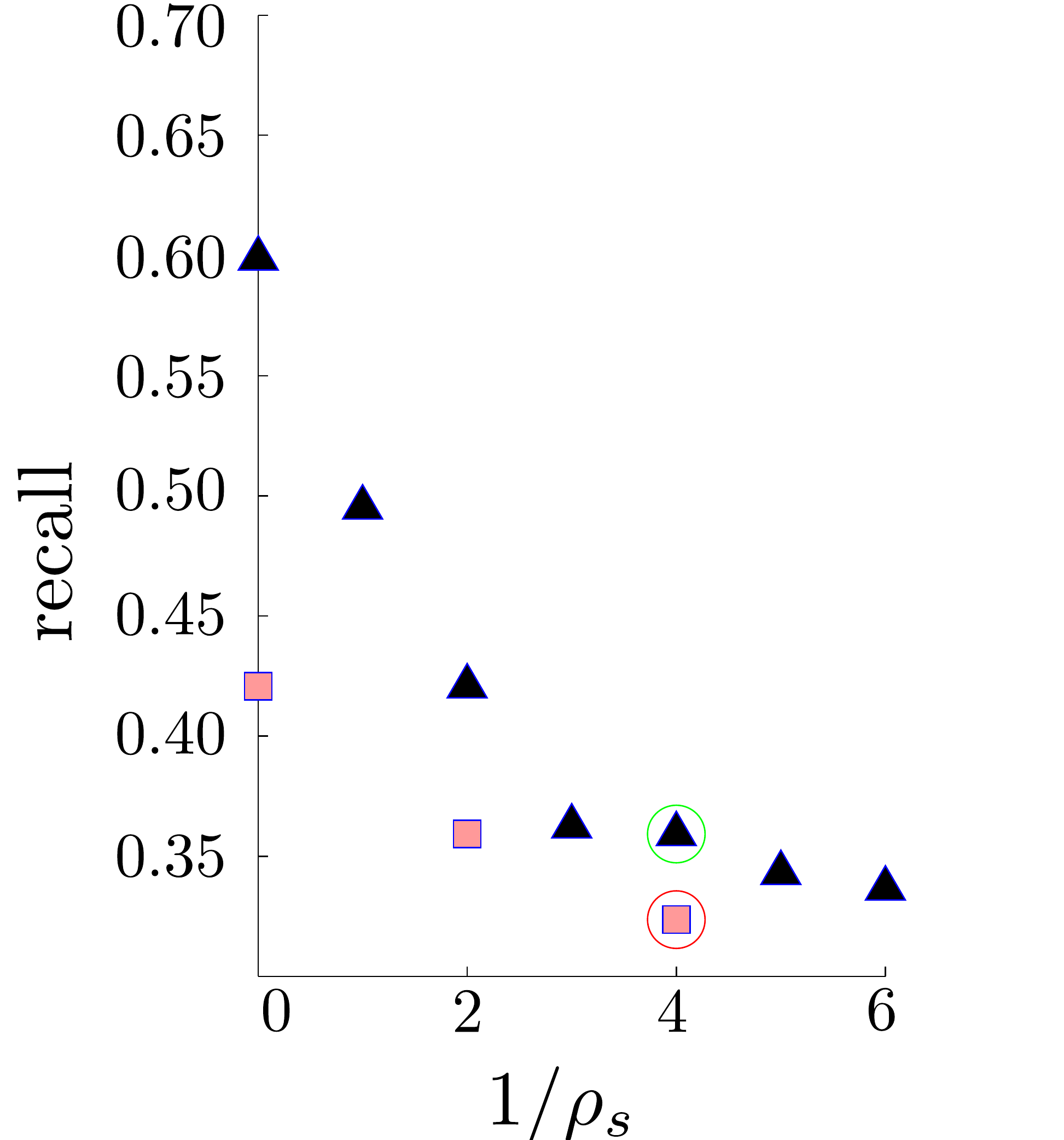}
	  }
    \subfloat[] { 
		    \includegraphics[width=2.1in]{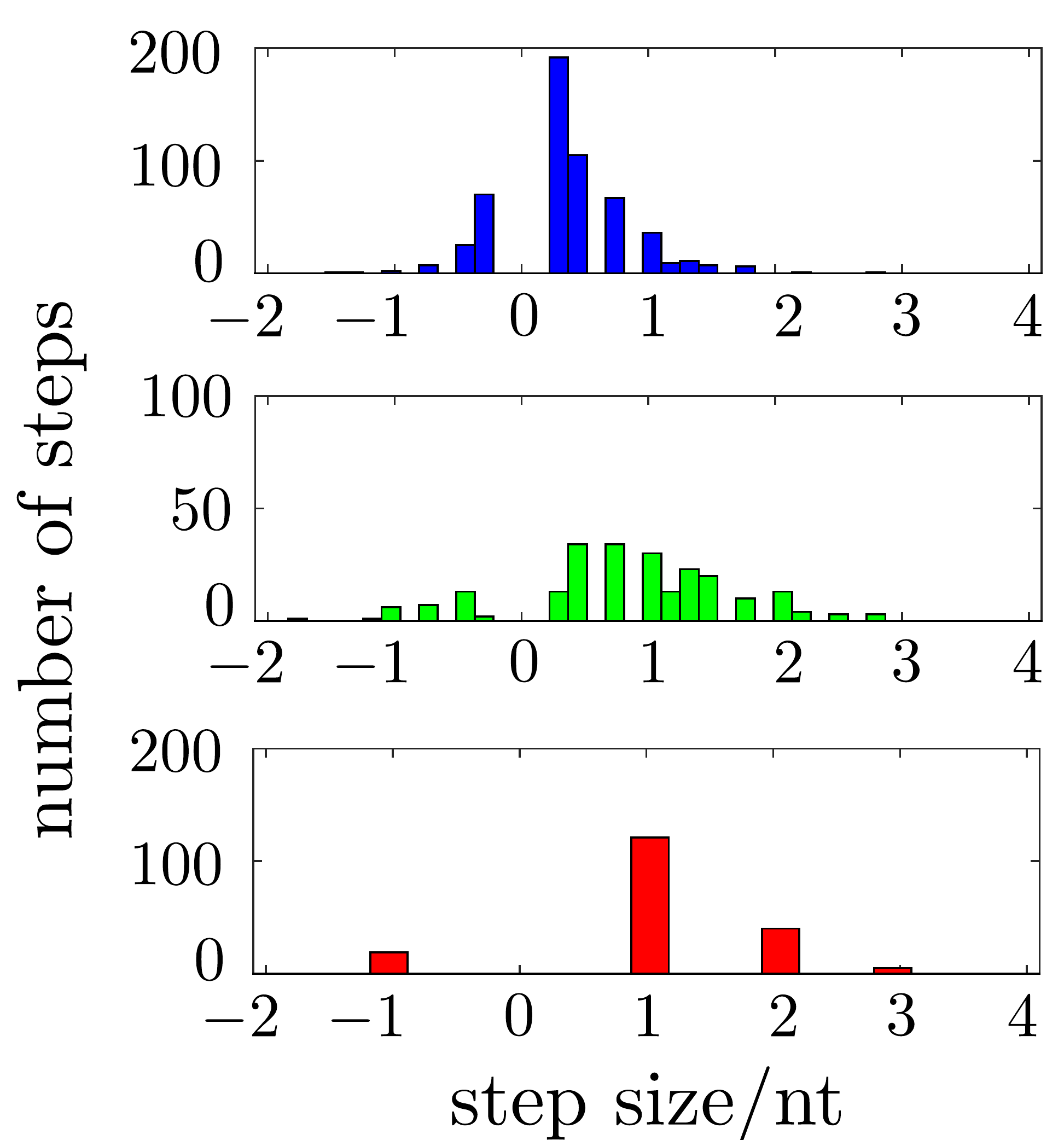}
		    \label{fig:labelpriorstepsizes}
	  } 
 	\caption{Prior terms and level grid regularize combinatorial clustering.
    (\ref{fig:effectoflabel}) Relative frequencies of correct steps among the
    number of detected steps (precision) as a function of prior potential
    strength $1/\rho_{S}$ ($\rho_{S}/\rho_{P}=0.08$ is kept constant) for
    simulated data using the intermediate scenario. Shown is the precision for
    clustering with a level grid of $1/4bp$ spacing (black triangles) and with a
    spacing of $1 bp$ (grey squares). (\ref{fig:labelpriorstepsizes}) step size
    histograms of detected steps with a label grid of $1/4bp$ without prior
    terms (blue), with a spacing of $1/4bp$ and prior terms (green) and with a
    spacing of $1bp$ and prior terms of the same strength (red). The computed
    precision corresponding to the three histograms is encircled with the
    respective color, Fig.(\ref{fig:effectoflabel}).} \label{fig:labelprior}
\end{figure*}

To analyze the impact of the prior terms and level grid spacing on step detection quality we performed CC with different prior potential strength and  level grid spacing (figure \ref{fig:effectoflabel}). We varied the prior regularization parameters $\rho_{S}$ and $\rho_{P}$ starting from $\rho_{S}=0$ to a maximum of $\rho_{S}=6$, while the jump-height prior parameter was varied simultaneously such that $\rho_{P}/\rho_{S}=12.5$ remained constant. By increasing the prior regularization parameters the precision of step detection can be  increased (triangles, fig.(\ref{fig:effectoflabel})). Furthermore, precision can be improved by choosing a level grid with a spacing of the simulated step size of $1bp$ (squares, fig.(\ref{fig:effectoflabel})).\\
The best result regarding the absolute number of correct steps and a small number of falsely
detected steps which lie outside a certain time window around an actual step
(Methods) was found for $\rho_{S}=4$, $\rho_{P}=50$ ( \ref{fig:effectoflabel}).\\
This increases the number of correct steps from $34 \%$ of the steps found at
vanishing prior potential to $48 \%$ for the $1/4 bp$ level grid analysis
(supplementary material). By increasing the spacing to $1bp$ the detection
precision can be improved even further to a fraction of $54 \%$ correct 
steps of the found steps.
\\
The prior potential strength of the smoothing term $\rho_{S}$ can not be
arbitrarily large since it would remove steps in favor of fewer large steps and
thus reduces the number of correctly recovered simulated steps.
The variation of the prior term also effects step size histograms (figure \ref{fig:labelpriorstepsizes}). When no prior terms are present ($\rho_{S} = \rho_{P} = 0$) the detected step-size is oftentimes smaller than the simulated step-height (figure \ref{fig:labelpriorstepsizes}, upper panel). Optimization of the regularization parameters as well as an increase in level spacing improves the precision of the EBS algorithm as shown in the histograms of detected step-sizes (figure \ref{fig:labelpriorstepsizes}, middle and lower panel).\\

\subsection*{EBS application to experimental data of $\varphi 29$ bacteriophage}

\begin{figure*}
	\centering
    \subfloat[] {	
		     \includegraphics[width=2.1in]{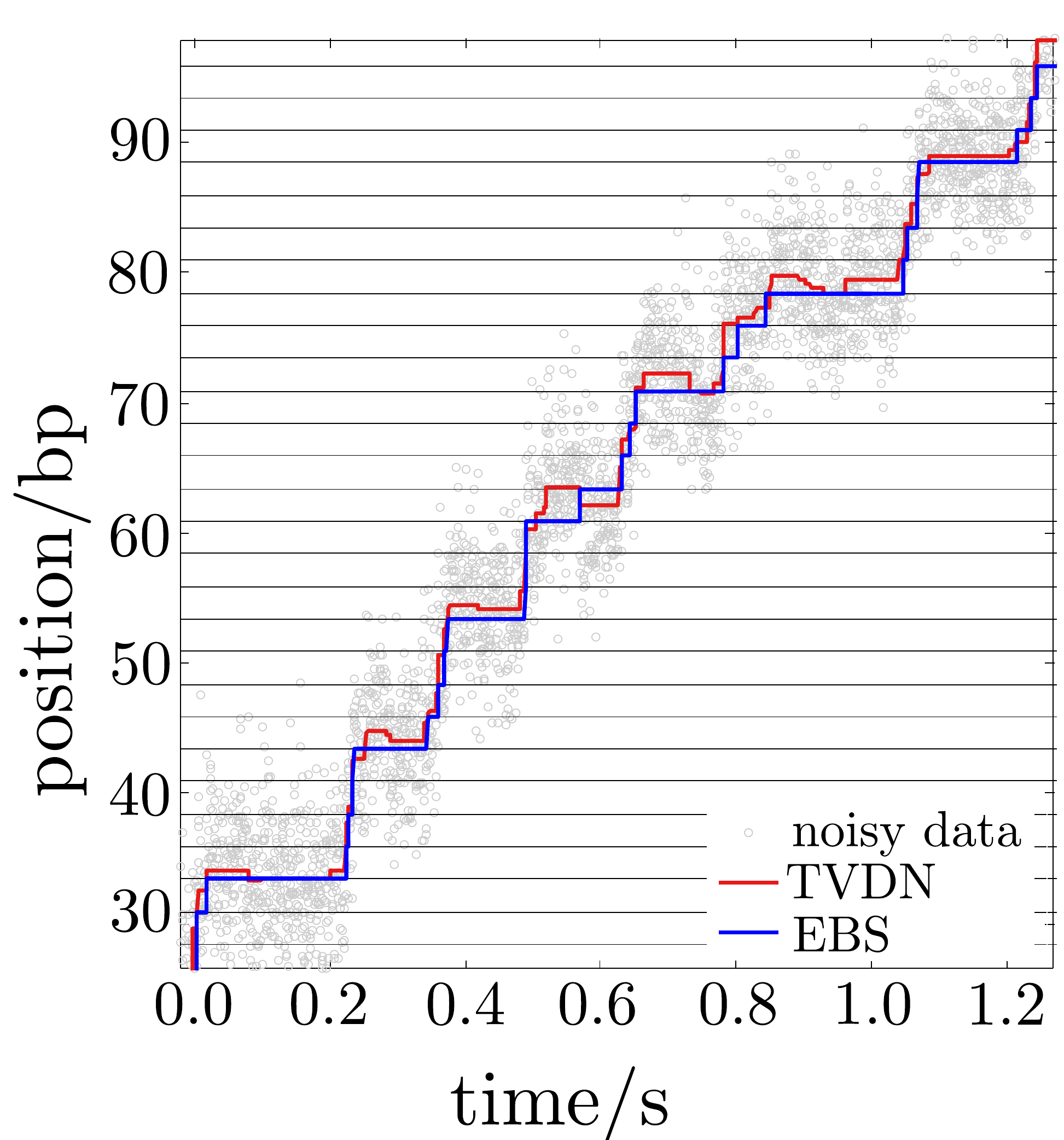}
			 \label{fig:phi29example1}
     }
     \subfloat[] {
          \includegraphics[width=2.1in]{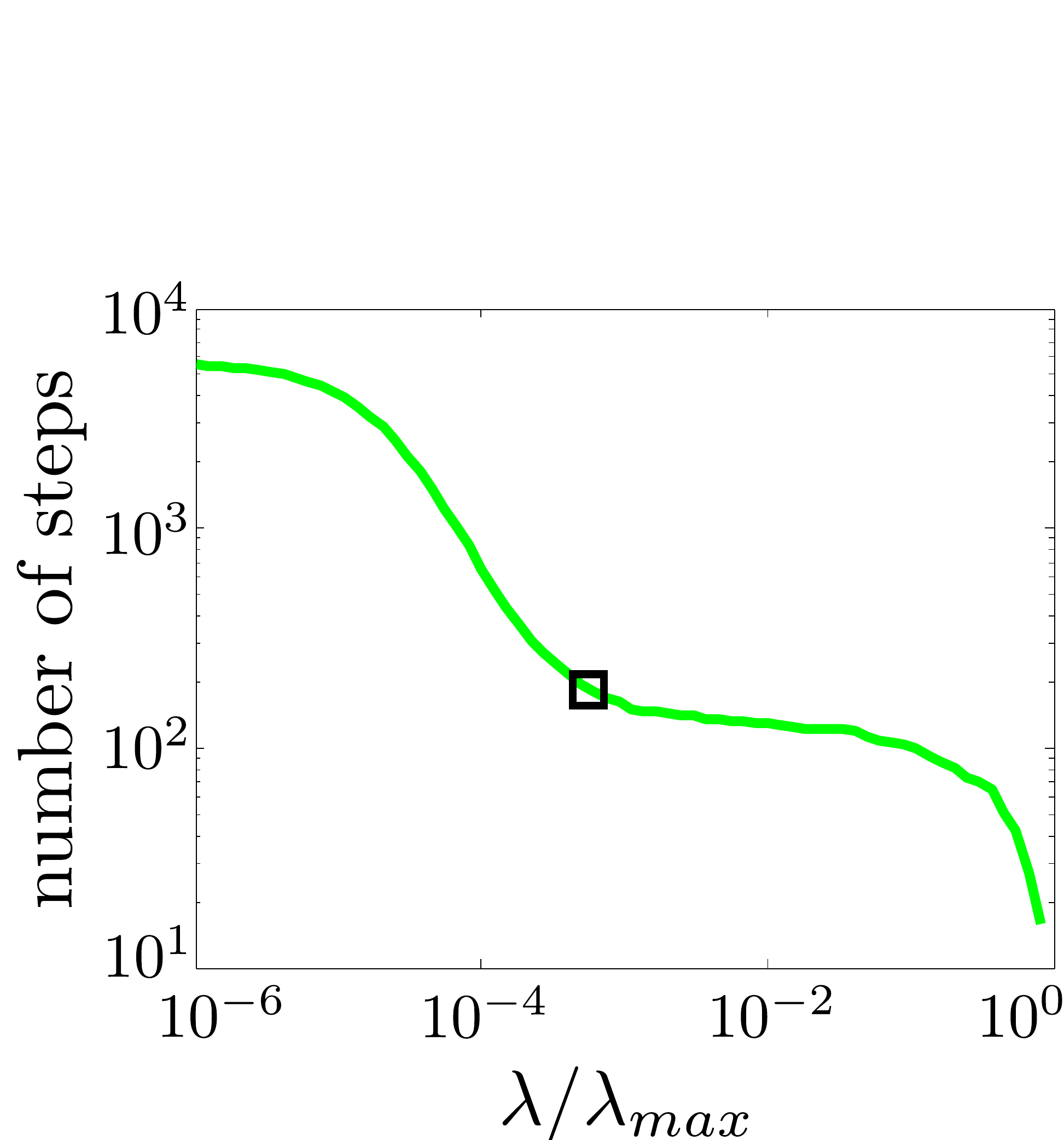}
		    \label{fig:phi29example2}
	   }
     \subfloat[] {
        \includegraphics[width=2.1in]{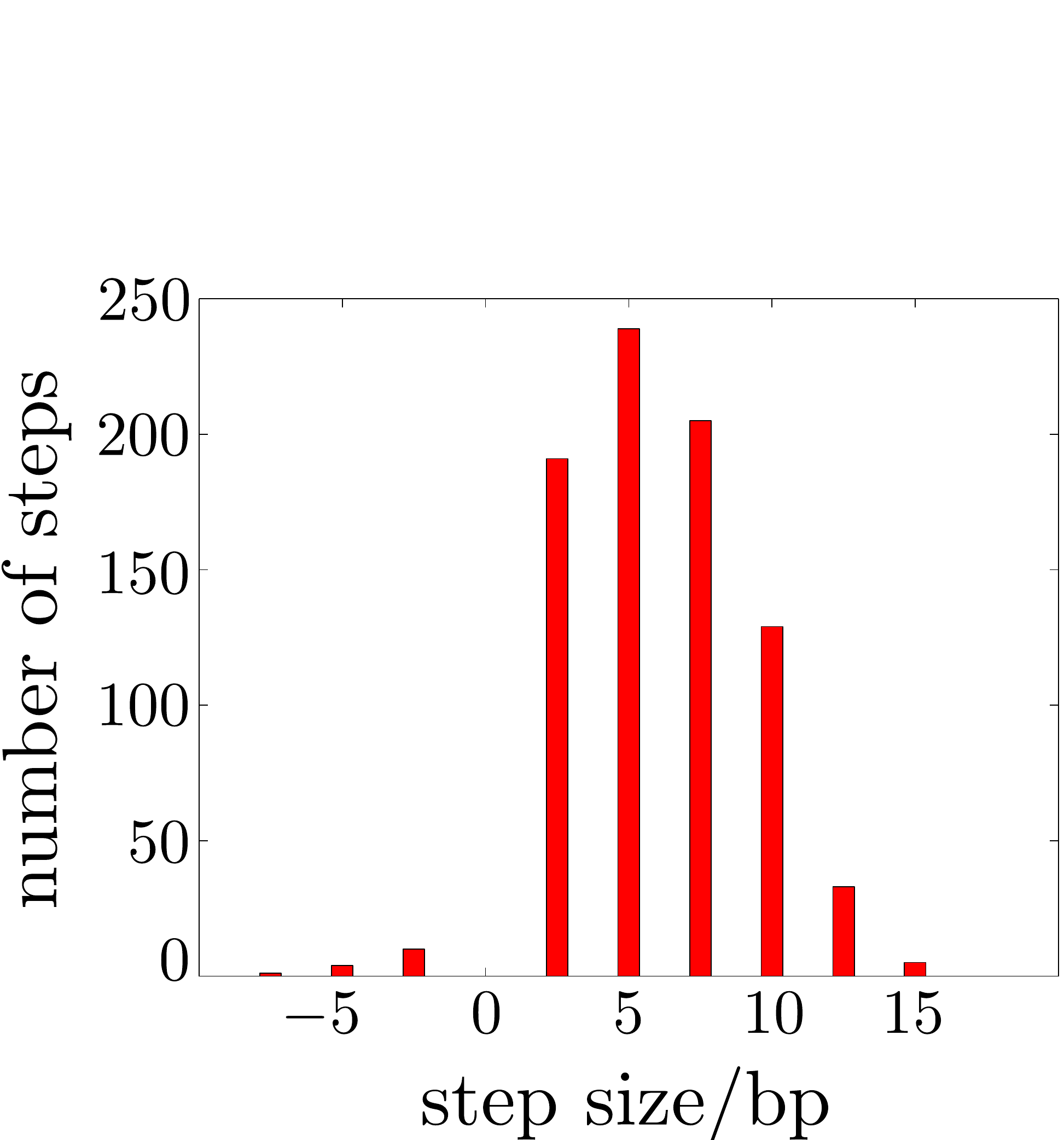}
		    \label{fig:phi29example3}
	    } 	
  \caption{Detected steps and intermediate results of EBS application to the
    $\varphi 29$ example (Results \& Discussion). Shown is the noisy data (grey
    circles, \ref{fig:phi29example1}), the corresponding
    $\lambda$-characteristic curve (green curve, Fig. \ref{fig:phi29example2}),
    TVDN with respect to optimal $\lambda_{h}$  (black box,
    \ref{fig:phi29example2}) and red signal, \ref{fig:phi29example1})), level
    grid for CC (black horizontal lines, \ref{fig:phi29example1})) and steps
    detected by EBS (blue signal, \ref{fig:phi29example1})).
    \ref{fig:phi29example3} shows the sum of detected step sizes of a set of 40
  $\varphi 29$ measurements.} \label{fig:phi29example}
\end{figure*}

The noisy $2.5kHz$ recording of a $\varphi 29$ bacteriophage (figure
\ref{fig:phi29example}) has the characteristic dependency of the number of
denoised steps on the $\lambda$ regularization parameter of TVDN (figure
\ref{fig:phi29example2}) and algorithm \ref{algo:lambda_heuristic} finds the
regularization parameter for optimal denoising, $\lambda_{h}$. Based on the TVDN
result (red signal, figure \ref{fig:phi29example1}) and the prior information,
that the $\varphi 29$ bacteriophage performs substeps of $2.5 bp$
\cite{phagettest}, a level grid is formed (black lines, figure
\ref{fig:phi29example1}). After combinatorial clustering the detected step
signal is obtained (blue signal, figure \ref{fig:phi29example1}).

\subsection*{Application of EBS to find pauses in experimental transcription data}

\begin{figure}[t!]
  \centering{
    \includegraphics[width=3.5in]{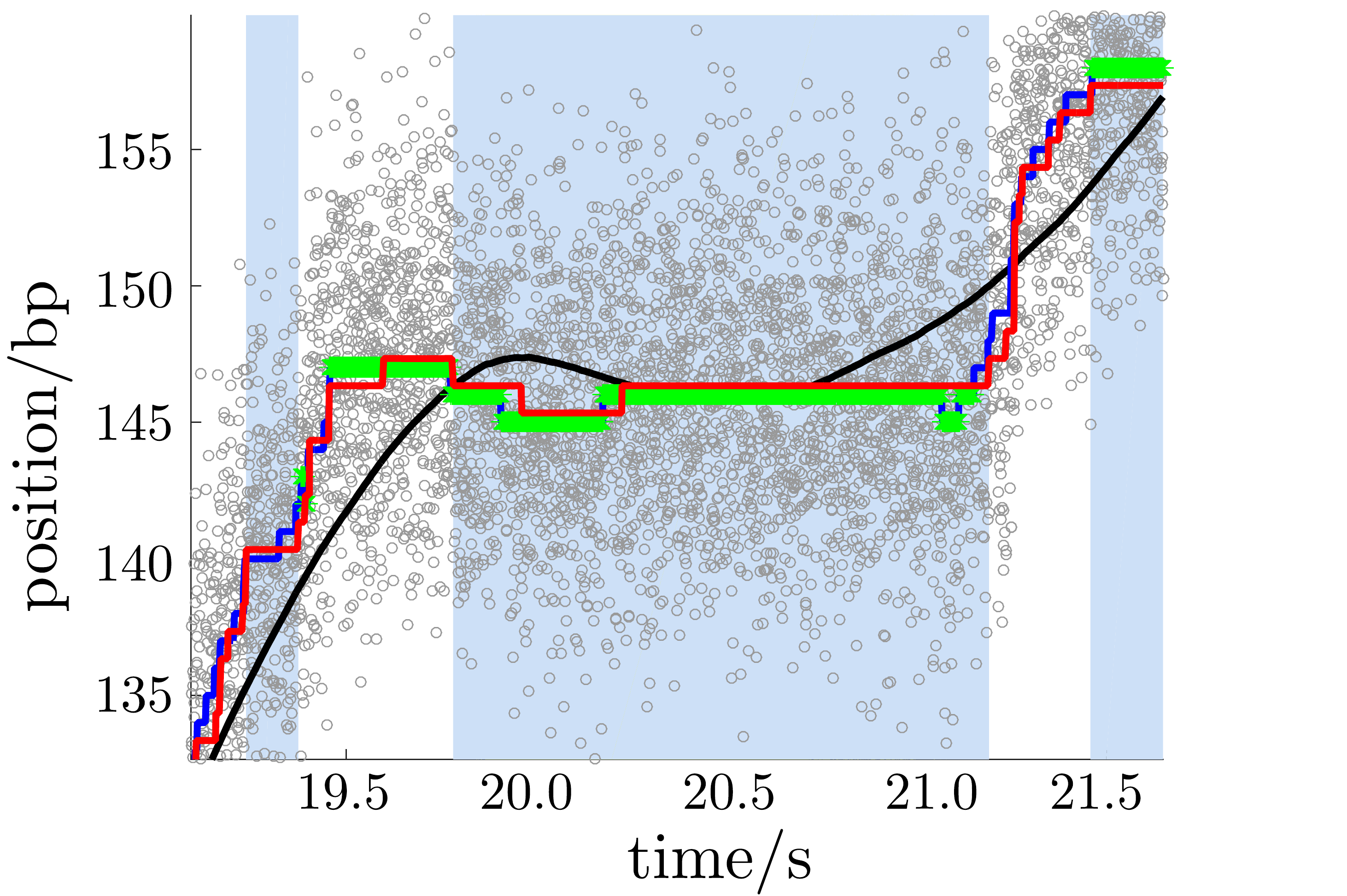}
  }
  \caption{Backtracked pause detection in simulated data. Shown is the noisy
  input signal (circles), the simulated step signal (blue), the Savitzky-Golay filtered
signal (black) and the detected step signal from EBS (red). Pauses in simulated
data are highlighted in green and paused regions in step detected data are
indicated by the blue shaded areas.} \label{fig:pausesOnSim}
\end{figure}

In the following we discuss the determination of pauses in experimental Pol II data as an example of further post processing of the detected steps and compare EBS based pause finding and SGVT on simulated data.\\
For the simulated Pol II steps dwell times are assigned to a pause when they lead to a backward step. The corresponding pause ends when a forward step brings Pol II back to the elongation state. For the detected steps this criterion also applies, however unlikely long dwells are also considered as pauses, since the algorithm will not perfectly find all steps present. Given the limited bandwidth ($1kHz$), high speed (saturating NTP concentration) of the enzyme and noise (standard deviation $\sim 10bp$) in the traces step detection performance should be similar to the fast scenario in our algorithm comparison. One can expect that mostly very fast steps are lost (figure \ref{fig:tvdnebsexamplemissingsteps}), i.e. fused to large steps. On the other hand that means that also short backtracks are likely to be skipped and instead a longer dwell time between two forward steps is returned by the algorithm.
Nevertheless, these longer dwells can be identified based on statistical hypothesis testing. Assuming that dwell times of forward stepping $\langle \tau_{forward} \rangle$ follow an exponential waiting time distribution, we calculate the mean dwell time of forward steps to estimate the probability distribution.
\begin{equation}
\langle \tau_{elongation} \rangle \sim \langle \tau_{forward} \rangle = \frac{1}{N} \sum_{i=1}^{N} \tau_{forward}
\end{equation}
Since not all backtracked pauses are discovered, this estimate of $\langle \tau_{elongation} \rangle$ also contains longer dwell times at a skipped pause. Thus $\langle \tau_{forward} \rangle$ can be larger than $\langle \tau_{elongation} \rangle$ and should be taken as an upper bound for the actual mean waiting time.\\
Furthermore we assume that forward steps obey an exponential distribution of the following form:
\begin{equation}
p(\tau) = \frac{1}{\langle \tau \rangle} exp(-\tau/{\langle \tau \rangle})
\end{equation}
Under these assumptions we can define a confidence level to discriminate between normal dwell times of elongation and unlikely long dwell times which are caused by undetected backtracks.\\
The confidence level can be adjusted by comparing recovered pauses to simulated backtracked pauses. A good compromise is found when most of the pauses are recovered and none or only very few of them are wrongly found.\\
To this end we simulated $10$ data sets with stepping rates and sampling frequency of the fast scenario and a computed noise amplitude of $\sim 6bp$. The simulated data is processed by EBS and the paused regions are identified according to the criterion described above. We also identify paused regions by SGVT with a threshold of two standard deviations of the pause peak as described in the methods section. SGVT sometimes  returns very short pauses which are not related to simulated ones and are presumably caused by high noise affecting the filtered signal. Thus we exclude pauses smaller than $10 ms$ in the SGVT analysis. Pauses found by EBS were always larger than $10 ms$ and thus there was no need for such an additional post-processing step. For each detected pause we identify if it is a correctly found one by checking whether it coincides with a simulated pause. We also take into account that, either two detected pauses  which are close but separated could overlap with a simulated pause, or that a single detected pause could cover two very close but separated simulated ones. Having identified how many pauses are correct, we can compute the recall  (i.e. the number of correctly found pauses divided by the number of simulated pauses), the precision (i.e. the number of correctly found pauses divided by the number of found pauses) and the false discovery rate (FDR, i.e. number of wrongly found pauses divided by the number of found pauses).
Moreover we compare the total cumulated length of all detected pauses to the total cumulated length of simulated ones. This value is relevant since a correct determination of the total length of pauses is important for determining pause-free velocities which are computed by excluding the paused intervals from the measured data. 
Table \ref{tab:overviewSGvsEBS} shows mean and standard deviation of recall, precision, FDR and total length for long ($t>t_{p}$) and short pauses ($t<t_{p}$) where the threshold for determining a long pause is $t_{p} = 0.8s$ (Results \& Discussion). Although in the fast scenario step detection performance is inappropriate for further dwell time analysis, finding pauses still works well, fig.(\ref{fig:pausesOnSim}) and table \ref{tab:overviewSGvsEBS}.\\

\begin{table}[!bht]
\begin{center}
\caption{Detection of short and long backtracks in simulated data by EBS and Savitzky-Golay (SG) filter. Shown is the number of correctly detected backtracks divided by the number of simulated backtracks (recall), the number of correctly detected backtracks divided by the number of found backtracked regions (precision) and the false discovery rate (FDR, number of false positives divided by number of found backtracked regions). Moreover, the total length of detected backtacks  divided by the total length of simulated backtracks is given. Backtracks with a detected duration $<10ms$ were excluded. The uncertainties for recall, precision and FDR are SEM.}
\label{tab:overviewSGvsEBS}
\begin{tabular}{lcccc}\hline
 & recall/$\%$ &  precision/$\%$ & FDR/$\%$ & total length/$\%$  \\ \hline
 \hline
short pauses: &  &  & & \\ \hline
SG filter & $38 \pm 7$  & $57 \pm 7$  & $43 \pm 8$  &  $70$\\
EBS & $61 \pm 4$  & $98 \pm 2$ & $8 \pm 2$ &  $91$ \\  \hline
long pauses: &  &  & & \\ \hline
SG filter & $98 \pm 1$  & $100$ & $0$  &  $94$ \\
EBS & $100$  & $100$ &  $0$ &  $113$ \\  \hline
\end{tabular}
\end{center}
\end{table}

\end{document}